\documentclass[preprint,sort&compress,12pt]{elsarticle}
\usepackage{comment}

\usepackage{xcolor}
 
\usepackage[algo2e]{algorithm2e}
\usepackage{amssymb}
\usepackage{amsthm}
\usepackage{amsmath}
\usepackage{mathrsfs}
\usepackage{mathtools}
\usepackage{textcomp}
\usepackage{gensymb}
\usepackage{graphicx}
\usepackage{subfig}
\usepackage{float}
\usepackage{caption}
\usepackage{tabu}
\usepackage{algorithm}
\usepackage{algpseudocode}
\usepackage{multirow}
\usepackage{enumerate}
\usepackage{array}

\usepackage[flushleft]{threeparttable}
\usepackage{lineno}
\usepackage{fullpage}
\usepackage{xcolor}
\usepackage{todonotes}
\usepackage{bbm}
\usepackage{listings}
\usepackage[colorlinks=true]{hyperref}
\usepackage{url}
\usepackage{siunitx}
\graphicspath{ {./figs/} }
\biboptions{comma,square}

\graphicspath{ {./figs/} }

\journal{Elsevier}

\newtheorem{theorem}{Theorem}
\theoremstyle{definition}

\theoremstyle{remark}
\newtheorem*{remark}{Remark}
 
\begin{document}
\begin{frontmatter}

\title{A Bi-fidelity Surrogate Modeling Approach for Uncertainty Propagation in Three-Dimensional Hemodynamic Simulations}



\author[ndAME,ndCICS]{Han Gao}
\author[iowa]{Xueyu Zhu\corref{1stcor}}
\author[ndAME,ndCICS]{Jian-Xun Wang\corref{1stcor}}

\address[ndAME]{Department of Aerospace and Mechanical Engineering, University of Notre Dame, Notre Dame, IN}
\address[ndCICS]{Center for Informatics and Computational Science, University of Notre Dame, Notre Dame, IN}
\address[iowa]{Department of Mathematics, University of Iowa, Iowa City, IA}
\cortext[1stcor]{Corresponding authors.\ \texttt{jwang33@nd.edu} (J.-X. Wang), \texttt{xueyu-zhu@uiowa.edu} (X. Zhu)}

\begin{abstract}
Image-based computational fluid dynamics (CFD) modeling enables derivation of hemodynamic information (e.g., flow field, wall shear stress, and pressure distribution), which has become a paradigm in cardiovascular research and healthcare. Nonetheless, the predictive accuracy largely depends on precisely specified boundary conditions and model parameters, which, however, are usually uncertain (or unknown) in most patient-specific cases. Quantifying the uncertainties in model predictions due to input randomness can provide predictive confidence and is critical to promote the transition of CFD modeling in clinical applications. In the meantime, forward propagation of input uncertainties often involves numerous expensive CFD simulations, which is computationally prohibitive in most practical scenarios. This paper presents an efficient bi-fidelity surrogate modeling framework for uncertainty quantification (UQ) in cardiovascular simulations, by leveraging the accuracy of high-fidelity models and efficiency of low-fidelity models. Contrary to most data-fit surrogate models with several scalar quantities of interest, this work aims to provide high-resolution, full-field predictions (e.g., velocity and pressure fields). Moreover, a novel empirical error bound estimation approach is introduced to evaluate the performance of the surrogate \emph{a priori}. The proposed framework is tested on a number of vascular flows with both standardized and patient-specific vessel geometries, and different combinations of high- and low-fidelity models are investigated. The results show that the bi-fidelity approach can achieve high predictive accuracy with a significant reduction of computational cost, exhibiting its merit and effectiveness. Particularly, the uncertainties from a high-dimensional input space can be accurately propagated to clinically relevant quantities of interest (e.g., wall shear stress) in the patient-specific case using only a limited number of high-fidelity simulations, suggesting a good potential in practical clinical applications.
\end{abstract}

\begin{keyword}
Uncertainty quantification \sep Cardiovascular simulation \sep Multi-fidelity \sep Random field \sep Hemodynamics \sep Surrogate modeling  
\end{keyword}
\end{frontmatter}


\section{Introduction}
\label{sec:intro}
Cardiovascular diseases (e.g., heart failure, stroke, vascular aneurysm) are the first leading cause of death and morbidity in the U.S., which poses a major healthcare concern~\cite{mozaffarian2015executive}. Hemodynamics information (e.g., blood flow velocity, pressure gradient, wall shear stress, viscous dissipation) can be used to improve diagnosis, treatment planning, and fundamental understanding of cardiovascular (patho)physiology. Such functional information is commonly obtained from computational fluid dynamics (CFD) simulations based on medical images, e.g., computed tomography (CT) and magnetic resonance imaging (MRI)~\cite{steinman2005flow}. However, the reliability of simulated hemodynamics largely depends on the boundary conditions and physiological/material parameters specified in the image-based CFD model, including inflow/outflow conditions, domain geometry, and mechanical properties, etc., which are usually uncertain or even unknown~\cite{steinman2018special,quarteroni2017cardiovascular}. For example, the patient-specific vessel geometry and inflow conditions typically extracted from anatomical images and flow imaging data (e.g., phase-contrast MRI) often contain large variations due to measurement noises and operator-based errors from the segmentation process. A recent international aneurysm CFD challenge showed that a wide variability exists in the model predictions of intracranial aneurysm wall shear stress from 26 participating teams, who were only provided with source three-dimensional (3-D) anatomical images~\cite{valen2018real,steinman2018special}. Moreover, some model parameters such as blood viscosity, vessel stiffness, and resistance of downstream vasculature are difficult or even impossible to measure. Rigorous assessment of confidence in model output predictions by considering the aleatory uncertainty (e.g., intrinsic randomness) and epistemic uncertainty (e.g., inter-patient or pathophysiological variations) in model input conditions is critical to push forward clinical translational applications of computational hemodynamics modeling. 

Uncertainty quantification (UQ) and sensitivity analysis (SA) of cardiovascular modeling have been gaining increasing attention in the past decade. There are numerous literature on examining the influence of variations in inflow/outflow boundary conditions~\cite{moyle2006inlet,marzo2009influence,karmonik2010temporal,troianowski2011three,jansen2014generalized,morales2015unraveling,sarrami2016uncertainty,brault2017uncertainty,bozzi2017uncertainty,bruening2018impact,madhavan2018effect,pirola2018computational,boccadifuoco2018impact}, segmented vascular geometry~\cite{moore1997computational,moore1999accuracy,cebral2005efficient,castro2006computational,hoi2006validation,omodaka2012influence,sankaran2015fast,sankaran2016uncertainty,nolte2019reducing,bruning2018uncertainty}, and mechanical properties of blood flow or vessel walls~\cite{lee2007relative,eck2017effects,pereira2013uncertainty,biehler2017probabilistic, steinman2013variability} on the simulated hemodynamics. However, the majority of these works focused on investigating the sensitivity of the model to its input factors using \emph{ad hoc} perturbation analysis but have yet to rigorously characterize and quantify the uncertainty distributions. One of the main challenges lies in the forward propagation of input uncertainty to model predictions, because this process usually requires a large number of repeated forward 3-D CFD simulations, which is computationally prohibitive for most non-trivial cases, particularly when considering complex (e.g., patient-specific) geometry or fluid-structure interaction (FSI) \cite{yu2013generalized, yu2016fractional}. For examples, S. Bozzi et al.~\cite{bozzi2017uncertainty} studied the effects of inflow variations in a 3-D ascending aorta model, where only 100 Monte Carlo samples were drawn (not sufficient enough to obtain converged statistics), since each model run took 56 hours on a cluster with 16 CPU cores and to conduct sufficient Monte Carlo simulations is nearly impossible. To tackle this challenge, people usually resort to surrogate modeling strategy, where a cost-effective emulator is built to replace the expensive CFD model to facility many-query applications. Surrogate models can be classified into two categories: (i) projection-based reduced-order model (ROM) and (ii) data-fit model~\cite{robinson2008surrogate,benner2015survey}. 

The essence of the projection-based ROM is to project the full-order governing partial differential equations (PDE), e.g., 3-D Navier-Stokes (NS) equations, onto a reduced subspace spanned by a group of basis functions, which can either be data-based basis such as proper orthogonal decomposition (POD) modes~\cite{benner2017model} or dictionary-based basis including polynomials~\cite{xiu2009efficient}, wavelets~\cite{le2004uncertainty}, and radial basis functions~\cite{bond2015galerkin}. It is expected that the reduced system after projection can be solved more efficiently. Note that the ROM here does not refer to reduced-dimension hemodynamic models such as one-dimensional (1D) models or lumped parameter (LP) models and this work is focused on 3-D full-field hemodynamic modeling. Manzoni et al.~\cite{manzoni2012model} developed a ROM using the reduced basis (RB) aimed at real-time blood flow simulations, and they also extended the framework to solve inverse problems in hemodynamics~\cite{lassila2013reduced}. Ballarin and Rozza~\cite{ballarin2016pod} proposed a monolithic ROM for parameterized FSI problems using POD-Galerkin projection and they applied the ROM to facilitate hemodynamics analysis in a 3-D patient-specific configuration of coronary artery bypass grafts~\cite{ballarin2016fast}. Chen et al.~\cite{chen2017reduced} have discussed the potential of using projection-based ROM for UQ applications. Nonetheless, projection-based ROM for parametric systems has emerged only recently~\cite{benner2015survey} and is far from mature for realistic hemodynamic applications due to its remaining challenges~\cite{lassila2014model,benner2017model}. Firstly, the stability and robustness issues are severe for hemodynamic systems with highly nonlinear behavior, large geometry variations, and turbulence complexity~\cite{lassila2014model}. How to improve stability is still an active research area~\cite{peherstorfer2018model,chaturantabut2010nonlinear}. Moreover, the speedup potential of the standard Galerkin-based ROM is largely limited when strong nonlinearity exists~\cite{chaturantabut2010nonlinear,peherstorfer2018model}. Furthermore, projection-based methods are code-intrusive, which poses great challenges to leveraging existing comprehensive hemodynamic solvers, e.g., SimVascular~\cite{updegrove2017simvascular}. 

As an alternative, data-fit surrogate models aim to build an empirical approximation of the full-order model using supervised learning from the full-fidelity simulation data (i.e., training data) at selected collocation points in the parameter space, which is thus non-intrusive to the code. There are many different ways to construct a data-fit approximation, including Gaussian process (GP)~\cite{kennedy2000predicting,atkinson2019structured}, radial basis~\cite{regis2007stochastic}, neural networks~\cite{sun2019surrogate,zhu2018bayesian,tripathy2018deep}, and polynomial chaos expansion (PCE)~\cite{xiu2002wiener, huan2013simulation, huan2014gradient, lei2015constructing}, among others. Data-fit surrogate models are preferable in hemodynamics modeling due to their non-intrusive nature and several prior studies have begun to emerge in the past a few years. Sankaran and Marsden developed an efficient forward and inverse UQ framework for 3-D blood flow simulations based on generalized PCE by sparse grid stochastic collocation methods~\cite{sankaran2010impact,sankaran2011stochastic}, and this approach has been applied for UQ analysis in various cardiovascular applications, including arterial growth and remodeling computations~\cite{sankaran2013efficient}, coronary blood flows~\cite{sankaran2016uncertainty,tran2019uncertainty}, single ventricle palliation~\cite{schiavazzi2016uncertainty}, and ascending thoracic aortic aneurysms~\cite{boccadifuoco2018impact}. However, the number of training samples required increases significantly when a relatively high-dimensional stochastic space is considered (even with adaptive sparse grid algorithms, e.g., Smolyak grid~\cite{ma2009adaptive}). Although several remedies have been proposed, e.g., a multi-resolution expansion strategy by partitioning the stochastic space~\cite{schiavazzi2017generalized} or using machine learning to accelerate the statistical convergence~\cite{sankaran2015impact}, the minimum number of required high-fidelity simulations is still beyond practical feasibility, particularly for high-dimensional UQ problems.

One promising strategy of efficient use of the computational budget for training is to combine the models with varying levels of accuracy and cost, which is known as a multi-fidelity method. Most of the efforts on developing multi-fidelity surrogate have been made from the statistical points of view, including GP-based multi-model approach (e.g., multi-fidelity co-Kriging model)~\cite{kennedy2001bayesian,huang2006sequential,le2013bayesian} and multi-level/multi-fidelity Monte Carlo (MLMC/MFMC) method \cite{giles2015multilevel,peherstorfer2018multifidelity}. The idea of using multi-fidelity models to facilitate UQ analysis in hemodynamics has been explored most recently. Biehler et al.~\cite{biehler2015towards,biehler2017probabilistic} developed an efficient UQ framework using a GP-based multi-fidelity scheme similar to Kennedy and O'Hagan's formulation~\cite{kennedy2001bayesian}, where the correction function from low- to high- fidelity solutions is approximated by a GP surrogate. Fleeter et al.~\cite{fleeter2019multi} started to exploit a stochastic framework that leverages widely-used reduced-dimension hemodynamic models (i.e., 1D and LP models) combined with 3-D high-fidelity model to formulate multi-level and multi-fidelity Monte Carlo estimators. Their results have shown promise towards efficient uncertainty propagation in large-scale hemodynamic problems.

In this work, we will develop a novel bi-fidelity UQ framework for 3-D hemodynamic simulations based on a recently proposed multi-fidelity stochastic collocation scheme~\cite{narayan2014stochastic,zhu2014computational,zhu2017multi}, aiming to efficiently reconstruct 3-D full-field hemodynamics information in a high-dimensional parametric setting. Contrary to prior multi-fidelity approaches, the low-fidelity model will be used to not only inform the global searches over the parameter space but also help the high-fidelity reconstruction. Moreover, a practical error bound estimation approach is proposed to assess the surrogate \emph{a priori}. The performance of the proposed methods are evaluated on a number of cardiovascular flow cases with both standardized and patient-specific arterial geometries, and different combinations of high- and low- fidelity models for hemodynamics are also discussed. This study focuses on the inflow uncertainty, including uncertain inflow rate, flow-split, and secondary flow patterns, modeled as spatial random fields.

The rest of the paper is organized as follows. The methodology and algorithm of the bi-fidelity surrogate modeling framework for uncertainty propagation in 3-D hemodynamics simulations are introduced in Section~\ref{sec:meth}. Several cardiovascular flow cases are investigated in Section~\ref{sec:result} to evaluate the performance of the proposed method, with regard to both accuracy and cost. The empirical error bound estimation of the bi-fidelity surrogate model is discussed in Section~\ref{sec:discussion}. Finally, Section~\ref{sec:conclusion} concludes the paper.

\section{Methodology}
\label{sec:meth}
\subsection{Problem formulation}
Blood flow in the cardiovascular system can be modeled using the steady incompressible Navier-Stokes equations under assumptions of rigid walls and Newtonian fluids. Here, we consider the following parameter-dependent formulation,
\begin{equation}
    \label{eq:ns}
     \left \{
    \begin{aligned}
    &\nabla \cdot \mathbf{u} (\mathbf{x},\mathbf{z}) = 0,  &\mathbf{x} \in \Omega_f,\quad &\mathbf{z} \in I_z,\\
    &\mathbf{u}(\mathbf{x},\mathbf{z})\cdot\nabla\mathbf{u}(\mathbf{x},\mathbf{z}) + \frac{1}{\rho}\nabla p - \nu\nabla^2\mathbf{u}(\mathbf{x},\mathbf{z}) + \mathbf{b}_f = 0, \quad   &\mathbf{x} \in \Omega_f,\quad  &\mathbf{z} \in I_z,
    \end{aligned} \right .
\end{equation}
where $\mathbf{x}$ is spatial coordinates in the 3-D fluid domain $\Omega_f \subseteq \mathbb{R}^3$, and $\mathbf{z}$ represents input parameters of the system, including parameters of inflow/outflow boundary conditions, geometry of the domain, and mechanical/material properties, etc. Note that the fluid density $\rho$ and viscosity $\nu$ can also belong to $\mathbf{z}$, although they are written explicitly in \eqref{eq:ns}. $I_z\subseteq \mathbb{R}^d$ denotes a $d$-dimensional parameter space, and $\mathbf{b}_f$ is the body force. The flow solutions, velocity $\mathbf{u}(\mathbf{x}, \mathbf{z})$ and pressure $p(\mathbf{x}, \mathbf{z})$, are functions of space and parameters, which can be uniquely determined with specified boundary conditions,
\begin{equation}
    \label{eq:bc}
    \mathcal{B}(p, \mathbf{u}, \mathbf{z}) = 0,\quad \mathbf{x}\in \partial\Omega_f,\quad \mathbf{z} \in I_z ,
\end{equation}
where $\mathcal{B}$ is a boundary operator and $\partial\Omega_f\subseteq \mathbb{R}^3$ represents the boundary region, which is time invariant under the rigid wall assumption. 

For a set of fixed input parameters $\mathbf{z}$ (i.e., one realization), the flow fields can be solved deterministically using mesh-based numerical discretization techniques, e.g., finite volume or finite element methods. To obtain accurate numerical solutions, a large-scale 3-D mesh (fine mesh) and sufficient numerical iterations (or time steps) are needed to spatially resolve the flow field and to achieve fully-converged solutions. This process can be seen as a \emph{high-fidelity} (HF) simulation, which is, however, computationally expensive and usually requires super-computing clusters. For UQ tasks, where input parameters are modeled as a finite-dimensional random variable $\mathbf{z}$ with a joint probability distribution density $P(\mathbf{z})$, forward propagation of the input uncertainty through the HF model necessitates numerous repeated model runs (e.g., for stochastic collocation methods or Monte Carlo sampling), which are often computationally prohibitive. On the other hand, various \emph{low-fidelity} (LF) models are available in hemodynamic computations, built by, e.g., reducing time-stepping, coarsening spatial/temporal discretization, or simplifying the physics (2-D, 1D, LP models), where the computational cost can be largely reduced by sacrificing the predictive accuracy. As mentioned above, a proper combination of the HF and LF simulations can potentially lead to a more efficient surrogate of the hemodynamic model by leveraging the efficiency of LF model and accuracy of HF model. This work investigates a bi-fidelity surrogate modeling strategy in the context of uncertainty propagation in hemodynamics, where input uncertainties are assumed to be adequately characterized and quantitatively represented with known probability distributions or random fields. Throughout this paper, we let $\mathbf{v}^H$ represent the HF hemodynamic solution, which is expensive to compute. Similarly, the LF CFD solution is represented by $\mathbf{v}^L$, which is cheap to compute.

\subsection{Bi-fidelity Surrogate Construction}
\subsubsection{Overview}
\begin{figure}[h]
    \centering
    {\includegraphics[width=1\textwidth]{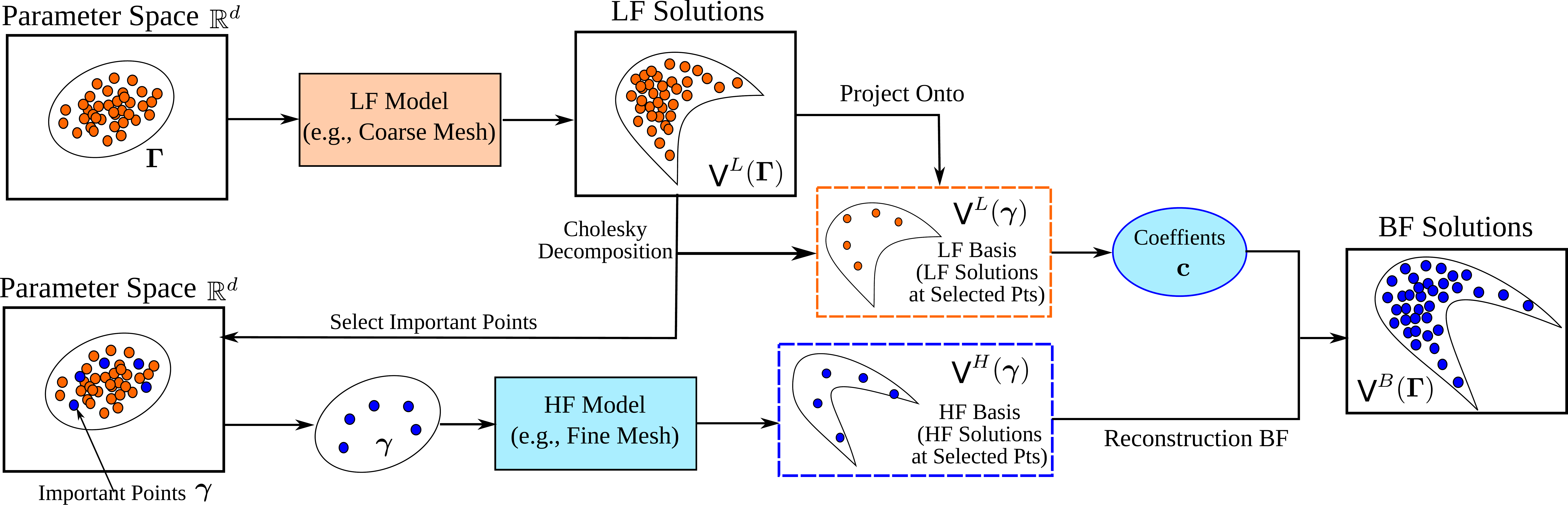}}
    \caption{Schematic diagram of the bi-fidelity surrogate construction in the context of uncertainty propagation.}
    \label{fig:sche}\vspace{-1.2em}
\end{figure}

The surrogate model is constructed based on both high- and low- fidelity solutions following a multi-fidelity stochastic collocation strategy proposed in~\cite{narayan2014stochastic,zhu2014computational}, where a low-rank approximation of the HF solutions is obtained with the assistance of LF simulations. Similar to other data-fit modeling approaches, the bi-fidelity surrogate is constructed based on an offline ``training" process. However, only a few HF solutions on a limited number of points selected from the parameter space $I_z$ are needed for ``training", and the selection of these ``important" points is informed by the cheaper LF simulations. After training, online evaluation of the bi-fidelity surrogate can be performed at any location of the parameter space solely based on the LF simulation, which could significantly reduce the computational cost, particularly when large numbers of online model queries are expected. The procedure for constructing the bi-fidelity surrogate is summarized as follows:
\vspace{-1.0em}
\begin{enumerate}
\item Offline training with HF and LF simulations
\begin{enumerate}
    \item Conduct LF simulations on a parameter set of $\Gamma \subset I_z$ with a sufficient number of points to fully cover the parameter space $I_z$.
    \item Select a subset of $\gamma \subset \Gamma$ containing a handful of important points, on which both HF and LF solutions are obtained as the HF and LF basis functions, respectively.    
\end{enumerate}
\item Online surrogate solution approximation on new parameter points
\begin{enumerate}
    \item For a new point $\mathbf{z} \in I_z$, conduct LF simulation, and the solution $\mathbf{v}^L(\mathbf{z})$ is projected onto the LF approximation space to obtain the coefficients $\mathbf{c}^L$.  
    \item Approximate the HF solution $\mathbf{v}^H(\mathbf{z})$ at the new parameter point $\mathbf{z}$ based on the HF solution basis and LF projection coefficients $\mathbf{c}^L$.  
\end{enumerate}
\end{enumerate}
A schematic of the bi-fidelity surrogate construction in the context of uncertainty propagation is shown in Fig.~\ref{fig:sche}.

\subsubsection{Offline training with selected high-fidelity simulations}
In the offline ``training" process, the first step is to explore the solution structure over the entire parameter space (or regions of interest) based on a large number ($M \gg 1$) of LF simulations, which are assumed to be cheap to conduct. Specifically, the LF model is run on a prescribed nodal set $\Gamma$ from the entire parameter space $I_z$, 
\begin{equation}
    \label{Point Selection}
    \Gamma = \{\mathbf{z}_1,...,\mathbf{z}_M\}\subset I_z.
\end{equation}
$\Gamma$ can be determined by sampling the characterized joint probabilistic density of parameters $\mathbf{z}$ using standard Monte Carlo method or more efficient sampling (or collocation) schemes, e.g., Latin hypercube sampling~\cite{stein1987large}, importance sampling~\cite{glynn1989importance}, or sparse grid collocation method~\cite{nobile2008sparse, ma2009adaptive}. The choice is not a restriction as long as the sampled points can sufficiently cover the regions of interest in $I_z$. On these sampled points, LF simulations are performed to obtain the LF snapshot matrix $V^L(\Gamma) = \left[\mathbf{v}^L(\mathbf{z}_1),...,\mathbf{v}^L(\mathbf{z}_M)\right]^T$.

The next key step is to select a subset of $m$ important points from $\Gamma$, i.e., 
\begin{equation}
    \gamma^m = \{\mathbf{z}_{1},...,\mathbf{z}_{m}\}\subset \Gamma,
\end{equation}
where the HF simulations will be performed and the corresponding HF snapshot matrix $V^H(\gamma^m) = \left[\mathbf{v}^H(\mathbf{z}_1),...,\mathbf{v}^H(\mathbf{z}_m)\right]^T$ is obtained. The size of $\gamma^m$ should be small ($\mathcal{O}(10) $) considering the high cost of the HF model. The solution snapshots on the $m$ selected important points can be seen as a low-rank approximation (i.e., basis) of the solution space, and the corresponding approximation spaces ($\mathbb{U}^H$ and $\mathbb{U}^L$) by the HF and LF basis functions are defined as, 
\begin{subequations}
    \begin{alignat}{2}
    \mathbb{U}^H(\gamma^m) = \mathrm{span}({V}^H(\gamma^m)) = \mathrm{span}\{\mathbf{v}^H(\mathbf{z}_1),...,\mathbf{v}^H(\mathbf{z}_m)\},\\
    \mathbb{U}^L(\gamma^m) = \mathrm{span}({V}^L(\gamma^m)) = \mathrm{span}\{\mathbf{v}^L(\mathbf{z}_1),...,\mathbf{v}^L(\mathbf{z}_m)\}.
    \end{alignat}
\end{subequations}
To select these important points, we followed a LF model informed strategy proposed in~\cite{narayan2014stochastic,zhu2014computational}, which is a greedy algorithm that iteratively adds a new node to the existing selected points such that the LF solution vector of the newly selected point is the furthest from the space spanned by the LF solutions of previously selected points in the parameter space. Namely, let $\gamma^{(k-1)} = \{\mathbf{z}_{1},...,\mathbf{z}_{k-1}\}$ be the existing subset of important points, and the next point will be selected by maximizing the distance between its LF solution and the existing subspace $\mathbb{U}^L(\gamma^{(k-1)})$ as,
\begin{subequations}
\label{eqn:maxmizer}
\begin{alignat}{2}
    \mathbf{z}_{k} = \arg\max_{\mathbf{z}\in\Gamma} \mathrm{d}\bigg(\mathbf{v}^L(\mathbf{z}), \mathbb{U}^L(\gamma^{(k-1)})\bigg), \\
    \gamma^{(k)} = \gamma^{(k-1)}\cup\{\mathbf{z}_k\},
\end{alignat}
\end{subequations}
where $d(\boldsymbol{v},W)$ is the distance function 
between the vector $\boldsymbol{v}\in \mathbf{v}^L(\Gamma)$ and subspace $W\subset \mathbb{U}^L(\Gamma)=\textrm{span}\{\mathbf{v}^L(\mathbf{z}_1),\cdots,\mathbf{v}^L(\mathbf{z}_M)\}$. This optimization can be accomplished by factorizing the Gramian matrix $G$ of LF solutions $V^L(\Gamma)$, and we choose the pivoted Cholesky decomposition as in~\cite{zhu2014computational},
\begin{equation}
    G = P^TLL^TP,
\end{equation}
where $L$ is a low-triangular matrix; $P$ is a permutation matrix such that $P\Gamma$ provides an order of ``importance", and the index corresponding to the first $m$ columns can be used to identify the important parameter points for HF simulations. Note that other factorization schemes, including the column-pivoting QR decomposition or full-pivoting LU decomposition, can also be used to achieve the same goal~\cite{narayan2014stochastic}. The Gramian matrix is defined by
\begin{equation}
    G = (G_{ij})_{1\leq i,j\leq M}, \ \ \ G_{i,j} = \langle\mathbf{v}^L(\mathbf{z}_i), \mathbf{v}^L(\mathbf{z}_j)\rangle^L,
\end{equation}
where $\langle \cdot,\cdot \rangle^L$ is the inner product defined in the LF solution space. In practice, we only need to compute the truncated Gramian matrix corresponding to the first $m$ important solution vectors. The implementation details of the important point selection approach are given in Algorithm~\ref{alg: pointselect}.
\begin{algorithm}[H]
\SetAlgoLined
\Begin{
 $V^L =[\mathbf{v}^L(\mathbf{z}_1),...,\mathbf{v}^L(\mathbf{z}_M)] $ \newline
 $\mathbf{d}[i] = (\mathbf{v}_i^L)^T\mathbf{v}_i^L$ for $i = 1,...,M$ \newline
 $P =$ \texttt{zeros}$(m,1)$, $L =$ \texttt{zeros}$(m,M)$\newline
 $k = 1$, 
 \While{$k \leq m$}{
  \textbf{1}. $P[k] =\arg\max \mathbf{d}[k:end]$;\newline
  \textbf{2}. Exchange $V^L[:,k]$ and $V^L[:,P[k]]$; Exchange $L[:,k]$ and $L[:,P[k]]$; Exchange $\mathbf{d}[k]$ and $\mathbf{d}[P[k]]$;\newline
  \textbf{3}. $\mathbf{r}(t) = (V^L(:,t))^TV^L(:,k) - \sum_{j=1}^{k-1}L(t,j)L(k,j)$ for $t=k+1,...,M$; \newline
  \textbf{4}. $L[k,k] = \sqrt{\mathbf{d}[k]}$;\newline
  \textbf{5}. $L[t,k] = \mathbf{r}[t]/L[k,k]$ for $t = k+1,...,M$;\newline
  \textbf{6}. $\mathbf{d}[t] = \mathbf{d}[t] - L^2[t,k]$ for $t = k+1,...,M$;\newline
  \textbf{7}. $k = k + 1$
 }
 $\gamma[:,t] = \Gamma[:,P[t]]$ for $t = 1,...,m$;\newline
 Form the truncated Gramian $G^L = LL^T$
}
\caption{Important Point Selection}
\label{alg: pointselect}
\end{algorithm}

\subsubsection{Online bi-fidelity construction for surrogate solutions}
Once the small subset of points $\gamma$ are selected, the HF and LF solutions on these important points (i.e., $V^H(\gamma^m)$ and $V^L(\gamma^m)$) can be used as the basis functions (i.e., low-rank approximation) to construct the HF and LF solution approximation spaces (i.e., $\mathbb{U}^H$ and $\mathbb{U}^L$), respectively. For a new point $\mathbf{z}\in I_z$, the HF and LF solutions can be reconstructed based on the corresponding basis. If we assume the reconstruction of HF and LF solutions share the same approximation rule, the approximated HF solution can be obtained at any desired location $\mathbf{z}$ by solely conducting the LF simulation. Specifically, we first simulate LF model on the new point $\mathbf{z}$ to obtain $\mathbf{v}^L(\mathbf{z})$, which is then projected onto the LF approximation space to obtain the projection coefficients $\mathbf{c}^L(\mathbf{z}) = [c_1,\hdots, c_m]^\top$. This can be simply achieved by the following equation:
\begin{equation}
    \mathbf{c}^L = G^{-1} (V^L(\gamma^m))^T\mathbf{v}^L(\mathbf{z}).
\end{equation}
Since we assume the HF and LF reconstructions share the same reconstruction coefficients, thus the HF solution at $\mathbf{z}$ can be approximated as follows:
\begin{equation}
    \mathbf{v}^H(\mathbf{z}) \approx \mathbf{v}^B(\mathbf{z}) = \sum_{k=1}^m c_k \mathbf{v}^H(\mathbf{z}_{k}),
\end{equation}
where ${\mathbf{v}}^B(\mathbf{z})$ is the bi-fidelity surrogate solution.

\subsubsection{ An empirical error bound estimation}
\label{sec:errorEstimate}
For practical applications of the bi-fidelity (BF) approach, it is useful to answer the following two questions: (1) whether the quality of a given LF model is good enough to build a reasonably accurate BF approximation? (2) If the LF model is good enough, how many HF samples are sufficient to obtain satisfactory results? In other words, \emph{a priori} assessment of the model quality and prediction error is of practical importance. Practical estimation of the error bound of the BF approach was proposed in \cite{hampton2018practical}, where a number of additional HF samples are required. In this subsection, we adopt an empirical alternative with ease of implementation, which is motivated from the following observation:
\begin{theorem}
\label{theorem:1}
Given the first $k+1$ pre-selected important points $\gamma^{k+1}$, the relative error between the bi-fidelity solution and the high-fidelity solution can be bounded for any point $\mathbf{z}_*\in\Gamma$ as follows:
\begin{equation}
\label{eq:eb_t}
\begin{split}
    \frac{||\mathbf{v}^H(\mathbf{z}_*) - \mathbf{v}^B(\mathbf{z}_*)||}{||\mathbf{v}^H(\mathbf{z}_*)||}
&\leq
\underbrace{
\frac{d^H(\mathbf{v}^H(\mathbf{z}_*),\mathbb{U}^H(\gamma^{k})))}{||\mathbf{v}^H(\mathbf{z}_*)||}
}_{\mathrm{\textbf{relative distance}}}
+ 
\underbrace{
\frac{|| P_{\mathbb{U}^H({\gamma^k})}\mathbf{v}^H(\mathbf{z}_*) -  \mathbf{v}^B(\mathbf{z}_*)||}{||\mathbf{v}^H(\mathbf{z}_*)||}
}_{\mathrm{\textbf{in-plane error}}}\\
&= \frac{d^H(\mathbf{v}^H(\mathbf{z}_*),\mathbb{U}^H(\gamma^{k}))}{||\mathbf{v}^H(\mathbf{z}_*)||}(1+\frac{\frac{|| P_{\mathbb{U}^H({\gamma^k})}\mathbf{v}^H(\mathbf{z}_*) -  \mathbf{v}^B(\mathbf{z}_*)||}{||\mathbf{v}^H(\mathbf{z}_*)||}}{ \frac{d^H(\mathbf{v}^H(\mathbf{z}_*),\mathbb{U}^H(\gamma^{k})))}{||\mathbf{v}^H(\mathbf{z}_*)||}} ),
\end{split}
\end{equation}
\end{theorem}
\noindent where $P_{\mathbb{U}^H({\gamma^k})}$ is the projection operator onto the subspace $\mathbb{U}^H({\gamma^k})$ and $d^H$ is distance function, which is defined as $d^H = \mathbf{v}^H - P_{\mathbb{U}^H}\mathbf{v}^H$. The proof is rather trivial and omitted here. The above error bound is rigorous but less useful, because for any given $\mathbf{z}_*$, we need to have the high-fidelity data $\mathbf{v}^H(\mathbf{z}_*)$ available. To address this issue, it is useful to introduce $R_s$, the model similarity, defined by the relative distance as:
\begin{equation}
\label{eqn:relativeDist}
R_{s}(\mathbf{z}) = \frac{d^H(\mathbf{v}^H(\mathbf{z}),\mathbb{U}^H(\gamma^{k}))}{||\mathbf{v}^H(\mathbf{z})||}/\frac{d^L(\mathbf{v}^L(\mathbf{z}),\mathbb{U}^L(\gamma^{k}))}{||\mathbf{v}^L(\mathbf{z})||},
\end{equation}
which characterizes the similarity between the LF/HF models. $R_{s}\approx 1$ indicates that LF model is informative for the purpose of the point selection. By the definition of $R_s$, \eqref{eq:eb_t} becomes,
\begin{equation}
\label{eq:eb_a}
    \frac{||\mathbf{v}^H(\mathbf{z}_*) - \mathbf{v}^B(\mathbf{z}_*)||}{||\mathbf{v}^H(\mathbf{z}_*)||}
\leq \frac{d^L(\mathbf{v}^L(\mathbf{z}_*),\mathbb{U}^L(\gamma^{k}))}{||\mathbf{v}^L(\mathbf{z}_*)||} R_s(\mathbf{z}_*) \big[1+\frac{\frac{|| P_{\mathbb{U}^H({\gamma^k})}\mathbf{v}^H(\mathbf{z}_*) -  \mathbf{v}^B(\mathbf{z}_*)||}{||\mathbf{v}^H(\mathbf{z}_*)||}}{ \frac{d^H(\mathbf{v}^H(\mathbf{z}_*),\mathbb{U}^H(\gamma^{k}))}{||\mathbf{v}^H(\mathbf{z}_*)||}}\big].
\end{equation}
Now, the first term of the right-hand side depends on the corresponding LF data $u^L(\mathbf{z}_*)$. To remove the dependence of HF data $u^H(\mathbf{z}_*)$ in the second term of right-hand side in \eqref{eq:eb_a}, we propose to use $\mathbf{z}_{k+1} \in \gamma_{k+1}$ as the test point to serve as an error surrogate of the BF approximation (based on the first $k$ pre-selected parameter point) in the entire parameter space. We \emph{conjecture} that, if the LF and HF models are similar ($R_s\approx 1$), there are constants $c_1$ and $c_2$, such that for the first $k+1$ pre-selected important points $\gamma_{k+1}$,
\begin{equation}
\label{eq:eb1}
\begin{split}
    \frac{||\mathbf{v}^H(\mathbf{z}_*) - \mathbf{v}^B(\mathbf{z}_*)||}{||\mathbf{v}^H(\mathbf{z}_*)||}
\leq
\frac{d^L(\mathbf{v}^L(\mathbf{z}_*),\mathbb{U}^L(\gamma^{k}))}{||\mathbf{v}^L(\mathbf{z})||}\big[c_1+c_2\frac{\frac{|| P_{\mathbb{U}^H({\gamma^k})}\mathbf{v}^H(\mathbf{z}_{k+1}) -  \mathbf{v}^B(\mathbf{z}_{k+1})||}{||\mathbf{v}^H(\mathbf{z}_{k+1})||}}{ \frac{d^H(\mathbf{v}^H(\mathbf{z}_{k+1}),\mathbb{U}^H(\gamma^{k}))}{||\mathbf{v}^H(\mathbf{z}_{k+1})||}} \big].
\end{split}
\end{equation}
In such way, \eqref{eq:eb1} only requires the LF data and a finite number of the first $k+1$ pre-selected HF samples, if $c_1$ and $c_2$ are determined properly.

It can be seen that besides $R_s\approx1$ (the LF and HF should be similar), the approximation quality of the BF approximation also depends on $R_{e}$, {the balance between the in-plane error and the relative distance,}
\begin{equation}
\label{eqn:rError}
    R_{e}(\mathbf{z}) = \frac{|| P_{\mathbb{U}({\gamma^k})}\mathbf{v}^H(\mathbf{z}) -  \mathbf{v}^B(\mathbf{z})||}{d^H(\mathbf{z},\mathbb{U}^H(\gamma^{k}))}.
\end{equation}
A large $R_{e}$ indicates that the in-plane error is dominant over the distance error. In this case, it is suggested to stop collecting new HF samples. With this definition, \eqref{eq:eb1} becomes 
\begin{equation}
\label{eq:eb2}
    \frac{||\mathbf{v}^H(\mathbf{z}_*) - \mathbf{v}^B(\mathbf{z}_*)||}{||\mathbf{v}^H(\mathbf{z}_*)||}
\leq \frac{d^L(\mathbf{v}^L(\mathbf{z}_*),\mathbb{U}^L(\gamma^{k}))}{||\mathbf{v}^L(\mathbf{z}_*)||}(c_1+c_2R_{e}(\mathbf{z}_{k+1})).
\end{equation}
Numerical experiments we have conducted in the Section~\ref{sec:discussion} support our conjectures above and \emph{indicate} that when $c_1$ and $c_2$ are set to be 1, if $R_{s} \approx 1$ and $R_{e} < 10$,  the BF approximation can usually deliver good results (better than the low-fidelity solutions).
\begin{remark}
We acknowledge that the error bound estimation is not rigorous. Nevertheless, it is a useful quantity to gauge the quality of the BF approximation in practice. 
\end{remark}

\section{Numerical Results}
\label{sec:result}
In this section, we present three vascular flow cases with both standardized and patient-specific geometries, where different types of high/low-fidelity (HF/LF) model pairs are studied to demonstrate the applicability, efficiency, and flexibility of the bi-fidelity (BF) approach for surrogate hemodynamic modeling. Specifically, a stenotic flow with standardized geometry is studied in subsection~\ref{sec:case1}, where simulations with converged/unconverged solutions are designed as the HF/LF models; subsection~\ref{sec:case2} presents a flow case of an idealized bifurcation aneurysm, where the simulations with 3-D/2-D geometries are used as the HF/LF models; lastly, a patient-specific case of cerebral bifurcation aneurysm is investigated in subsection~\ref{sec:case3}, where an HF/LF model pair with fine/coarse meshes is used to build the BF surrogate. To evaluate the accuracy of the BF surrogate model, the following error metric is defined,
\begin{equation}
    \label{eqn:errmetric}
    \mathrm{Relative\ Error} = 
    \sqrt{\frac{\sum_{i=1}^N||\mathbf{v}^H(\mathbf{z}_i)- \mathbf{v}^B(\mathbf{z}_i)||_{L^2(\Omega_f)}}
    {\sum_{i=1}^N||\mathbf{v}^H(\mathbf{z}_i)||_{L^2(\Omega_f)}}},
\end{equation}
where $N$ is the number of test samples in parameter space; $\mathbf{v}^B$ and $\mathbf{v}^H$ are BF and HF solutions, respectively. In each case, the BF surrogate is applied to perform the forward propagation of inflow uncertainties, and the results are benchmarked against those from the HF-based Monte Carlo (MC) simulations.

In this study, all CFD simulations are conducted based on the open-source CFD platform, OpenFOAM. The continuity and momentum equations for incompressible laminar flows were solved using the SIMPLE (semi-implicit method for pressure-linked equations) algorithm~\cite{anderson2016computational}. Collocated grids were used and the Rhie and Chow interpolation was used to prevent the pressure--velocity decoupling~\cite{rhie1983numerical}. Spatial derivatives were discretized with the finite volume method using the second-order central scheme for both convection and diffusion terms. A second-order implicit time-integration scheme was used to discretize the temporal derivatives. All the mesh files were generated by using ANSYS ICEM software.

\subsection{Idealized Stenosis Model (Case 1)}
\label{sec:case1}
\begin{figure}[htb]
    \centering
    \subfloat[Coarse mesh for LF model]{\includegraphics[width=0.5\textwidth]{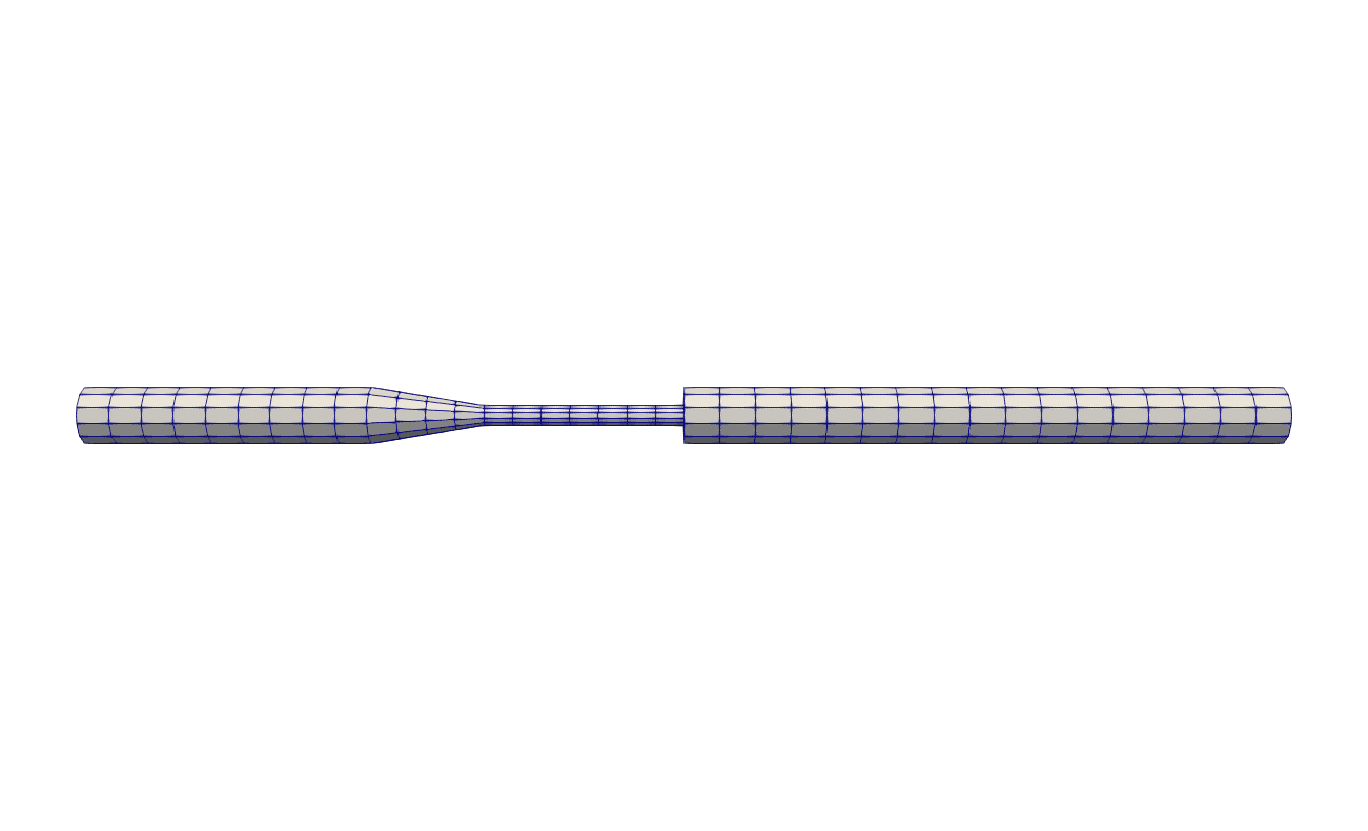}}
    \subfloat[Fine mesh for HF model] {\includegraphics[width=0.5\textwidth]{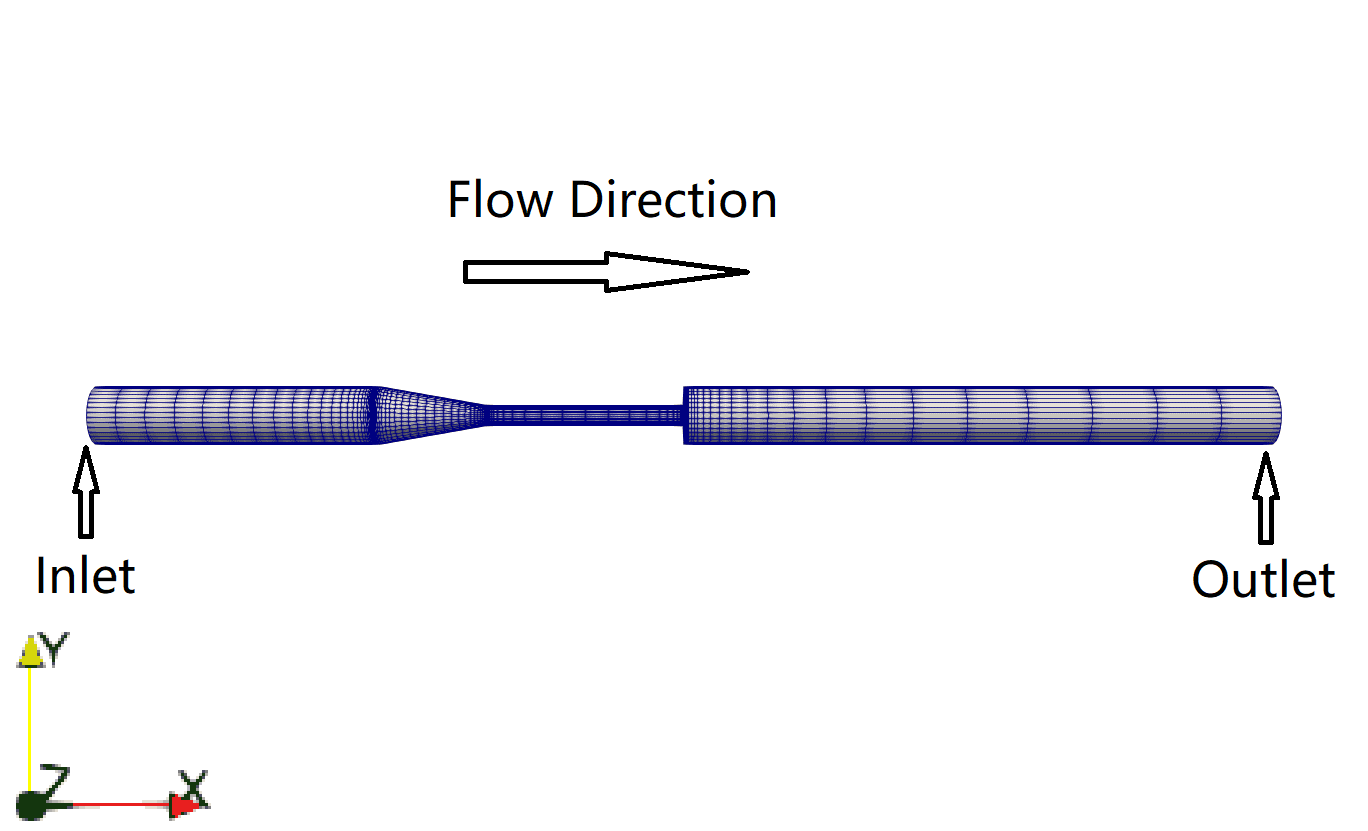}}
    \caption{The 3-D asymmetric nozzle with (a) coarse and (b) fine computational meshes.}
    \label{fig:case1geo}
\end{figure}

In the first example, we evaluate inflow uncertainties in an idealized stenosis model. Stenotic flow, as a classic hemodynamic problem, has been extensively studied in the cardiovascular community, since it is related to many cardiovascular diseases, e.g., arteriosclerosis, stroke, and heart attack~\cite{berger2000flows}. Notably, the trans-stenotic pressure drop computed from the image-based model can be incredibly valuable in clinical practice (as evidenced by Heartflow, Inc., recently valued 1.5 billion). However, the credibility of the model prediction largely relies on the accuracy of inflow boundary conditions, and thus quantification of the uncertainties associated with the inlet is crucial. Here, an idealized stenotic vessel geometry (3-D asymmetric nozzle) is considered, which was originally developed by the U.S. Food and Drug Administration (FDA) as a benchmark problem for a CFD round-robin study~\cite{stewart2012assessment, stewart2013results}. The nozzle geometry is scaled down to the coronary dimension ($D_{in} = 3\times 10^{-3}$ m)~\cite{stiehm2017numerical}, as shown in Fig.~\ref{fig:case1geo}.   
The setting of the baseline CFD simulation follows~\cite{stiehm2017numerical}, where a unidirectional, uniform inflow velocity profile of $\mathbf{u}_{in} = [0.184, 0, 0]$ m/s is prescribed, with the Reynolds number $\mathrm{Re} \approx 167$. For the HF simulation, a high-resolution structured mesh with 66,861 cells (Fig.~\ref{fig:case1geo}b) is used to obtain the fully converged solution with sufficient iterations (2000 iterations). Mesh convergence study is conducted to ensure the mesh quality and solution accuracy of the HF model. To construct the corresponding LF model, the fine mesh is downsampled to a coarse one with only 3,359 cells, and unconverged solution with insufficient iterations (30 iterations) is adopted to reduce the computational cost. In this setting, the LF simulation has a remarkable speedup compared to the HF simulation by significantly sacrificing predictive accuracy. Specifically, the cost of a single LF model run is about $0.6$ CPU seconds, while the counterpart HF simulation roughly takes $3000$ CPU seconds.

To model the uncertainty in the inlet boundary condition, a stationary Gaussian random field $f(\mathbf{x})$ is introduced to the streamwise ($x-$) component of the inflow velocity, which can be expressed as:
\begin{equation}
\begin{split}
    f(\mathbf{x}) \sim \mathcal{GP}(0, K(\mathbf{x}, \mathbf{x}')), \ K(\mathbf{x}, \mathbf{x}') = \sigma_0^2\exp(\frac{|\mathbf{x}-\mathbf{x}'|}{2l^2}),  
\end{split}
\label{GaussianField}
\end{equation}
where $K(\mathbf{x},\mathbf{x}')$ is the exponential kernel function, $\sigma_0$ and $l$ define the standard deviation and length scale of the random field, respectively. The random field can be expressed in a compact form using Karhunen-Loeve (K-L) expansion~\cite{tipping1999probabilistic},
\begin{equation}
    f(\mathbf{x}) = \sum_{i=1}^{n_{k}\to\infty} \sqrt{\lambda_i}\phi_i(\mathbf{x})\omega_i,
    \label{klexpansion}
\end{equation}
where $\lambda_i$ and $\phi_i(\mathbf{x})$ are eigenvalues and eigenfunctions of the kernel $K$; $\omega_i$ is an uncorrelated random variable with zero mean and unit variance. Usually, the KL expansion can be truncated with a finite number ($n_{k}$) of KL basis to approximate the stochastic process. In this nozzle case, a Gaussian random field with $l = 5\times 10^{-3}$ m and $\sigma_0 = 0.02$ m/s is imposed on the streamwise velocity, and the first three ($n_k = 3$) K-L modes are used, capturing $96\%$ energy of the random field. Therefore, the parameter space of this case has three dimensions, i.e., $\mathbf{z} = [\omega_1, \omega_2, \omega_3]^T \in I_z \subset \mathcal{R}^3$.

The LF simulations are conducted on $M = 1000$ parameter points (i.e., $\Gamma$), which are sampled from a multivariate Gaussian distribution $\mathcal{N}(\mathbf{0},\mathbf{I})$ to well cover $I_z$, where $\mathbf{I}$ is an identity matrix. The HF simulations are only conducted on $m = 6$ important points, selected from $\Gamma$ based on Algorithm~\ref{alg: pointselect}. To test the performance of the BF surrogate model, N = 600 test points are independently sampled from the same distribution $\mathcal{N}(\mathbf{0},\mathbf{I})$ within the parameter space $I_{z}$ for uncertainty propagation. Note that the Latin hypercube sampling method (LHS)~\cite{stein1987large} is applied for all sampling tasks. 

\begin{figure}[htp]
  \centering
    \subfloat[Pressure]{\includegraphics[width=0.432\textwidth]{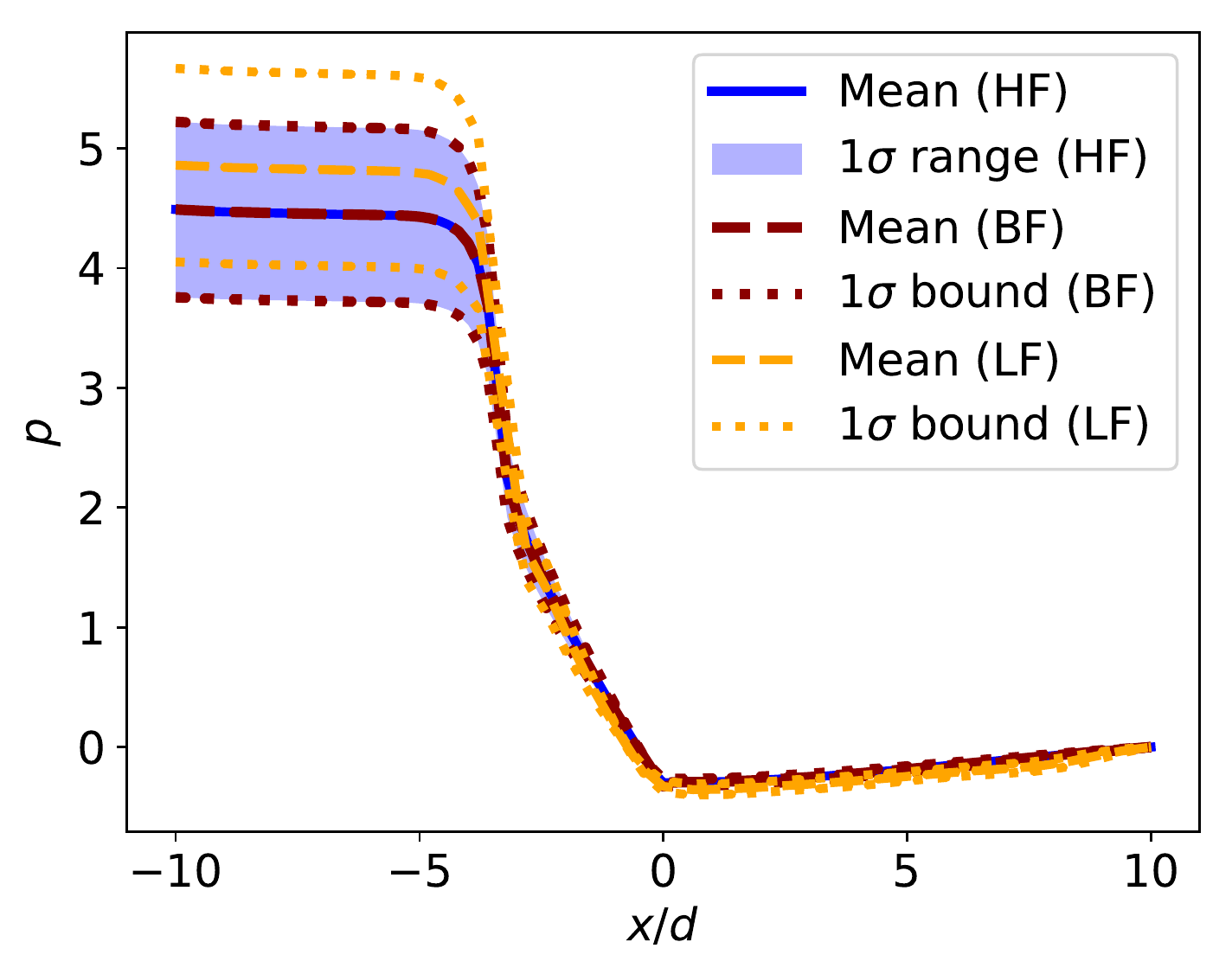}}
    \subfloat[Velocity magnitude] {\includegraphics[width=0.45\textwidth]{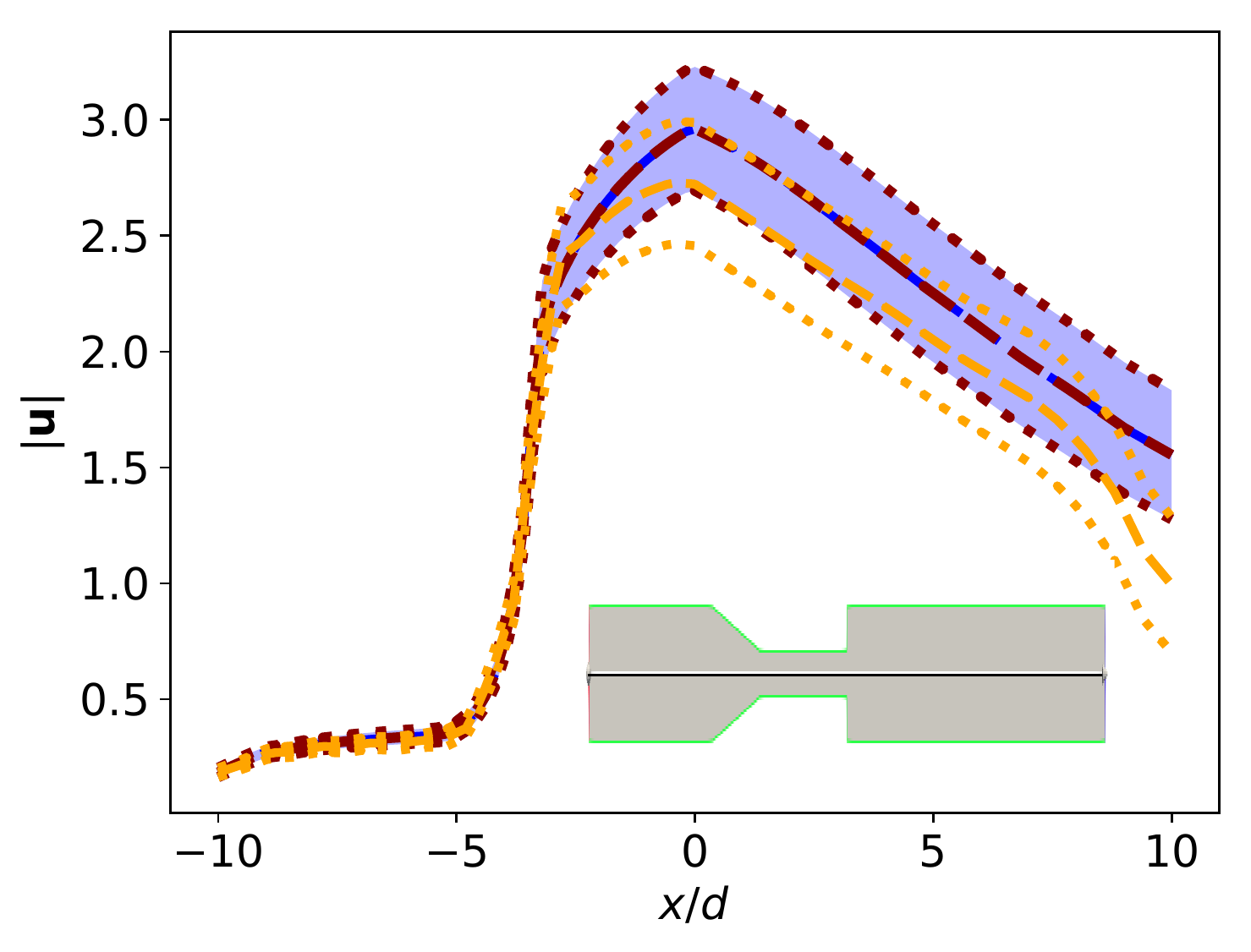}}
    \caption{ The (a) pressure and (b) velocity magnitude profiles along the center line with $1 \sigma$ uncertainty intervals. The position of the center line is showed in the schematic diagram.}
    \label{fig:uqcenterlinecase1}
\end{figure}

The propagated uncertainty from the inlets to the computed pressure drop and velocity magnitude profiles along the centerline of the nozzle is studied. The means and uncertainty ($1\sigma$) ranges of the centerline pressure and velocity magnitude are plotted in Fig~\ref{fig:uqcenterlinecase1}, where the HF (blue), LF (orange), and BF (red) solutions are compared against each other.   
The flow from inlet undergoes a gradual constriction, narrow throat, and sudden expansion regions before reaching the outlet, and thus the velocity magnitude significantly increases across the converging region and gradually decreases after the expansion (Fig.~\ref{fig:uqcenterlinecase1}b). The variation of nozzle walls leads to a highly nonlinear pressure drop (Fig.~\ref{fig:uqcenterlinecase1}a). We can see that the perturbation of inlet causes the scattering of centerline pressure and velocity profiles. The pressure is largely scattered near the inlet region, while the velocity magnitude notably varies after the sudden expansion. In the narrow region, both the pressure and velocity are less scattered since the flow is restricted. These features of the pressure and velocity distribution can be accurately captured by the BF surrogate model as both the means and $1 \sigma$ envelope bounds obtained by the BF model are almost overlapped with the ones from HF-based MC simulations (ground truth), while the LF model over-predicts the pressure near the inlet and under-predicts the velocity magnitude near the outlet. Moreover, the uncertainty of the most important QoI, trans-stenotic pressure drop, can also be accurately captured by the BF model, where almost 100-times speedup is gained over the HF-based MC simulations with 600 
samples (details see Appendix A).

\begin{figure}[htb]
\centering
\includegraphics[width=0.5\textwidth]{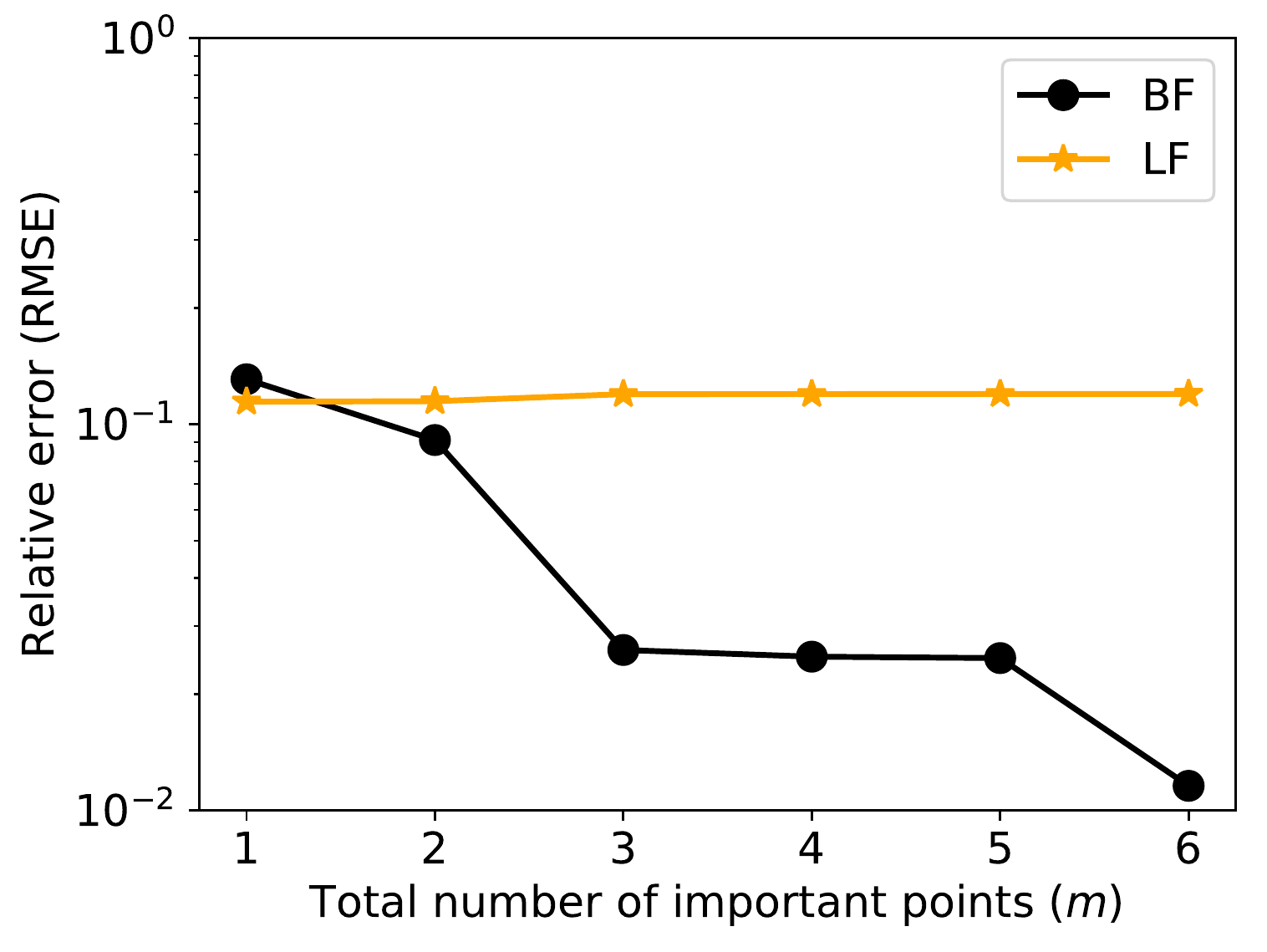}
\caption{The relative root mean squared error (RMSE) of the BF model (black) over 600 test parameter points with respect to the number of important points (HF simulations) used for BF surrogate construction in test case 1. The corresponding RMSE of the one with LF basis (orange) are plotted for comparison.}
\label{fig:convergence1}
\end{figure}

To better evaluate the BF surrogate approximation compared to the LF baseline, the decay of the relative error (\ref{eqn:errmetric}) of both BF and LF models over 600 test points are computed with respect to the number of HF solutions ($m$) used for constructing the BF surrogate, which is shown in Fig.~\ref{fig:convergence1}. 
It can be seen that the LF model baseline originally has a $10\%$ error, while this error can be reduced by more than $80\%$ with only three HF solutions based on the BF approach. When HF solutions on six important points are used, the relative error of the BF surrogate is reduced by one order of magnitude, demonstrating a fast decay rate. Note that the error here is evaluated on 600 test points over the entire parameter space.    

\subsection{Idealized Bifurcation Aneurysm Model (Case 2)}
\label{sec:case2}
Vascular aneurysm refers to pathological dilatation of the vessel wall, which is an abnormal biological response (a.k.a., growth remodeling of vessel walls) caused by certain unusual hemodynamic conditions (e.g., low or rapidly changed wall shear stress)~\cite{sforza2009hemodynamics}. Aneurysms are more likely to be formed at vessel bifurcations, where the flow pattern is complex and wall shear stress is usually low or oscillates~\cite{weir2002unruptured}. In the second test case, we investigate an idealized bifurcation aneurysm model, whose geometry and boundary condition are adopted from an experimental benchmark work~\cite{cole2001flow}.   
\begin{figure}[htb!]
    \centering
    \subfloat[2-D Geometry for LF  model]{\includegraphics[width=0.45\textwidth]{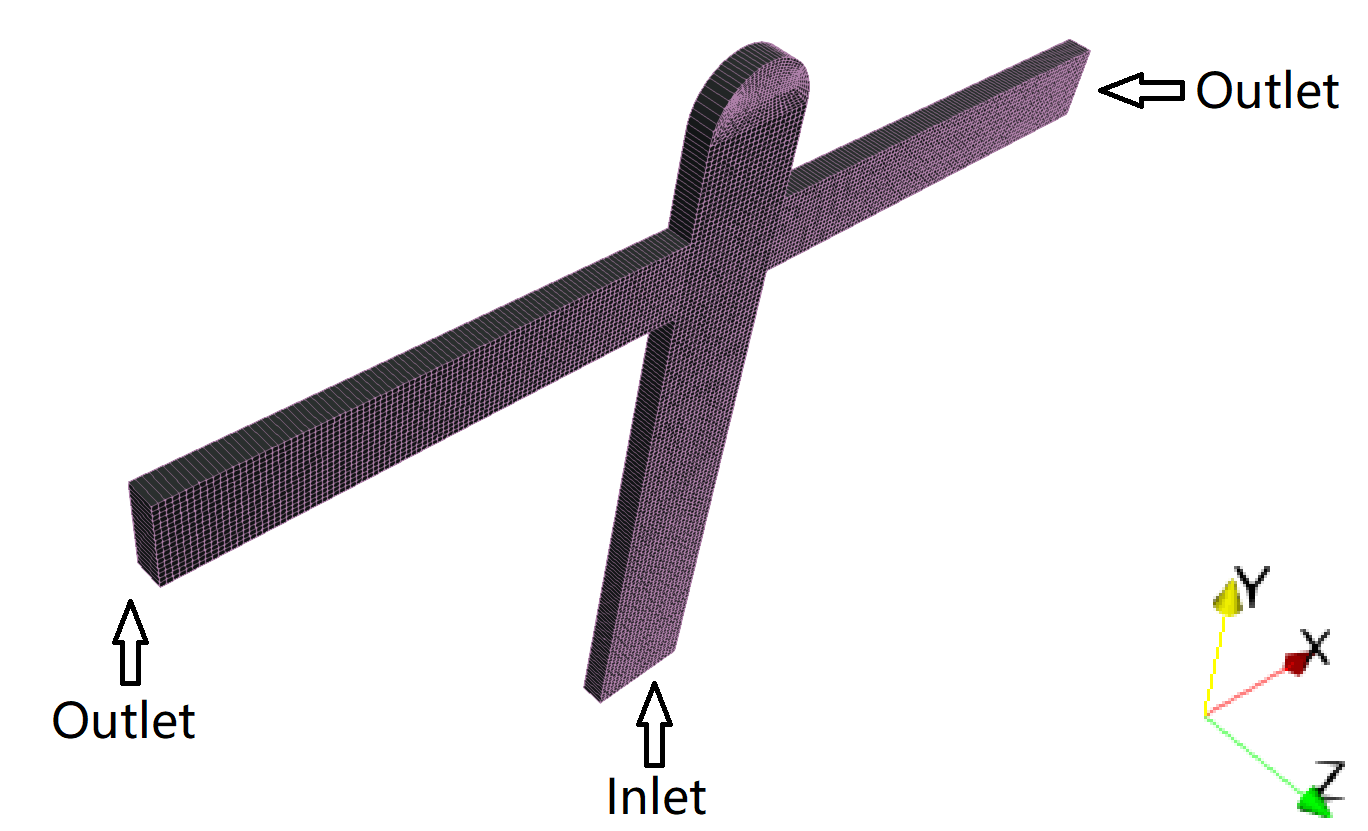}}
    \subfloat[3-D Geometry for HF model] {\includegraphics[width=0.45\textwidth]{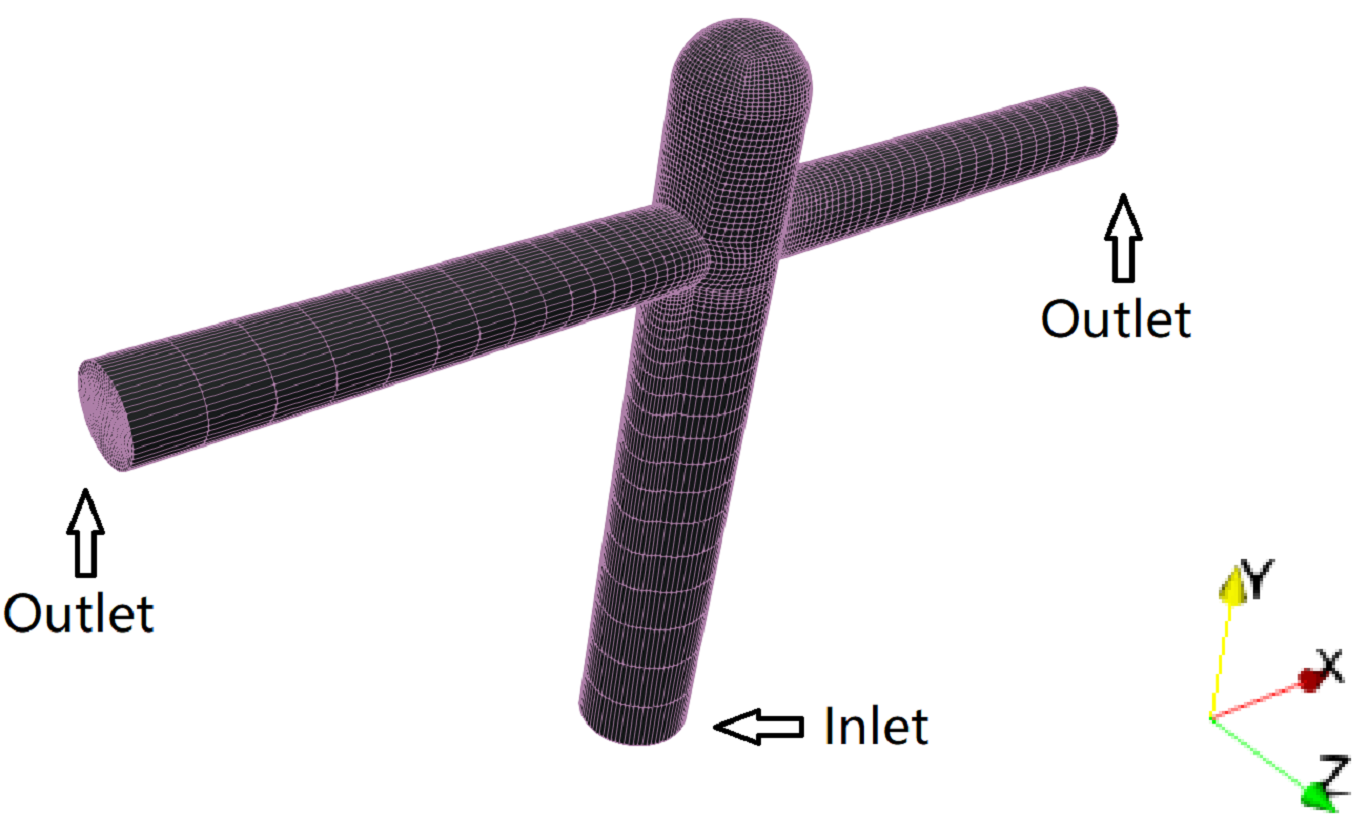}}
    \caption{The idealized bifurcation aneurysm geometry with (a) 2-D and (b) 3-D computational meshes.}
    \label{fig:case2geo}
\end{figure}
As shown in Fig.~\ref{fig:case2geo}, the model has a perfect ``T" shape, where a $90^{\circ}$ bifurcation has a symmetric placement of outflow tubes. The dome at the end of the input tube represents an idealized terminal aneurysm. In this case, the simplified 2-D counterpart (Fig.~\ref{fig:case2geo}a) of the 3-D geometry (Fig.~\ref{fig:case2geo}b) is used to formulate the LF model, which is much cheaper to evaluate. Structured meshes are generated for both 3-D and 2-D geometries with 109,641 cells and 6,681 cells, respectively. Sufficient iterations are performed in both LF and HF simulations to obtain fully converged solutions. The computational costs of the HF and LF models are about $23\sim27$ CPU seconds and $3,800\sim4,000$ CPU seconds, respectively.

Volumetric flow rate $q$ is specified at the inlet and outlet to impose flow boundary conditions. To mimic an uneven flow distribution to the two outflow brunches in reality (e.g., due to geometrical asymmetry), a non-dimensional flow-split parameter $\alpha \in [0,1]$ is introduced,
\begin{equation}
    \begin{split}
        q_{outL} &= \alpha q_{in},\\
        q_{outR} &= (1-\alpha) q_{in},
    \end{split}
    \label{eqn:case2uq}
\end{equation}
where $q_{in}$ is the inlet volume flow rate, and $q_{outL}, q_{outR}$ are volume flow rates at left and right outlets. Generally, it is difficult to determine how the blood flow is distributed among different distal branches clinically, and the flow-split is often estimated based on Murray's law~\cite{sherman1981connecting}. However, such estimation is merely a rough one and contains large uncertainties. Hence, the flow-split parameter $\alpha$ can be modeled as a random variable. Besides, the viscosity $\nu$ of blood is also assumed to be uncertain in this test case and varies from $7\times 10^{-6}$ $\mathrm{m^2/s}$ to $2.1\times 10^{-5}$ $\mathrm{m^2/s}$ independently from $\alpha$. Therefore, two independent random parameters are considered here and the parameter space has two dimensions, i.e., $\mathbf{z} = [\alpha, \nu] \in I_z \subseteq \mathbb{R}^2$. The setting for the baseline simulation follows Valencia's work~\cite{valencia2005simulation}, where the inlet volume rate $q_{in}$ is $1.42 \times 10^{-5}$ $\mathrm{m^3/s}$, blood viscosity $\nu$ is $1.4\times 10^{-5}$ $\mathrm{m^2/s}$, and flow-split $\alpha$ is set to be 0.5 based on Murray's law. The Reynolds number of the baseline case is around $143$. We assume the two uncertain parameters are uniformly distributed in the given intervals. To explore the parameter space, $100$ points are sampled from a multi-uniform distribution $[0,1] \mathbf{\times} [7\times 10^{-6}, 2.1\times 10^{-5}]$ based on the LHS method to form $\Gamma$. A subset of $m = 25$ important points are selected from $\Gamma$, where HF solutions are collected to construct the BF surrogate. A test set of 250 parameter points are sampled from $I_z$ independently from the same distribution. 

\begin{figure}[htb!]
\centering
\subfloat[$x$-centerline pressure]{\includegraphics[width=0.335\textwidth]{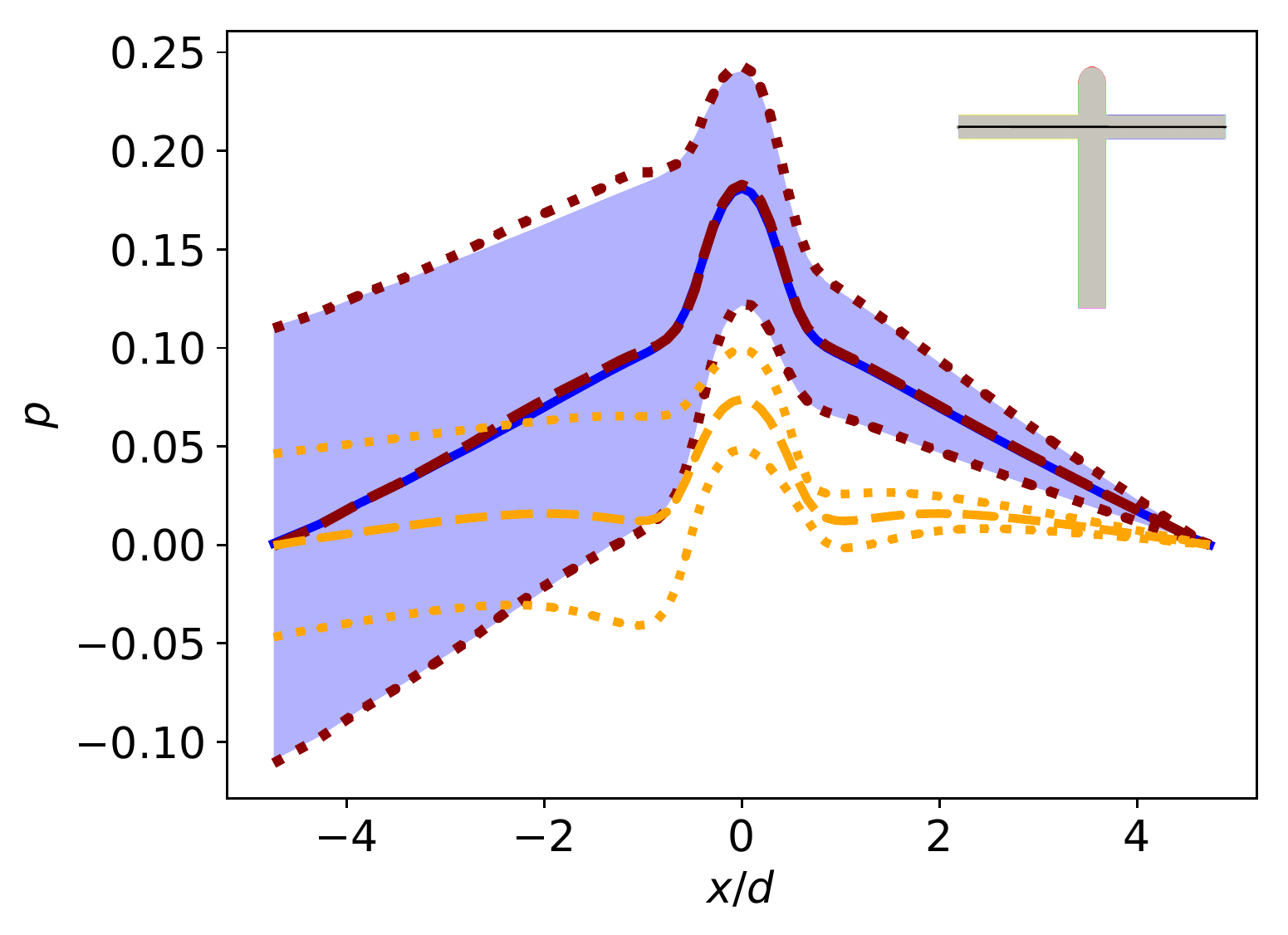}}
\subfloat[$y$-centerline pressure]{\includegraphics[width=0.325\textwidth]{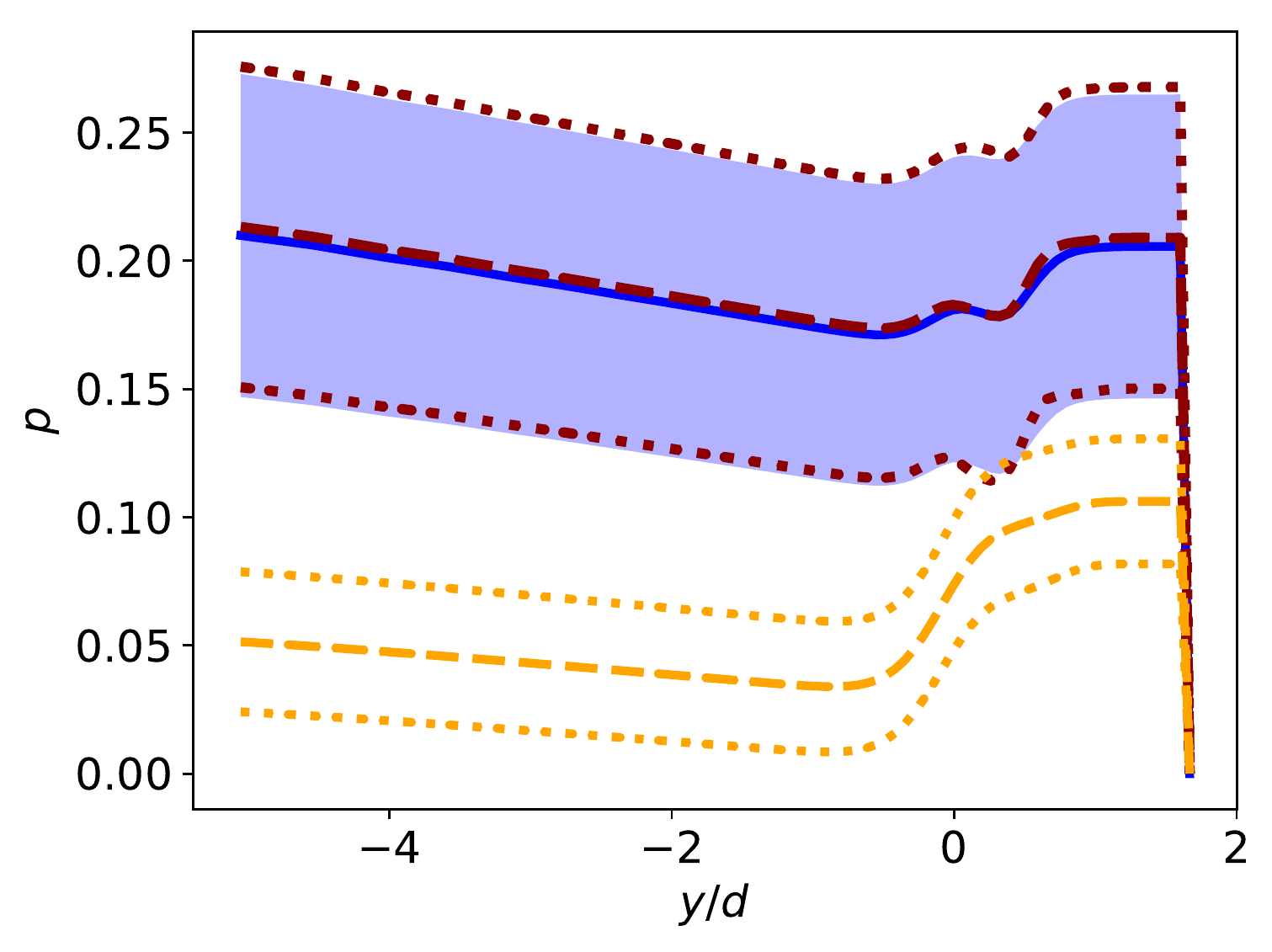}}   
\subfloat[$z$-centerline pressure]{\includegraphics[width=0.325\textwidth]{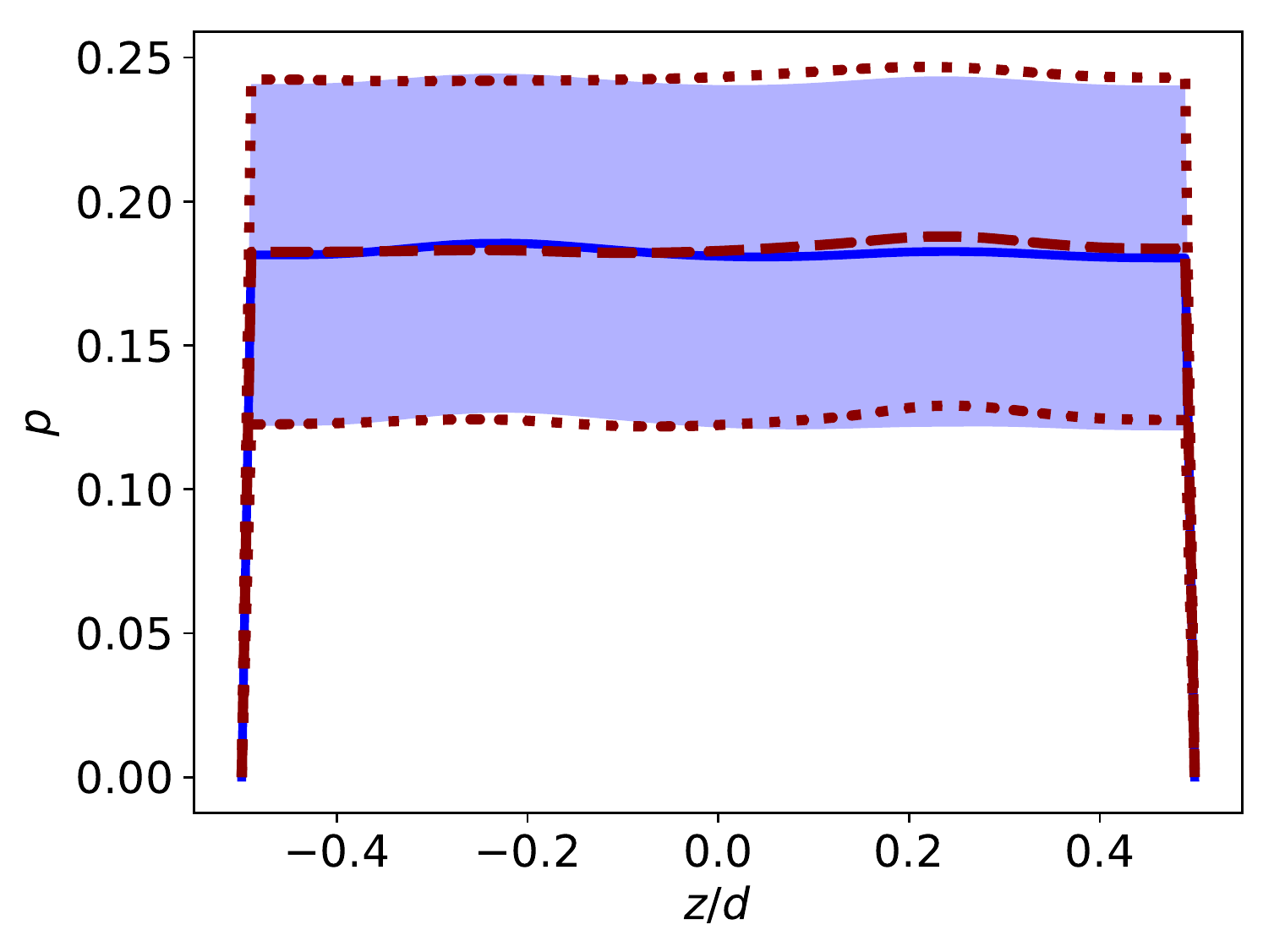}}\\
\subfloat[$x$-centerline velocity]{\includegraphics[width=0.33\textwidth]{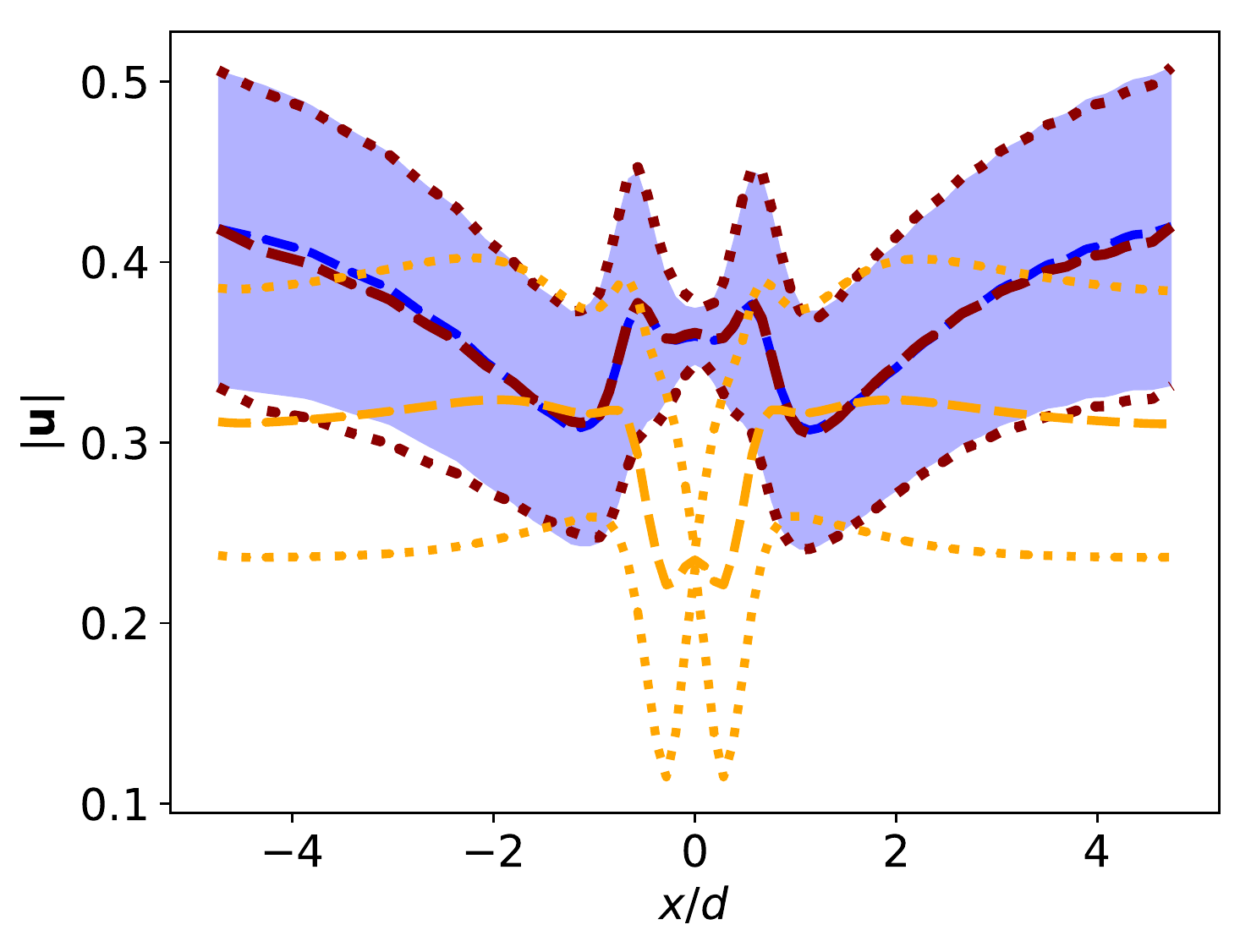}}
\subfloat[$y$-centerline velocity]{\includegraphics[width=0.33\textwidth]{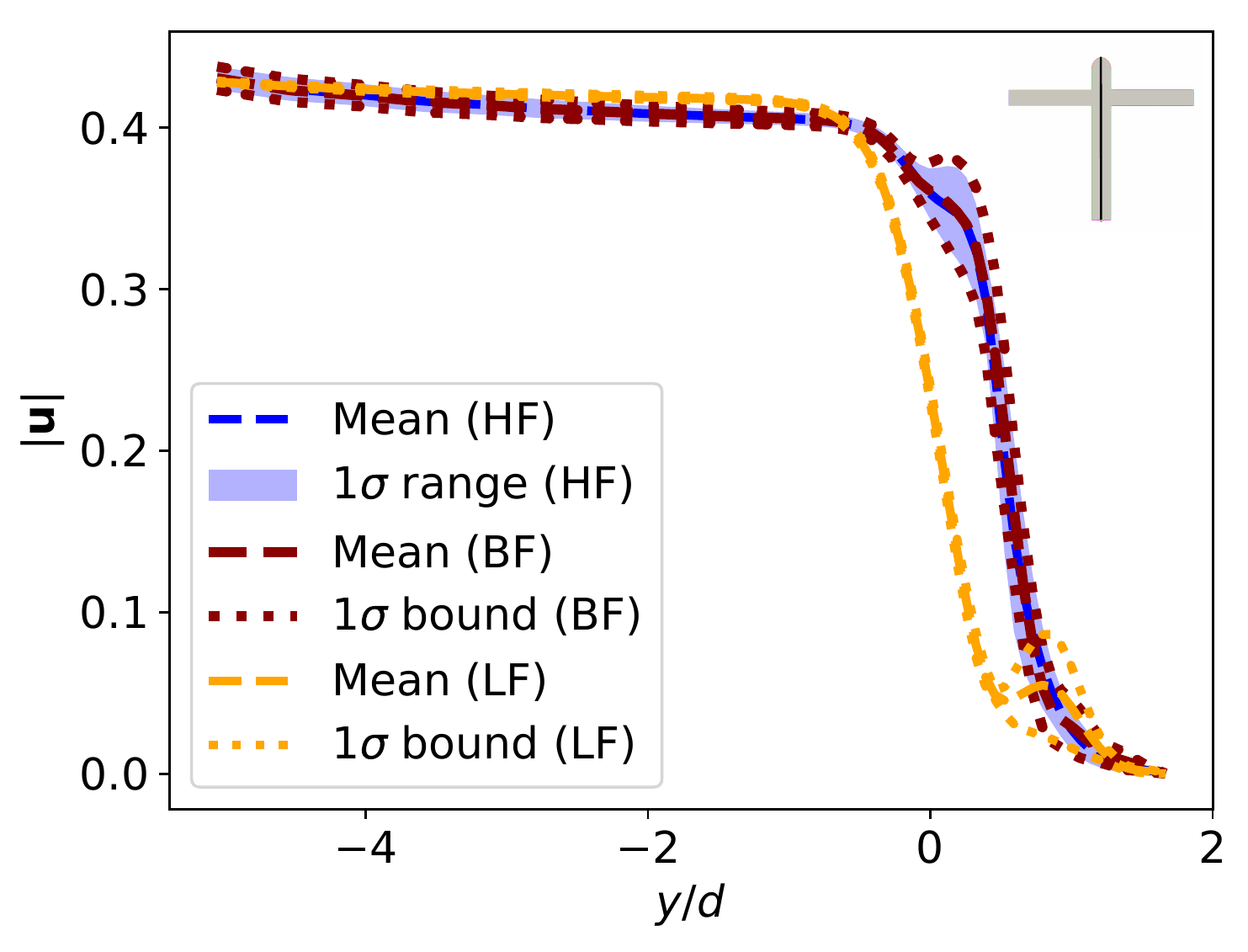}}
\subfloat[$z$-centerline velocity]{\includegraphics[width=0.33\textwidth]{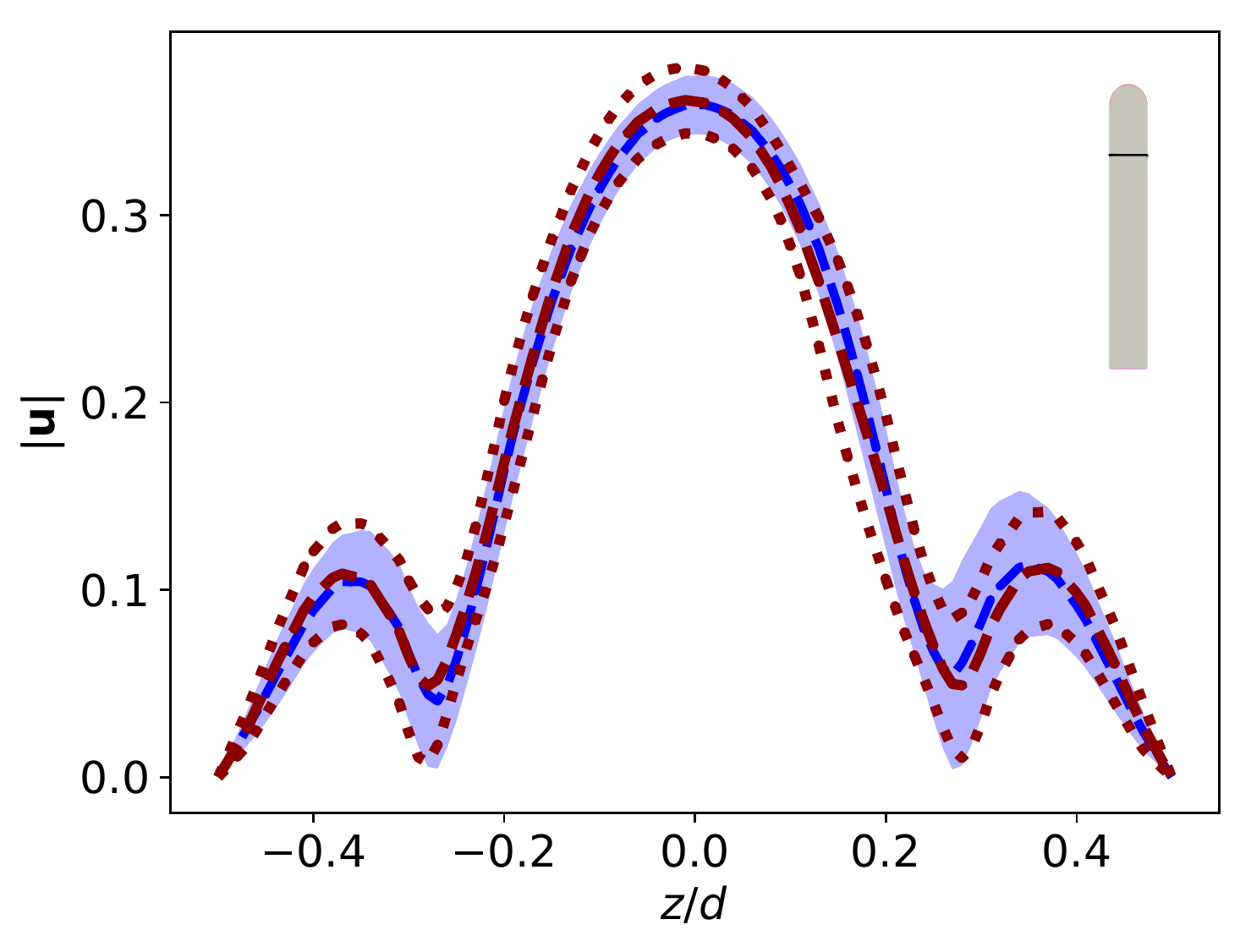}}
\caption{The (a-c) pressure and (d-f) velocity magnitude profiles along $x$-, $y$-, and $z$- centerlines with $1 \sigma$ uncertainty intervals, and the solution along $z$-centerline is only available in HF and BF results.}
\label{fig:uqcenterlinecase2}
\end{figure}
We plot the mean centerline pressure and velocity magnitude profiles with the $1\sigma$ uncertainty intervals in Fig.~\ref{fig:uqcenterlinecase2}. The three columns of sub-panels show the profiles along the centerlines in $x$, $y$, and $z$ directions, respectively. In $x$-direction (Figs~\ref{fig:uqcenterlinecase2}a and~\ref{fig:uqcenterlinecase2}d), the profiles of the mean and variance of the velocity magnitude is nearly symmetric along the $x$-centerline since the flow split ratio $\alpha$ is uniformly sampled from $[0, 1]$. However, the pressure uncertainty is more asymmetrically distributed, as the profiles are largely scattered at the left outlet and shrink to a constant at the right outlet, which is because the reference pressure is imposed at the right end of the tube. As for the main flow direction, i.e., $y$-direction (Figs~\ref{fig:uqcenterlinecase2}b and~\ref{fig:uqcenterlinecase2}e), the flow velocity profiles are less scattered in general, while the pressure shows a relatively large uncertainty. We also can see that the uncertainty close to the dome is very small since the flow is stagnated, while a larger uncertainty is observed in the cross-neck region where flow recirculations occur. The LF model fails to accurately capture these propagated uncertainties, and the LF-predicted mean and uncertainty intervals significantly deviate from the corresponding HF solutions. In contrast, the BF surrogate has remarkably better performance since the BF-predicted results agree with the HF ground truth well. It is worthy to note that the missing flow information of $z$-direction in solutions of the LF model with 2-D geometry can be well recovered by the BF surrogate, as shown in Figs~\ref{fig:uqcenterlinecase2}c and~\ref{fig:uqcenterlinecase2}f.

A notable feature of the proposed BF surrogate model is the capability of providing full-field predictions with the HF resolution since HF solutions are used as the basis for reconstruction. To extensively evaluate the performance of the BF surrogate model, the maximum values of the BF-predicted field quantities, including velocity, vorticity, and wall pressure fields, are calculated. We want to examine how well the extreme values of flow variables over the entire computational domain can be captured by the BF surrogate, which is usually a challenging task in surrogate modeling.
\begin{figure}[htb]
  \centering
    \subfloat[The max velocity magnitude]{\includegraphics[width=0.327\textwidth]{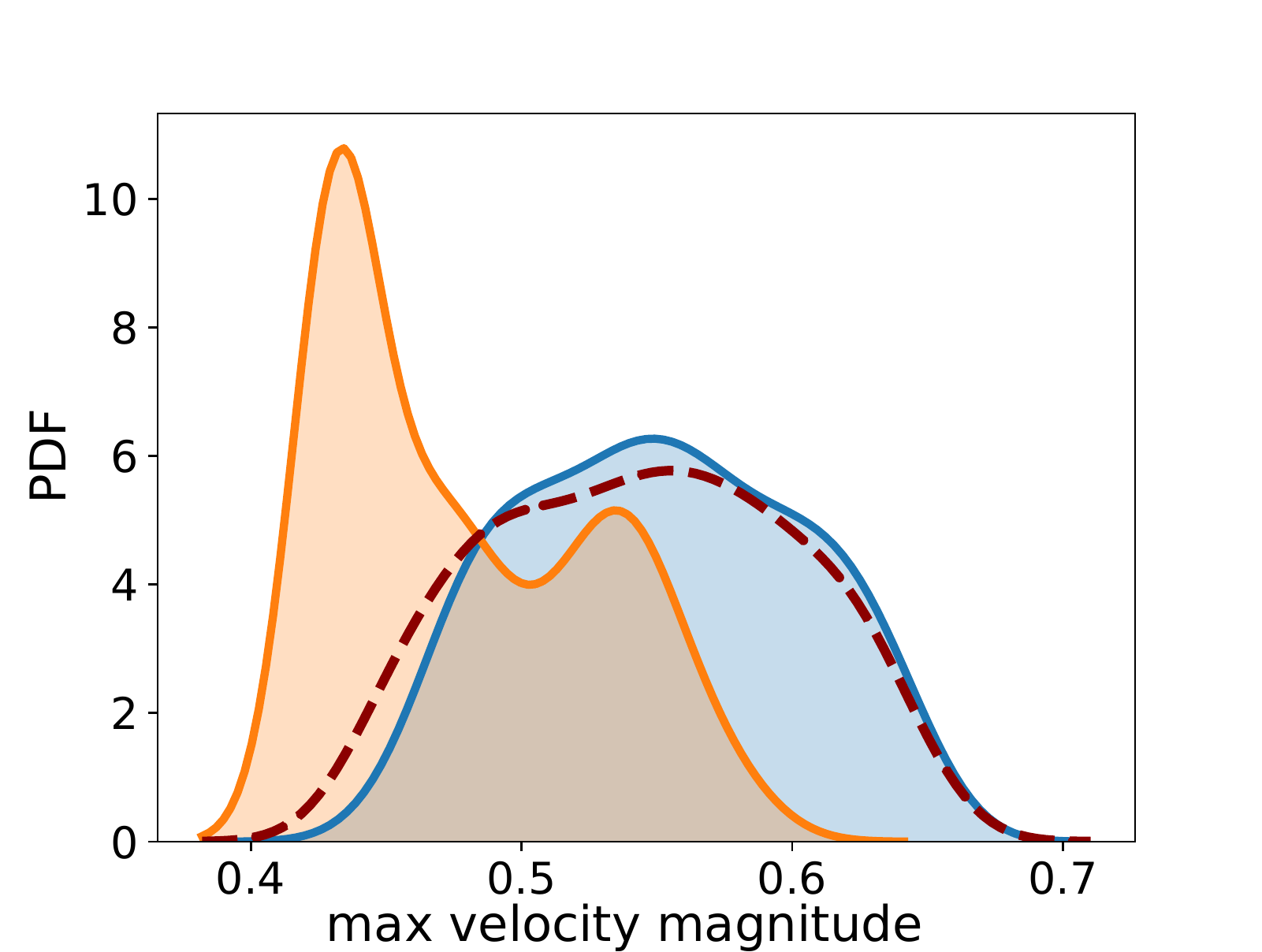}}
    \subfloat[The max vorticity magnitude] {\includegraphics[width=0.335\textwidth]{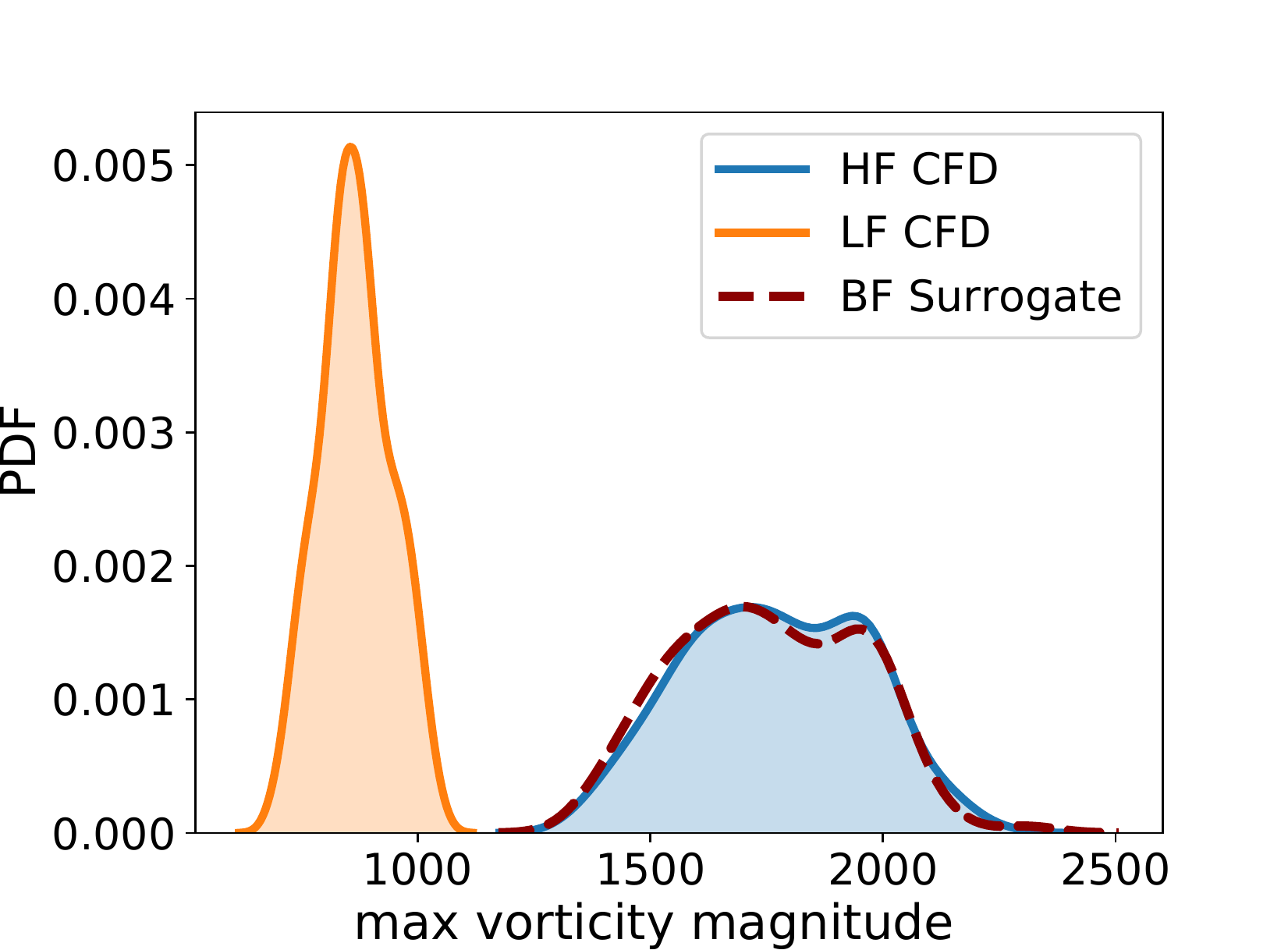}}
    \subfloat[The max wall pressure] {\includegraphics[width=0.327\textwidth]{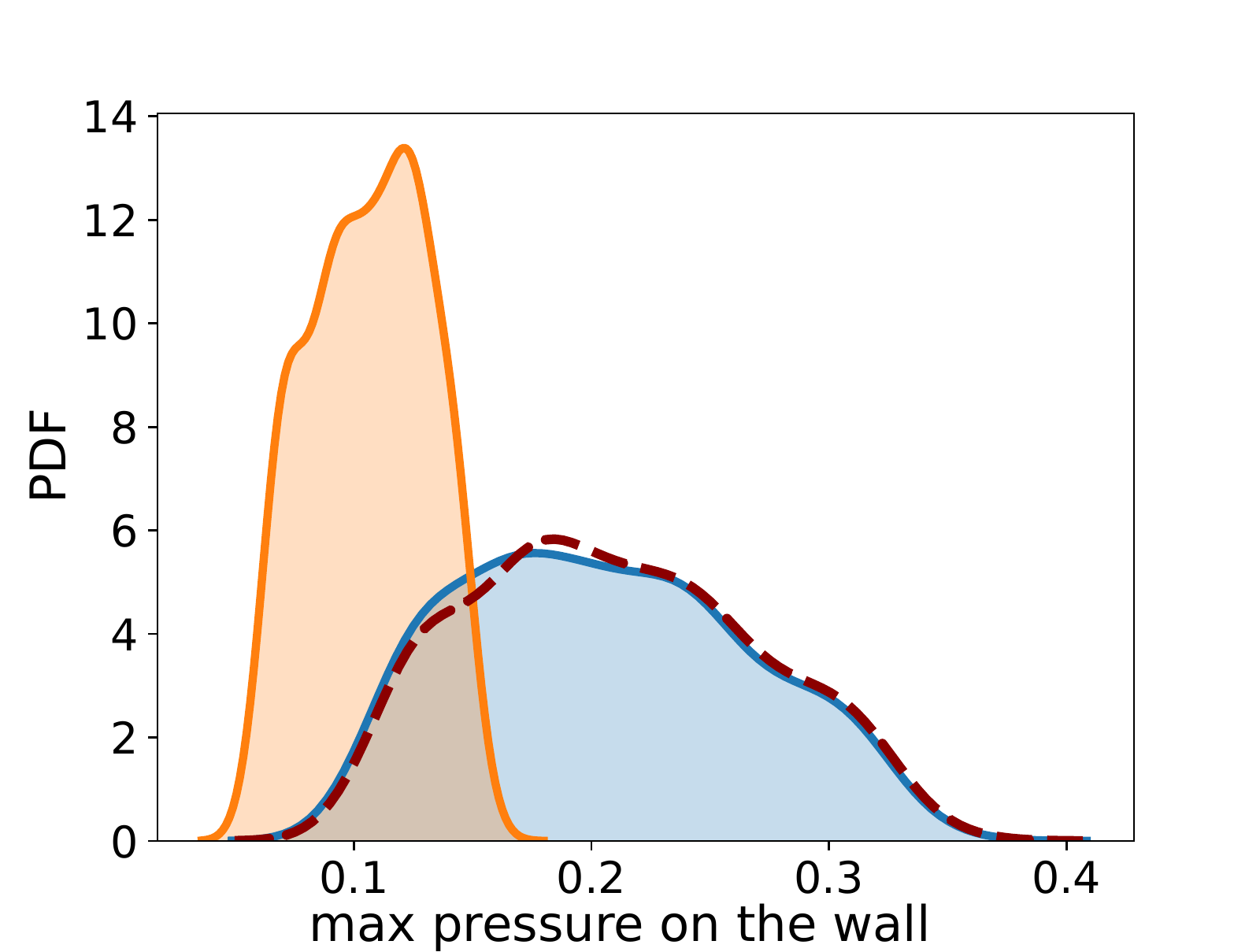}}
    \caption{The probability density functions (PDF) of maximum velocity magnitude, vorticity magnitude and pressure on the wall over 250 test points.}
    \label{fig:pdfcase2}
\end{figure}
The probability distributions of the maximum values of flow velocity magnitude, vorticity magnitude, and wall pressure propagated by the LF, BF, and HF models over the test set are shown in Fig.~\ref{fig:pdfcase2}. The LF model under-predicts the means of all the maximum values, and the shapes of the distributions are notably different from the ones propagated by the HF model. In contrast, the BF surrogate can accurately propagate the uncertainty as the BF-based uncertainty distributions are almost overlapped with the HF ground truth. Moreover, since only 25 HF solutions are used to construct the BF surrogate, the speedup is around 10 for propagating 250 MC samples (details see Appendix A). 

\begin{figure}[htb]
  \centering
  \includegraphics[width=0.5\textwidth]{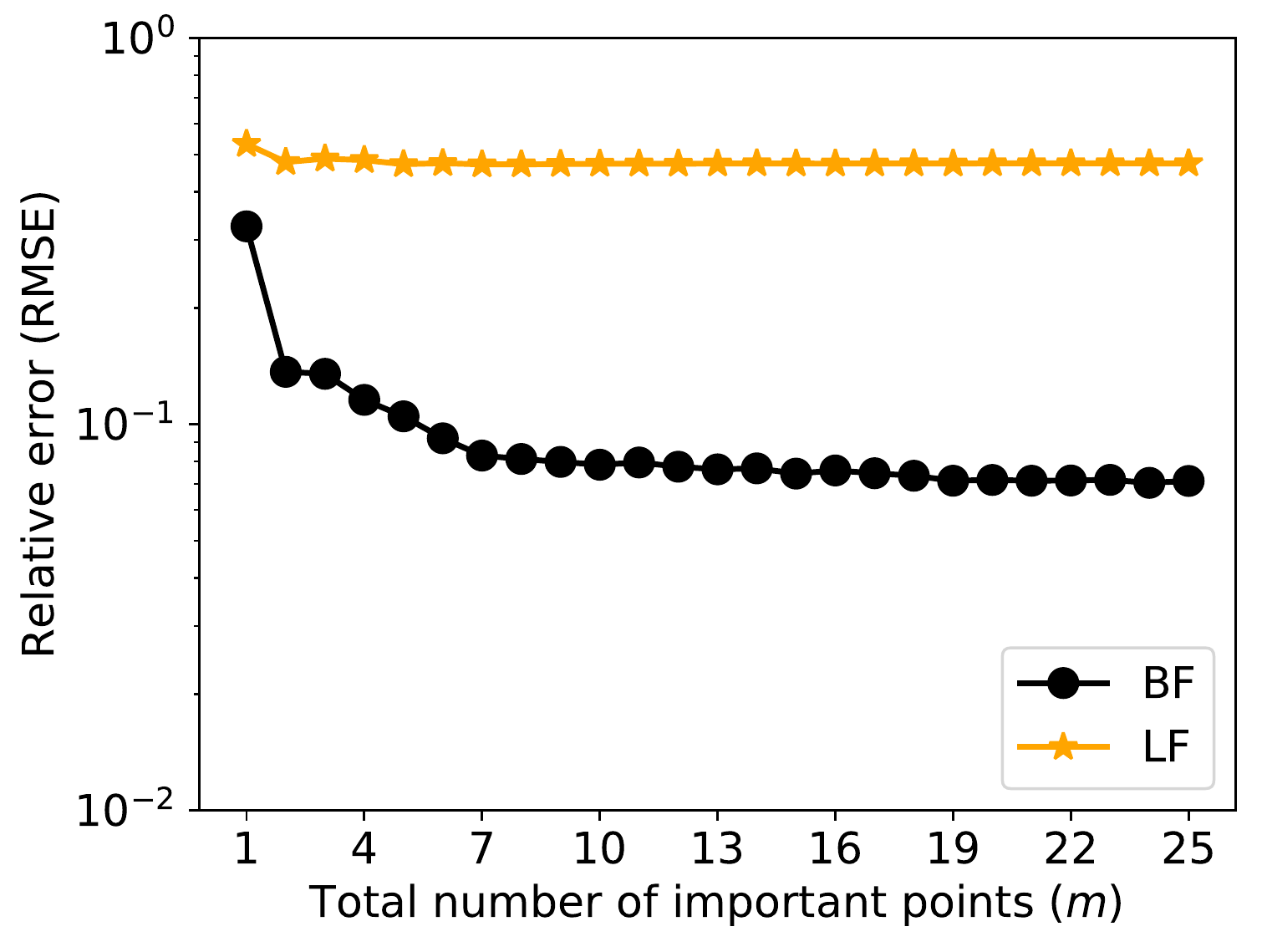}
    \caption{The relative root mean squared error (RMSE) of the BF model (black) over 250 test samples with respect to the number of important points (HF simulations) in test case 2. The corresponding RMSE of the one with LF basis (orange) are plotted for comparison.}
    \label{fig:convergence2}
\end{figure}
Figure~\ref{fig:convergence2} shows the relative errors (RMSE) of the BF surrogate model with respect to the number ($m$) of HF simulations used for reconstruction. The relative error of the LF model is also plotted for comparison, which is relatively large (nearly $100\%$) due to the geometry simplification from 3-D to 2-D configuration. Using BF strategy, only with $4\sim5$ HF simulations on important parameter points, the RSME error can be significantly reduced to the error level less than $10\%$. The convergence rate of RMSE is fast in the first a few HF samples, and the error can continuously decrease by increasing total of important points (HF samples) but saturates after $m>10$. How to determine the optimal number of HF simulations to be conducted will be discussed later on in Section~\ref{sec:discussion}.

\subsection{Patient-specific Cerebral Aneurysm Model (Case 3)}
\label{sec:case3}
As mentioned above, aneurysms are lesions of the arterial wall, and such pathological dilatation occurring at intracranial arteries may cause serious consequences, e.g., aneurysm rupture and intracranial hemorrhage, associated with high mortality and morbidity rates~\cite{kaminogo2003incidence}. The formation, progression, and rupture of a cerebral aneurysm involve complex pathological processes. Hemodynamics is known to be a major factor involved in these processes~\cite{sforza2009hemodynamics}, and accurately quantifying the hemodynamics is significantly important for improving the prognosis, diagnosis, and treatment planning of cerebral aneurysms and their ruptures. Nonetheless, the reliability of the model-based hemodynamic predictions largely depends on the boundary conditions, which often have large uncertainties. For example, the inflow velocity field obtained from phase-contrast MR images usually has a low resolution and contains measurement noises. In this subsection, a real-world application of the BF model on patient-specific aneurysm is investigated. Namely, the BF surrogate is applied to propagate a high-dimensional inflow uncertainty in a 3-D cerebral aneurysm model with a real patient-specific geometry.  

\begin{figure}[htp]
    \centering
    \subfloat[Coarse mesh for LF model]{\includegraphics[width=0.45\textwidth]{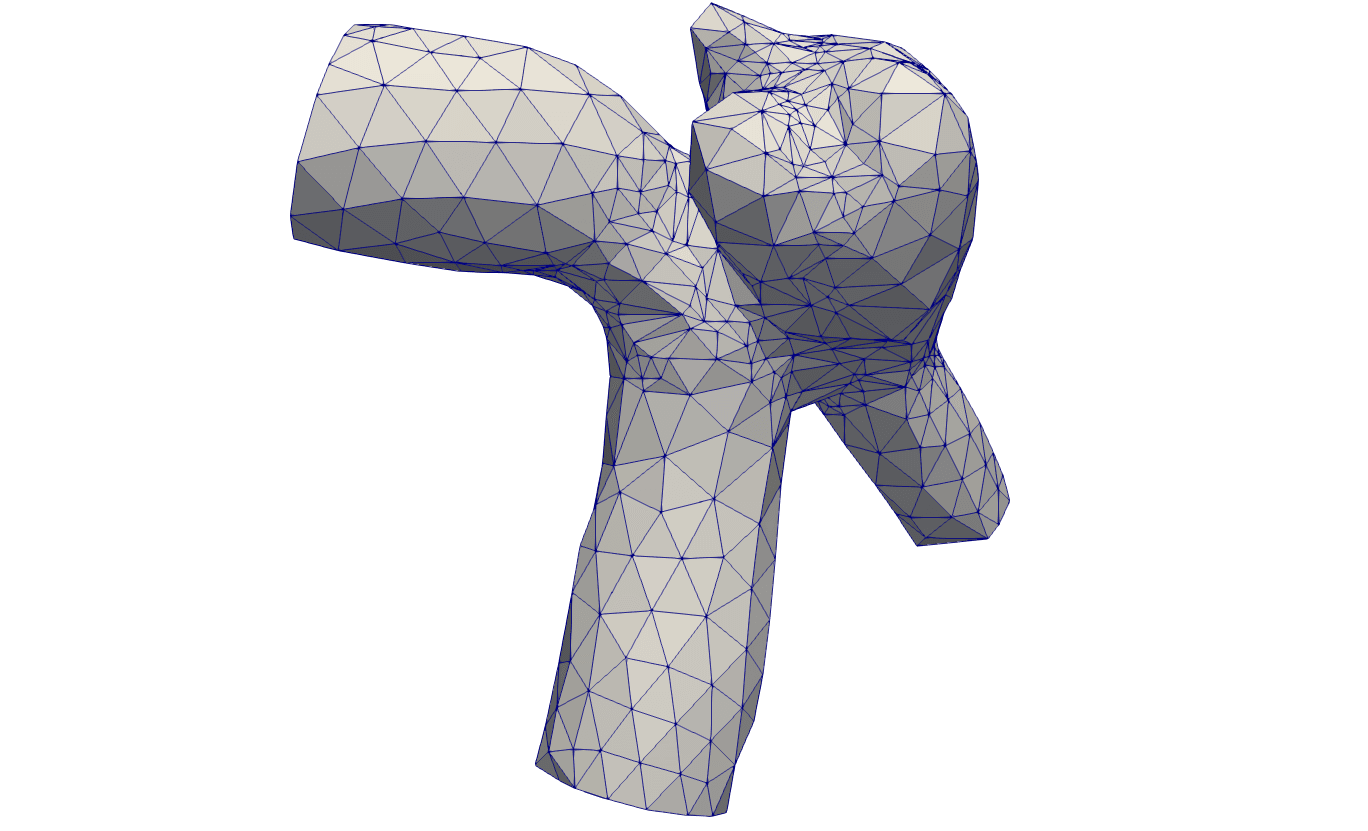}}
    \subfloat[Fine mesh for HF model] {\includegraphics[width=0.45\textwidth]{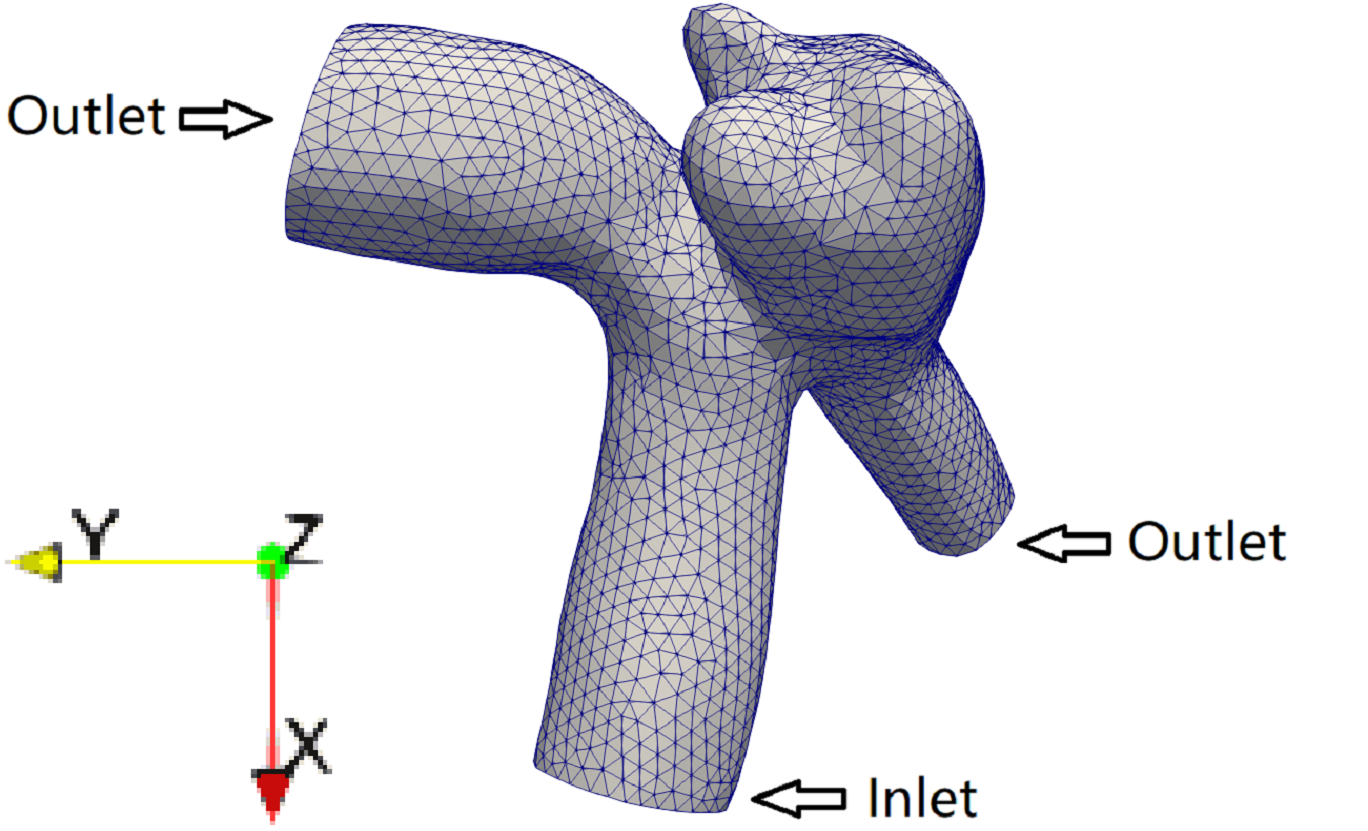}}
    \caption{The patient-specific MCA aneurysm geometry with (a) coarse and (b) fine computational meshes}
    \label{fig:case3geo}
\end{figure}
As cerebral aneurysms are frequently located near arterial bifurcations in the circle of Willis~\cite{sforza2009hemodynamics}, here a middle cerebral artery (MCA) aneurysm is studied, which is one of the five MCA geometries used in the 2015 International Aneurysm CFD Challenge~\cite{valen2018real}. A spatially-resolved unstructured mesh with 51,648 cells is used to build the HF model, while the LF model adopts a coarse mesh with only 4,464 cells. In both HF and LF simulations, sufficient numerical iterations are performed to achieve the convergence. The computational cost for one HF model evaluation is about 1,110 CPU seconds, while one LF simulation only takes 2 CPU seconds. The baseline CFD simulation adopts the setting given in the 2015 International Aneurysm CFD Challenge~\cite{valen2018real}. Specifically, a uniform inflow velocity profiles with a constant value of $0.509$ m/s is prescribed normal to inlet plane, and no secondary flow is imposed, i.e., in-plane velocity is zero. The Reynolds number is $Re = 345$, based on the inlet diameter $D_{in} = 2.77 \times 10^{-3}$ m. The zero-traction outlet and no-slip wall boundary conditions are applied.  

The inflow uncertainty is introduced in a similar way as in section~\ref{sec:case1}, where the uncertainties are modeled as Gaussian random fields. However, here we not only consider the uncertainty in streamwise direction but also introduce uncertain secondary flows. Namely, three Gaussian scalar random fields $f_x(\mathbf{x}), f_y(\mathbf{x})$, and $f_z(\mathbf{x})$, expressed by the KL expansion, are added to the three components of baseline velocity inlet $\mathbf{u}_{in}^{base}$ as,   
\begin{equation}
\mathbf{u}_{in} = \mathbf{u}_{in}^{base} + [f_x(\mathbf{x}), f_y(\mathbf{x}), f_z(\mathbf{x})]\\
\label{case3velocity}
\end{equation}
In this case, the randomness of velocity in all directions is modeled by a stationary Gaussian process with the length scale $l = 2 \times 10^{-3}$ and variance $\sigma_0^2 = 0.01$. Three K-L modes are used to cover $90\%$ energy of each random field. Therefore, a nine-dimensional uncertain parameter space is defined in this test case, and any parameter point $\mathbf{z}$ can be written as,
\begin{equation}
    \mathbf{z} = [ 
    \underbrace{\omega_1, \omega_2, \omega_3}_{\mathrm{x\ streamwise}},
    \underbrace{\omega_4, \omega_5, \omega_6}_{\mathrm{y\ in-plane}},
    \underbrace{\omega_7, \omega_8, \omega_9}_{\mathrm{z\ in-plane}}
] \in I_z \subseteq \mathbb{R}^9
    \label{eqn:paraspacecase3}
\end{equation}
where $\omega_i, i = 1\cdots9$ are independent and identically distributed (i.i.d.) random variables. $M = 2000$ points are sampled independently from a multivariate Gaussian distribution $\mathcal{N}(\mathbf{0},\mathbf{I})$ to form $\Gamma$, and $m = 40$ important ponts are selected from $\Gamma$ to conduct HF simulations for BF surrogate construction. To evaluate the BF model and propagate uncertainties, 600 MC samples are independently drawn from $I_z$ to form the test set.

The inflow uncertainties can be propagated to the simulated hemodynamic fields through the HF, BF, and LF models on the MC samples in the test set. To better visualize the uncertainty propagation results for the local hemodynamic information, we present contour plots of the mean and standard deviation (std) fields of the ensembles of velocities, wall pressures, and wall shear stresses (WSS). 
\begin{figure}[htb]
  \centering
    \subfloat[HF mean]{\includegraphics[width=0.32\textwidth]{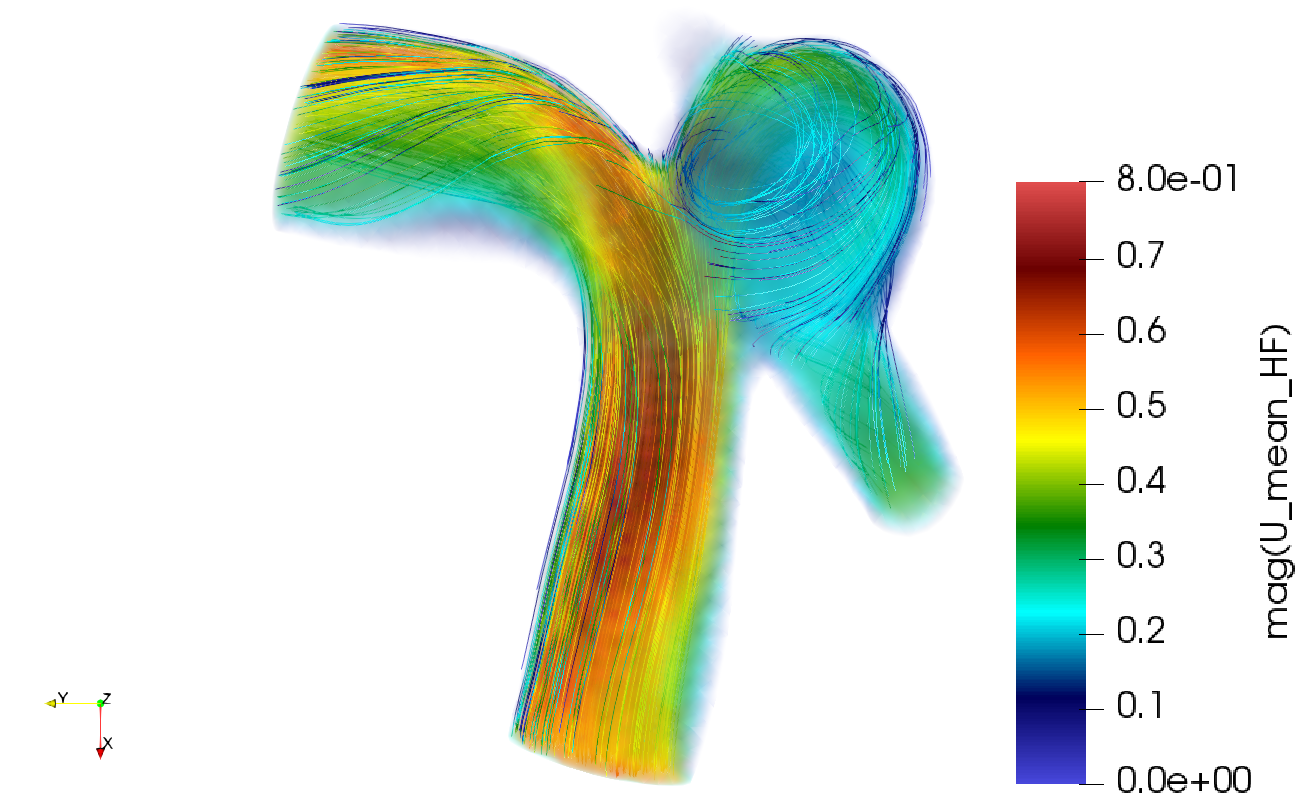}}
    \subfloat[BF mean] {\includegraphics[width=0.32\textwidth]{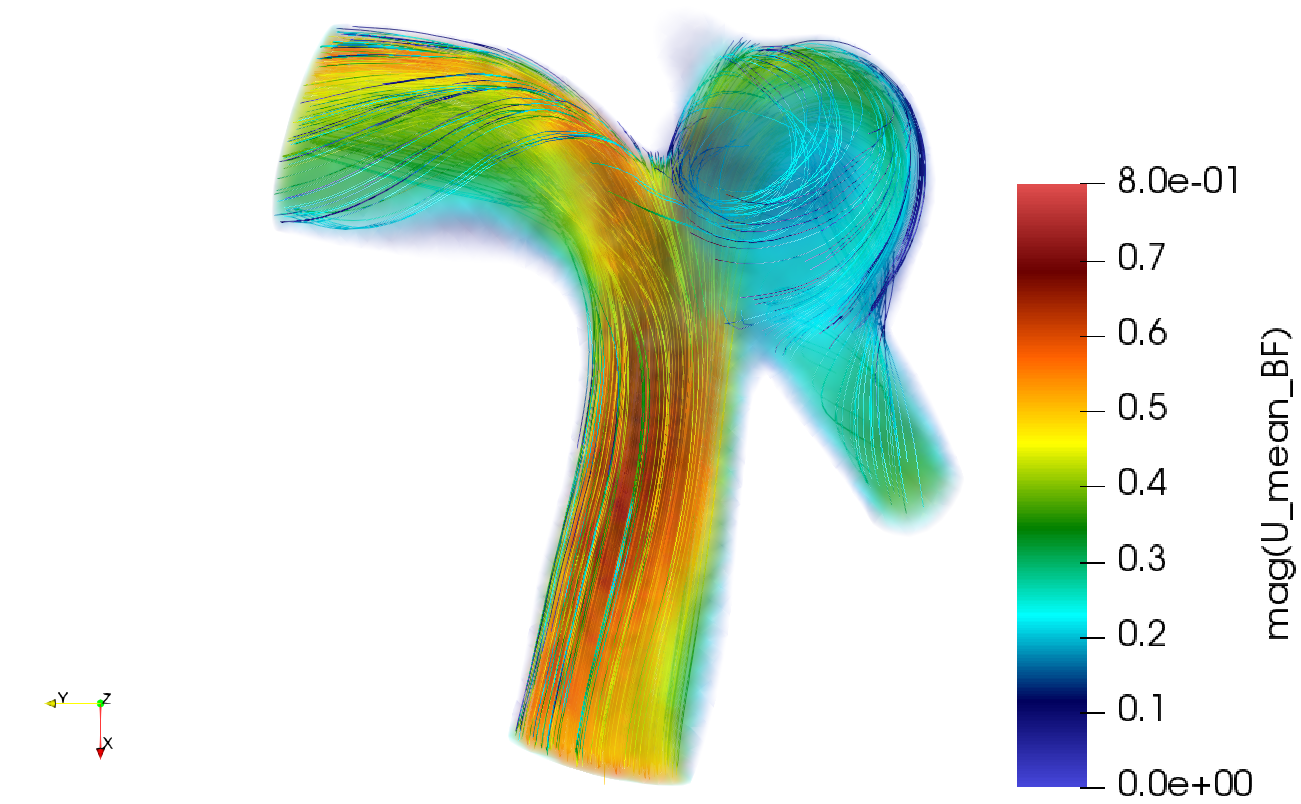}}
    \subfloat[LF mean] {\includegraphics[width=0.32\textwidth]{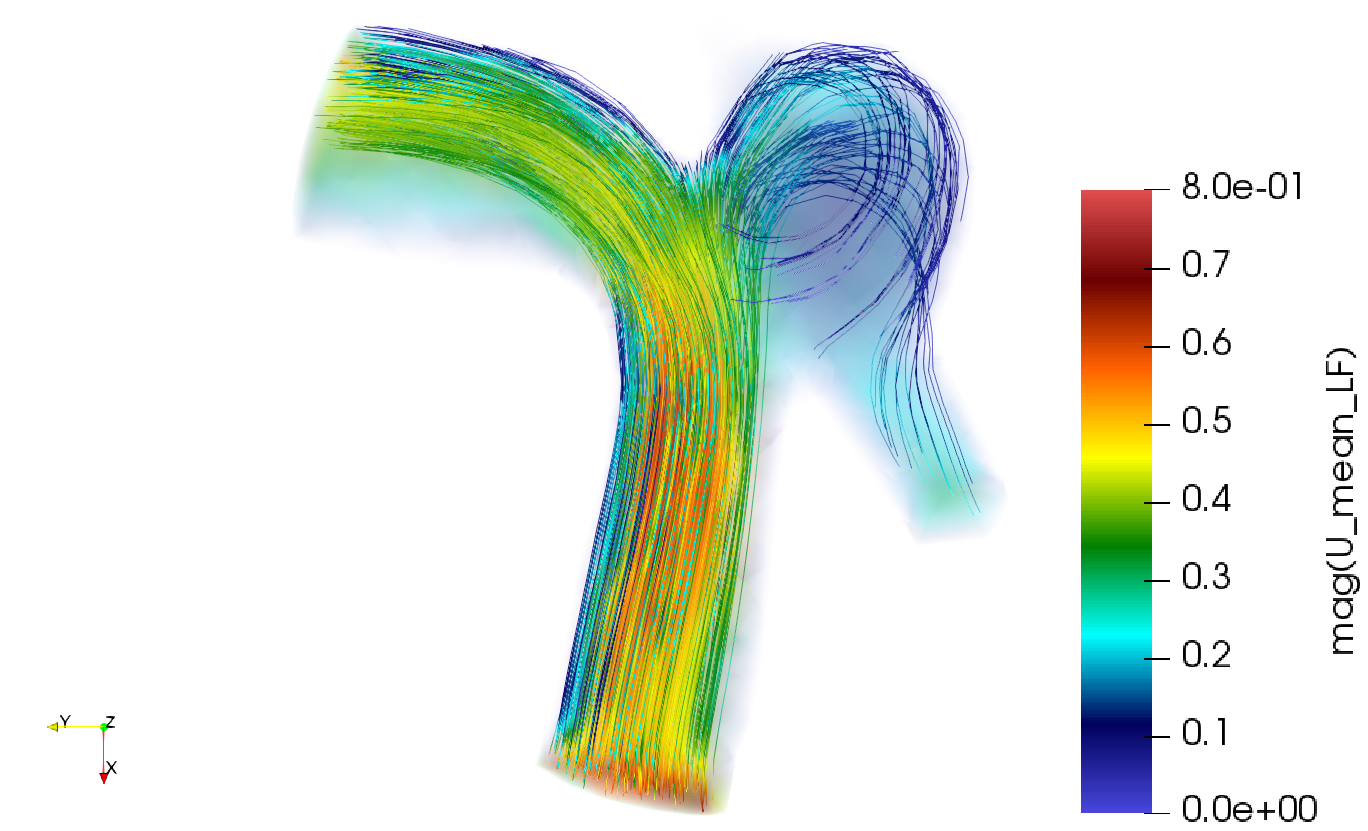}}\\
    \subfloat[HF std]{\includegraphics[width=0.32\textwidth]{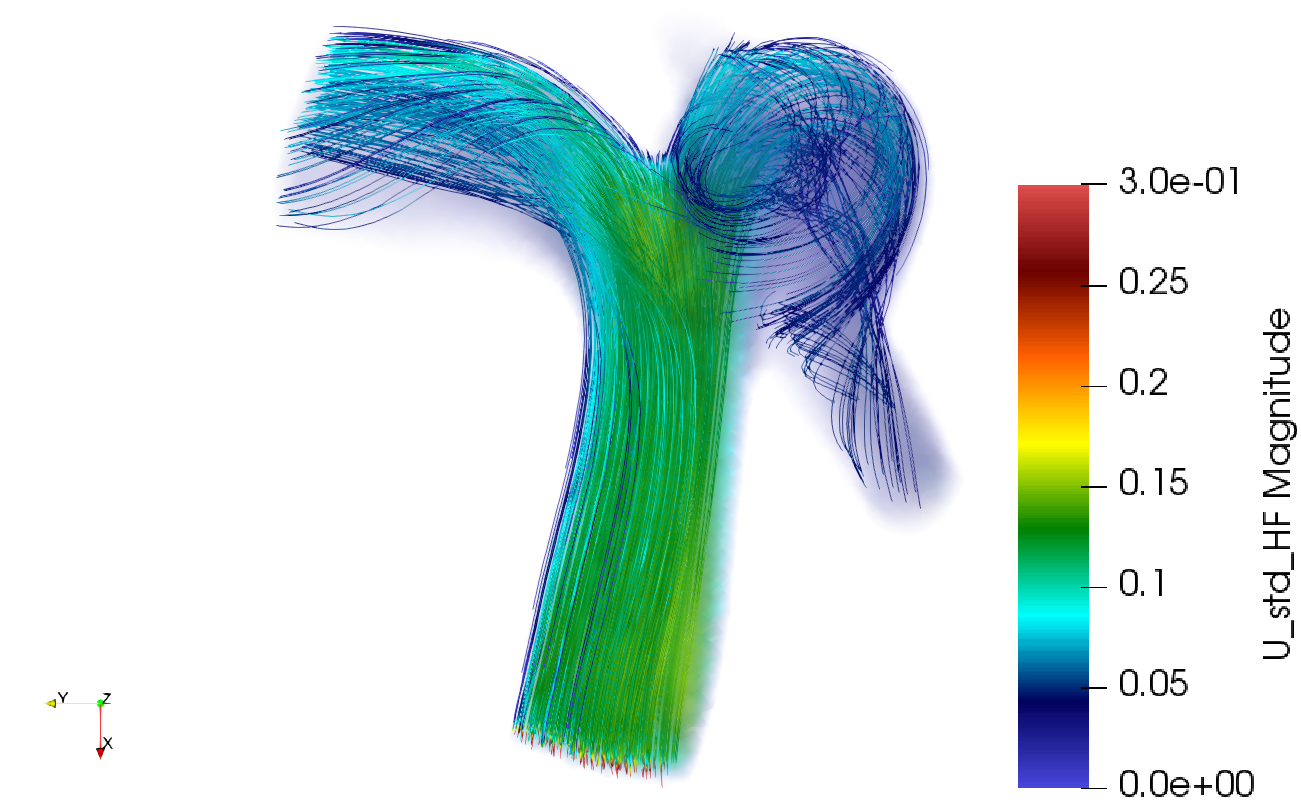}}
    \subfloat[BF std] {\includegraphics[width=0.32\textwidth]{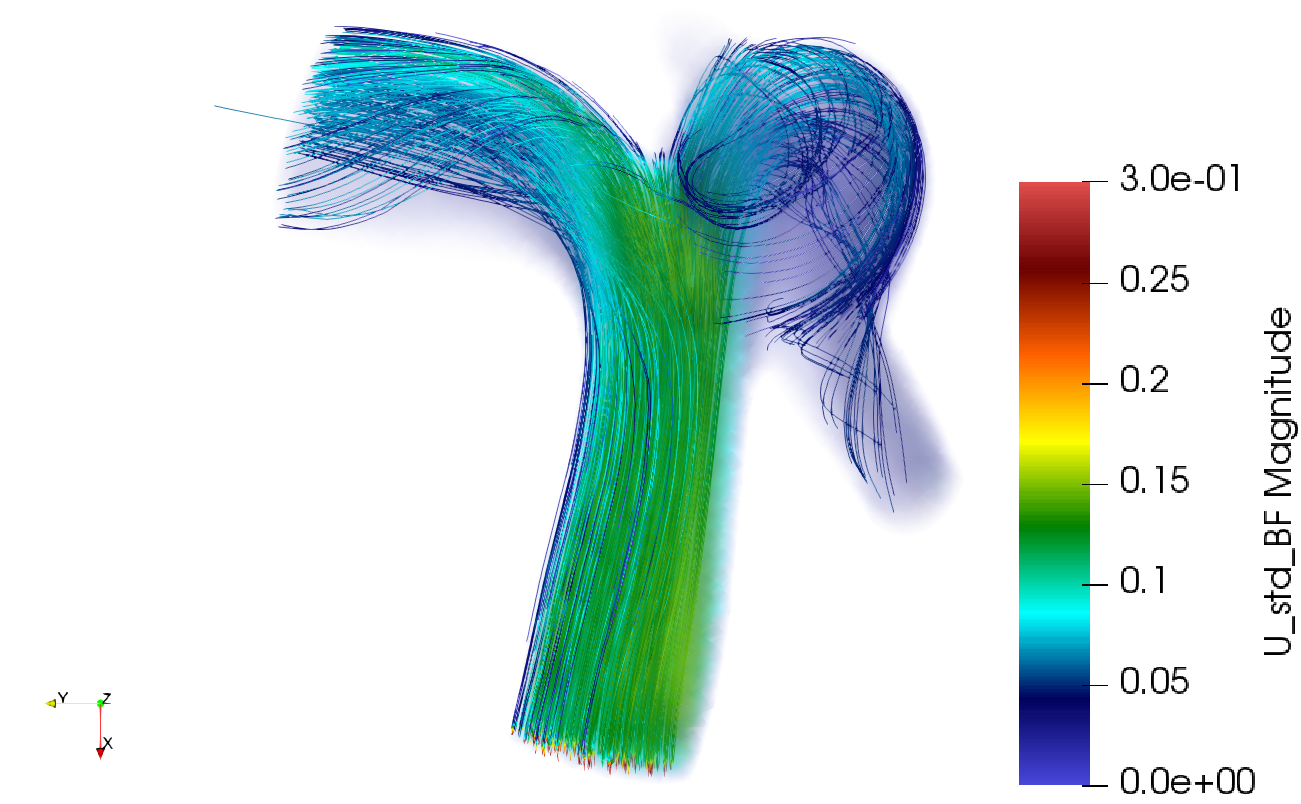}}
    \subfloat[LF std] {\includegraphics[width=0.32\textwidth]{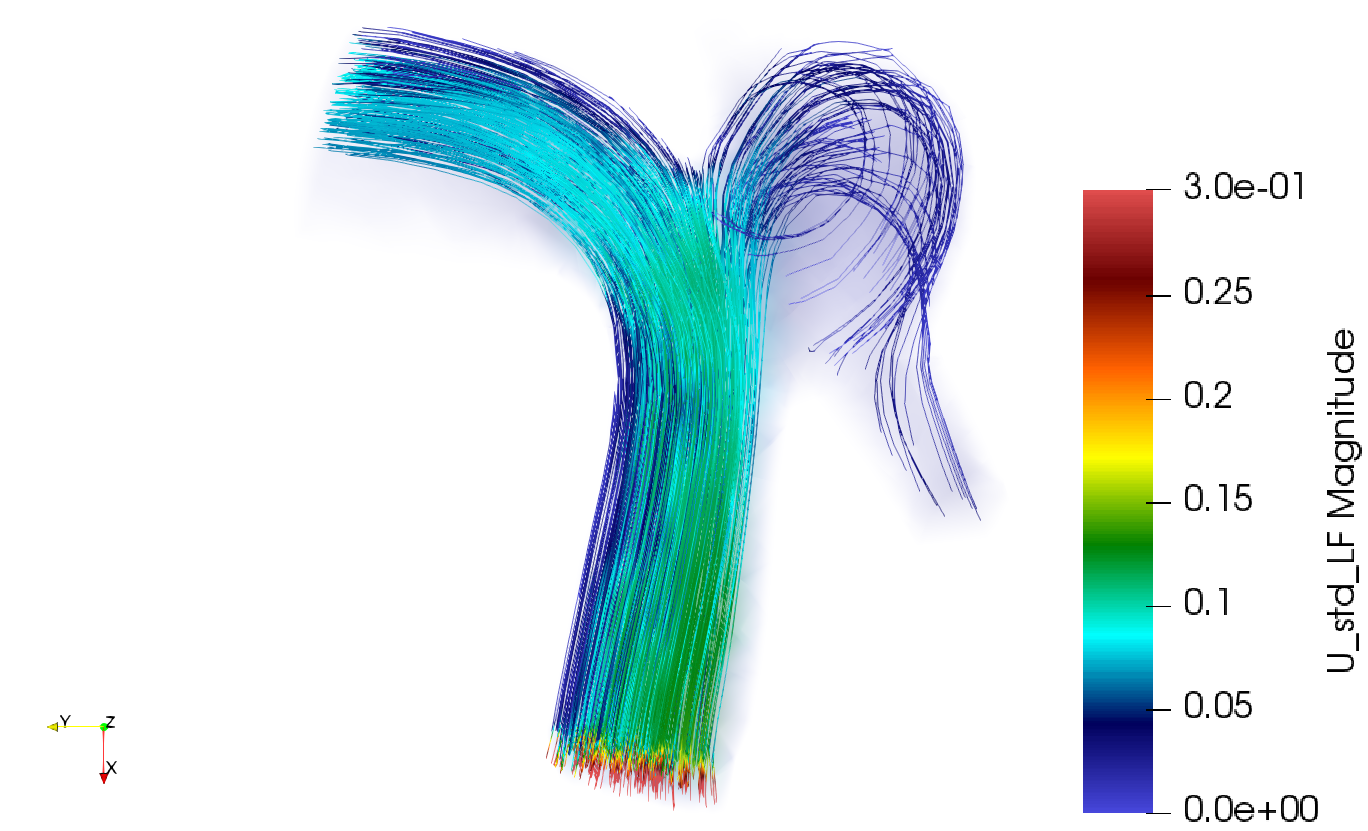}}
    \caption{Volume rendering of the (a-c) mean and (d-f) standard deviation of the internal velocity fields over the test set. Flow streamlines are also plotted with each velocity contour.}
    \label{fig:velocitycontourcase3}
\end{figure}
The mean and std values of internal velocity ensemble are shown in Fig.~\ref{fig:velocitycontourcase3}, where the streamlines are also plotted to visualize the mean flow field. We can see that the flow pattern is very complex, especially in the aneurysm region. A large recirculation and several vortical structures are observed within the aneurysm (Fig.~\ref{fig:velocitycontourcase3}a). The outflows from the aneurysm to two bifurcation arms remain strongly helical and the vortex core line is eccentric along the radial directions of the arteries. The complexity of the intra-aneurysmal flow patterns is largely caused by the geometric complexity in the patient-specific case, and thus the extrapolation from an idealized aneurysm model is usually not sufficient~\cite{sforza2009hemodynamics}. The LF-predicted mean velocity field (Fig.~\ref{fig:velocitycontourcase3}c) has a large discrepancy compared to the HF ground truth (Fig.~\ref{fig:velocitycontourcase3}c), particularly within the aneurysm, where the vortex intensity is notably under-predicted and different flow patterns are observed. In contrast, the BF model accurately captures the mean flow pattern and the prediction shows significant improvement (Fig.~\ref{fig:velocitycontourcase3}b). Figure~\ref{fig:velocitycontourcase3}d shows the std field of velocities obtained by the HF model, suggesting a large scattering of velocity magnitude due to the inflow perturbations. In the aneurysm, the std value is nearly $50\%$ of the local velocity magnitude, showing that the inflow has large effects on the propagated velocities. The BF surrogate model can accurately propagate these inflow uncertainties and the BF-predicted std contour (Fig.~\ref{fig:velocitycontourcase3}e) well agrees with the HF result (Fig.~\ref{fig:velocitycontourcase3}d), while the LF model largely under-estimates the uncertainty, especially within the aneurysm region (Fig.~\ref{fig:velocitycontourcase3}f).

\begin{figure}[htb]
  \centering
    \subfloat[HF mean]{\includegraphics[width=0.32\textwidth]{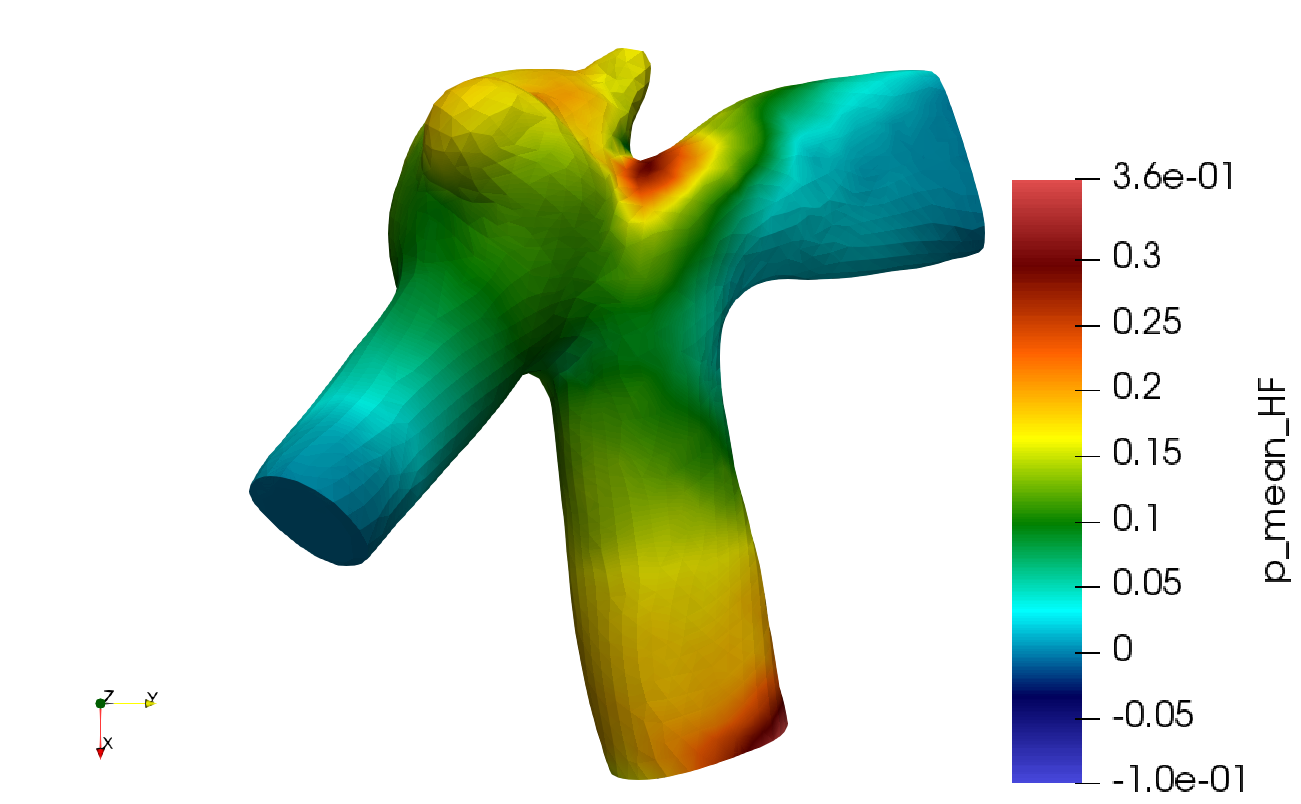}}
    \subfloat[BF mean] {\includegraphics[width=0.32\textwidth]{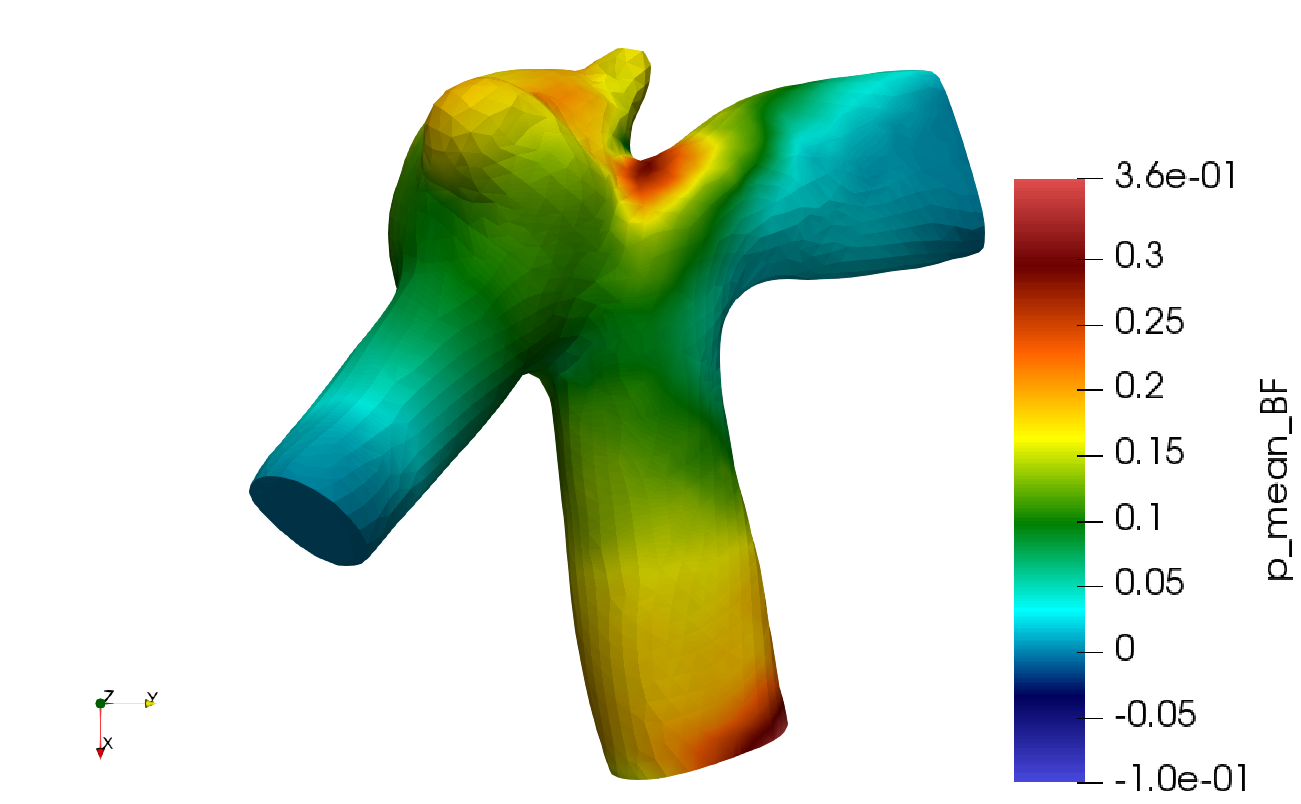}}
    \subfloat[LF mean] {\includegraphics[width=0.32\textwidth]{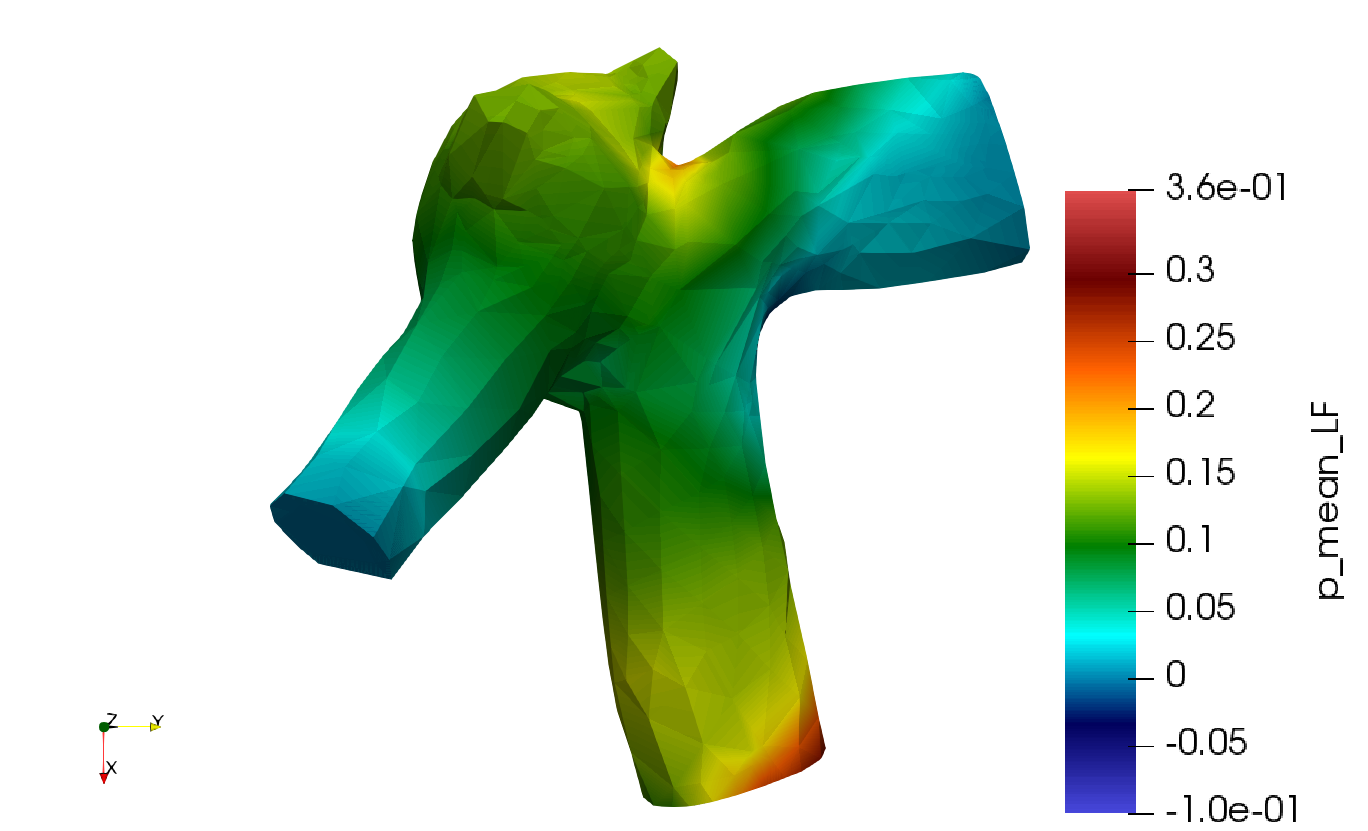}}\\
    \subfloat[HF std]{\includegraphics[width=0.32\textwidth]{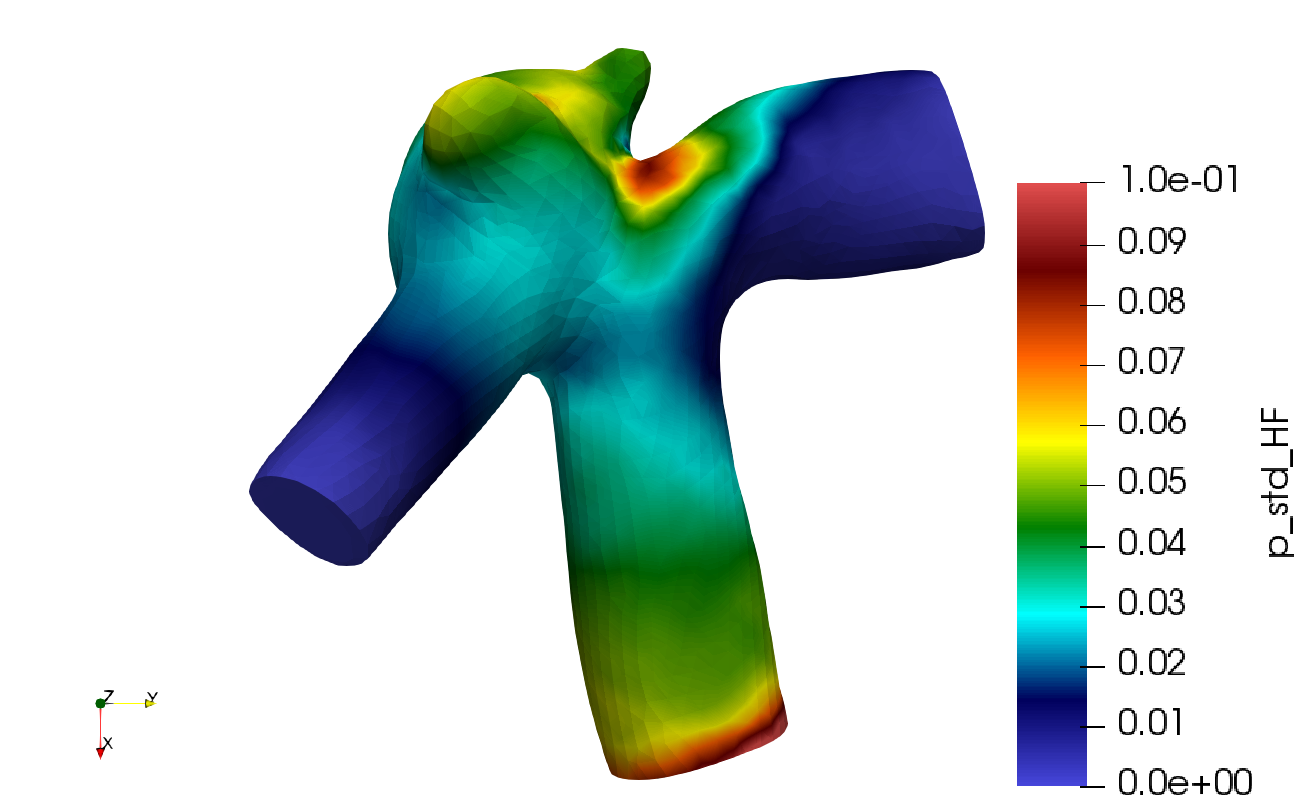}}
    \subfloat[BF std] {\includegraphics[width=0.32\textwidth]{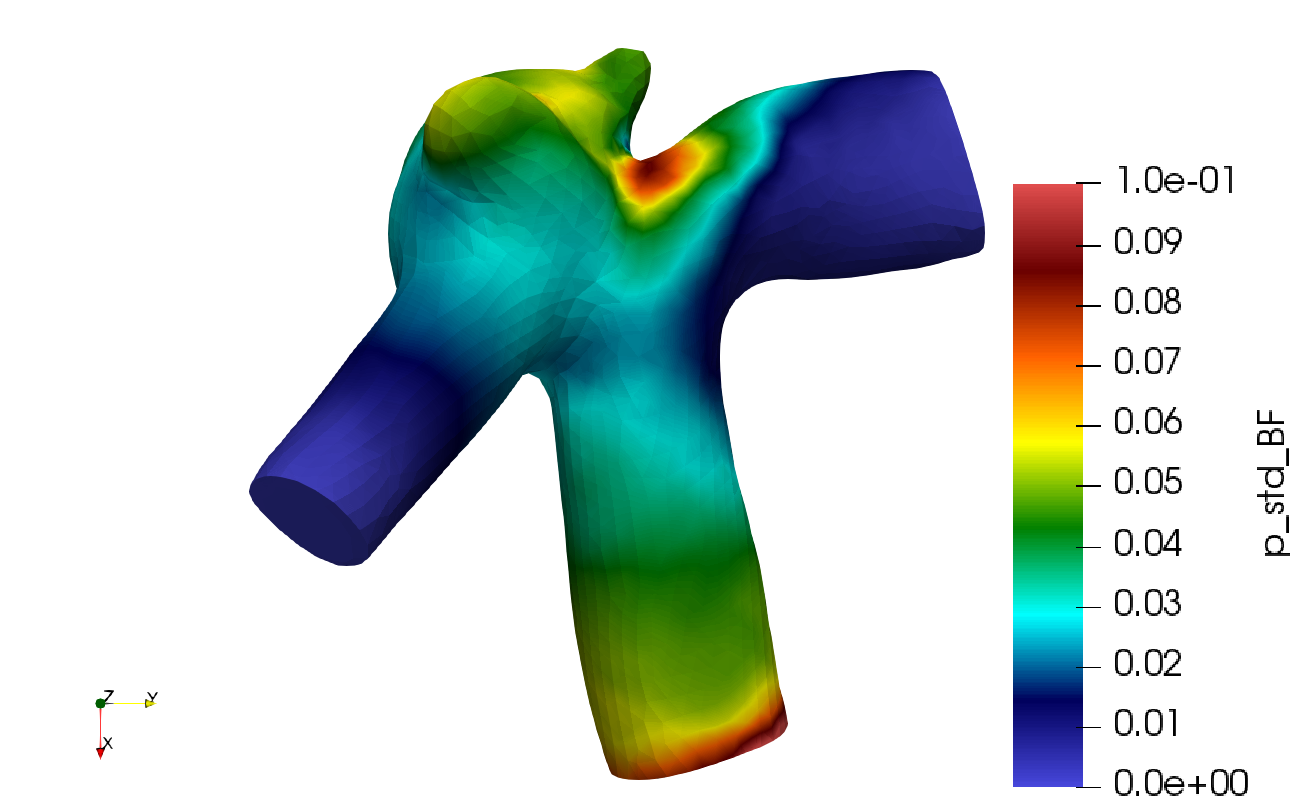}}
    \subfloat[LF std] {\includegraphics[width=0.32\textwidth]{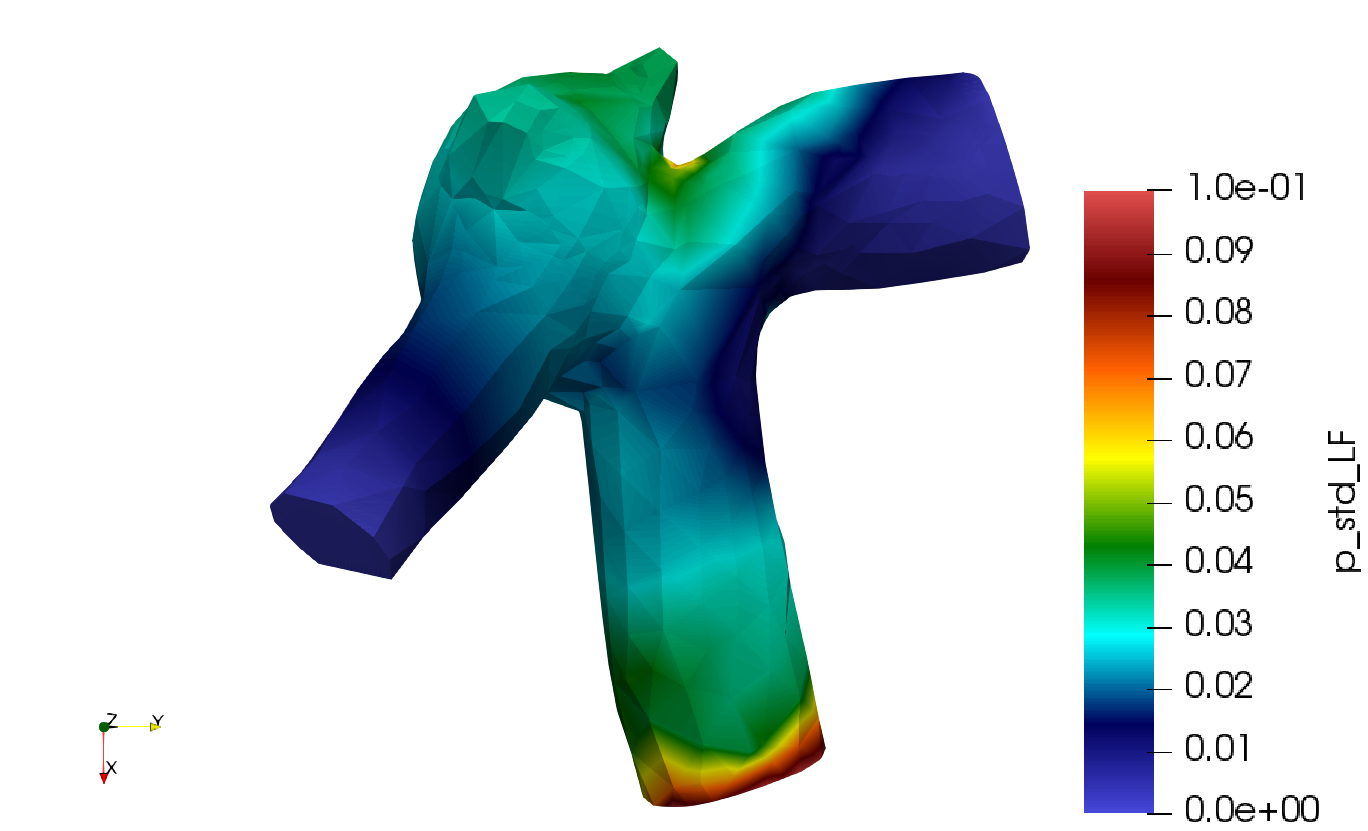}}
    \caption{Contours of the (a-c) mean and (d-f) standard deviation of the pressures on the vessel wall over the test set.}
    \label{fig:pressurecontourcase3}
\end{figure}
The mean and std values of the wall pressures of the test ensemble are plotted in Fig.~\ref{fig:pressurecontourcase3}. The average wall pressure is maximum around the stagnation point at the junction between the aneurysm and right bifurcation arm (Fig.~\ref{fig:pressurecontourcase3}a). Similarly, the BF surrogate can accurately predict the mean pressure patterns on the wall (Fig.~\ref{fig:pressurecontourcase3}b), which shows a significant improvement over the LF solutions (Fig.~\ref{fig:pressurecontourcase3}c). Moreover, the uncertainties of the wall pressure fields are captured by the BF surrogate very well as the contour plot of the std field (Fig.~\ref{fig:pressurecontourcase3}e) is almost identical to the HF ground truth (Fig.~\ref{fig:pressurecontourcase3}d), while the LF under-estimates the large variance near the junction region of the bifurcation (Fig.~\ref{fig:pressurecontourcase3}f).        
 
\begin{figure}[htb!]
  \centering
    \subfloat[HF mean]{\includegraphics[width=0.33\textwidth]{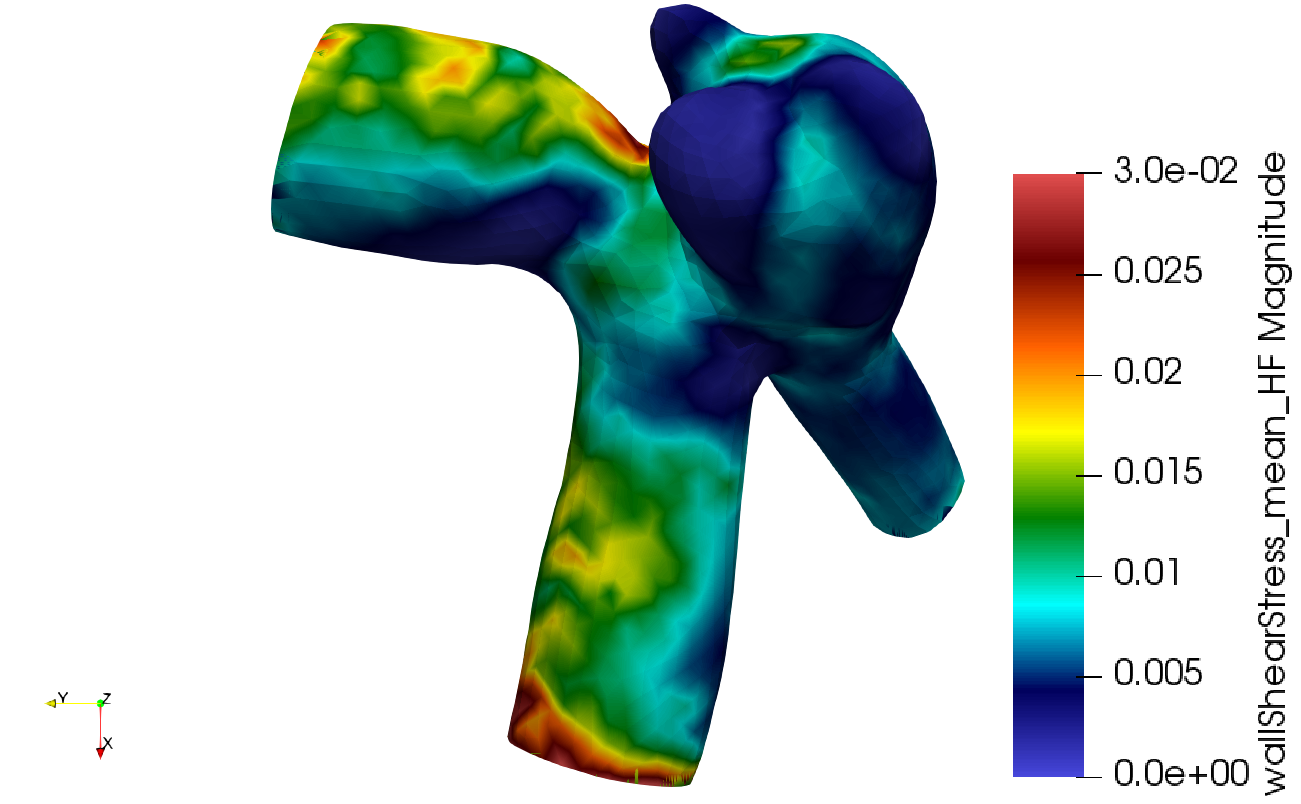}}
    \subfloat[BF mean] {\includegraphics[width=0.33\textwidth]{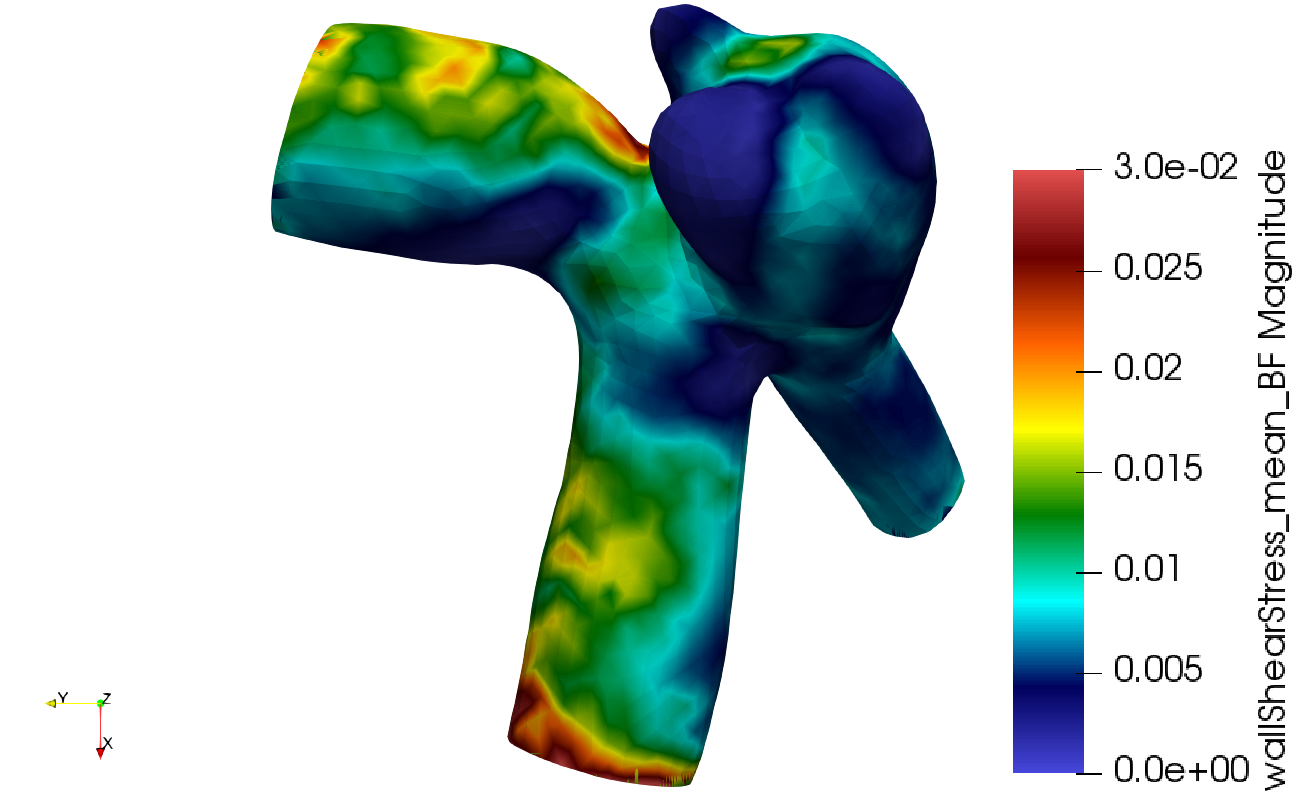}}
    \subfloat[LF mean] {\includegraphics[width=0.33\textwidth]{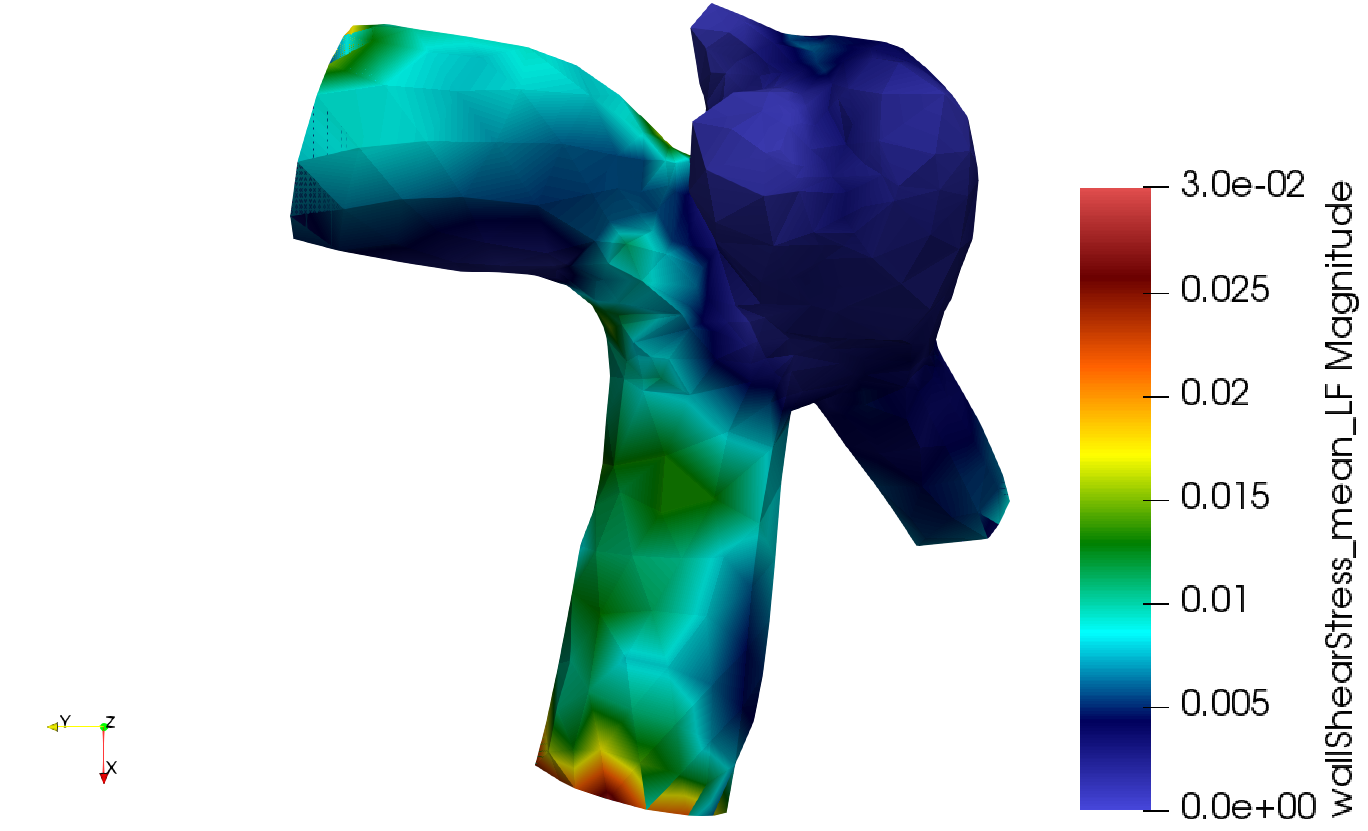}}\\
    \subfloat[HF std]{\includegraphics[width=0.33\textwidth]{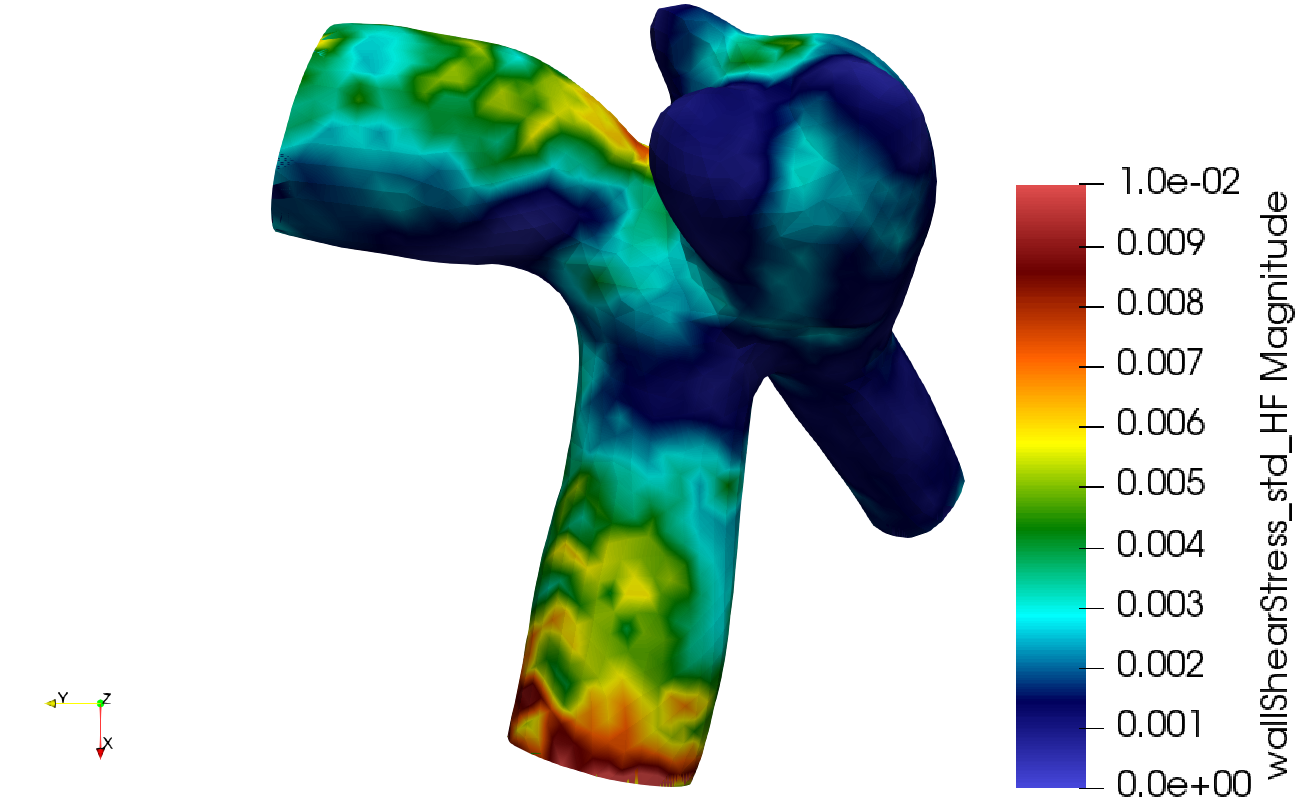}}
    \subfloat[BF std] {\includegraphics[width=0.33\textwidth]{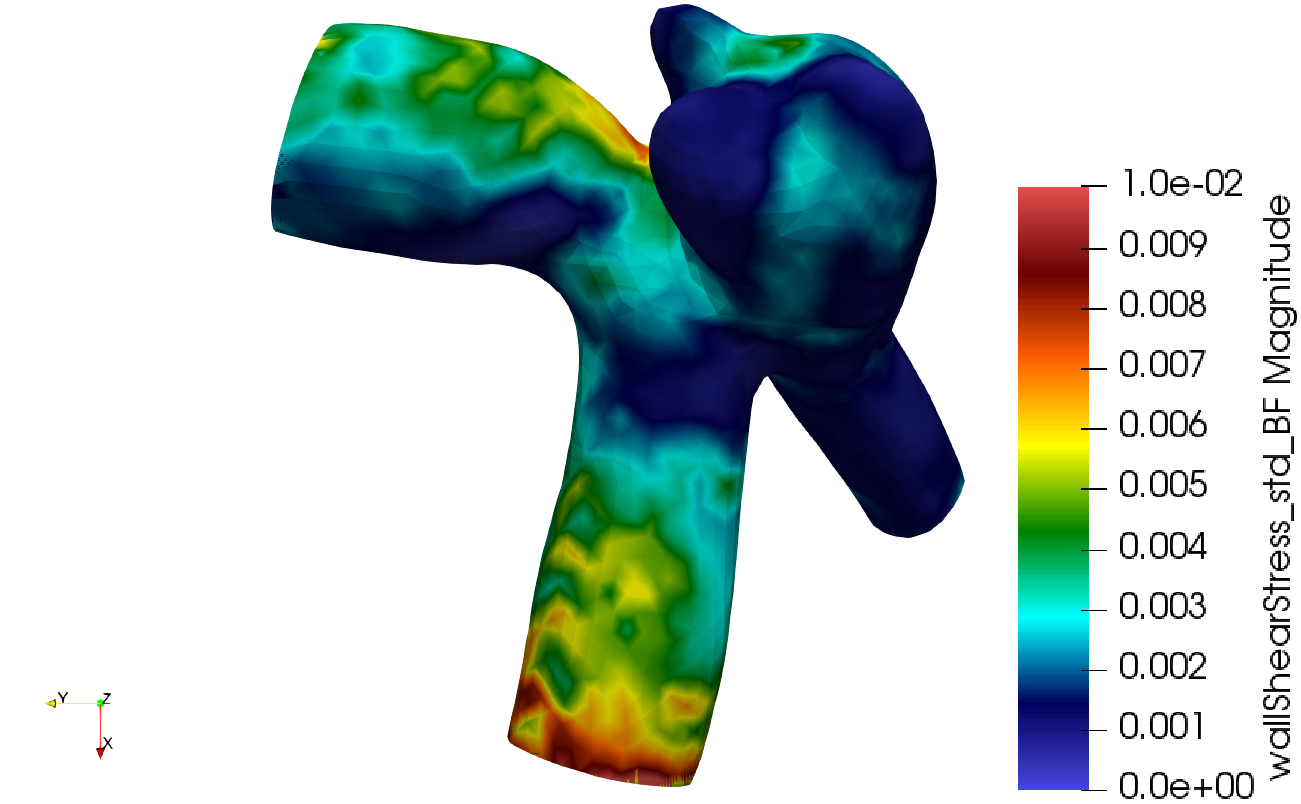}}
    \subfloat[LF std] {\includegraphics[width=0.33\textwidth]{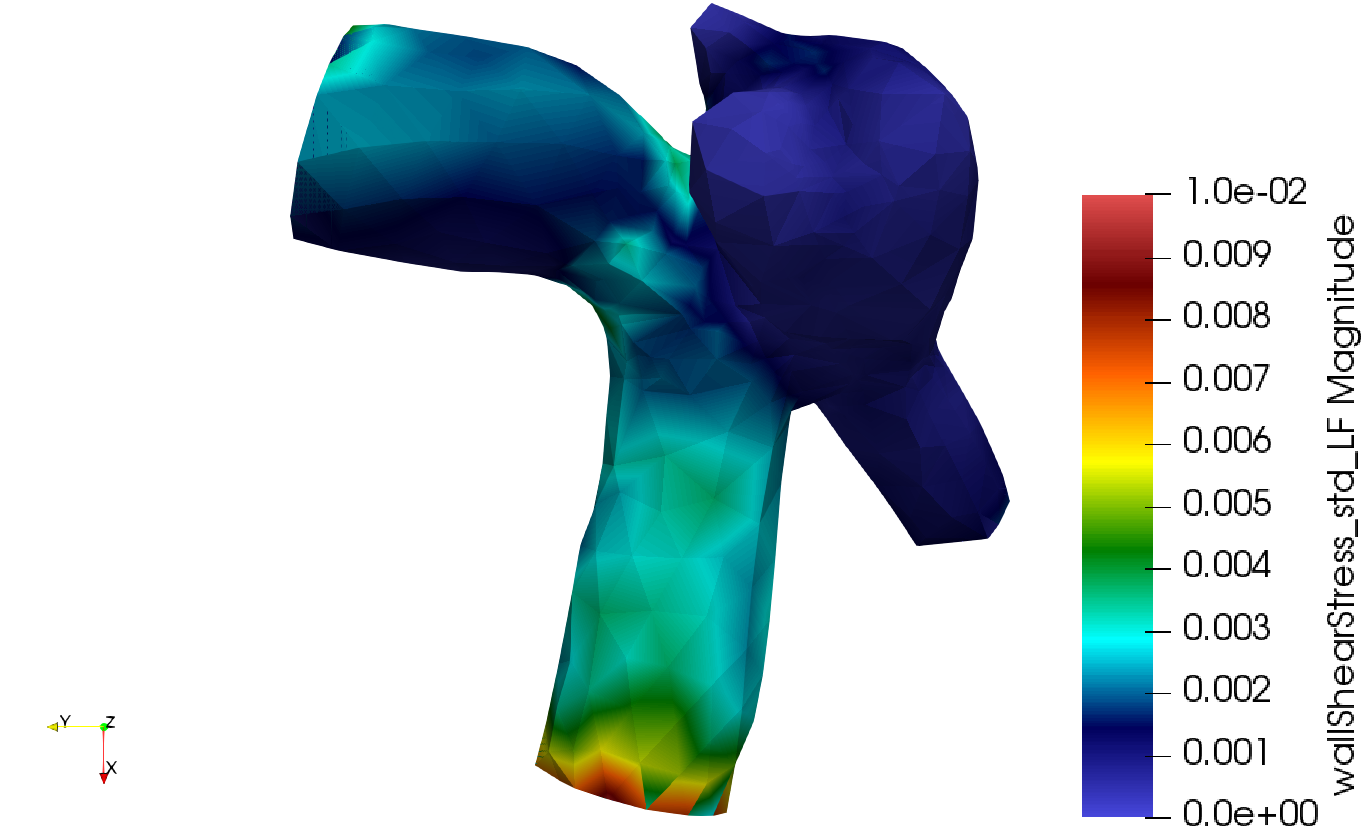}}
    \caption{Contours of the (a-c) mean and (d-f) standard deviation of the wall shear stress (WSS) magnitudes over the test set.}
    \label{fig:wsscontourcase3}
\end{figure}
Wall shear stress (WSS) is one of the most critical hemodynamic factors to the growth and rupture of cerebral aneurysms since the endothelial cells of vascular walls are capable to sense WSS and lead to the growth remodeling of vessel structures~\cite{drexler1999endothelial}. Therefore, an accurate prediction of the WSS field with quantified uncertainties is extraordinarily crucial in this case. Fig.~\ref{fig:wsscontourcase3} shows the mean and std fields of WSS results obtained by HF, BF, and LF models. The mean WSS field exhibits a very complex pattern, where the high and low WSS values are irregularly distributed on the entire vascular domain. In general, we can see that the WSS is relatively low over the aneurysm region, where the backflows, recirculation, and flow stagnation are more likely to happen. Compared to the HF solutions (Fig.~\ref{fig:wsscontourcase3}a) LF-predicted WSS results miss many details and features (Figure~\ref{fig:wsscontourcase3}c), particularly in the aneurysm region. Moreover, the propagated uncertainties are significantly under-estimated by the LF model. In contrast, the BF surrogate model shows significant improvement, and both the mean and std contours (Figs.~\ref{fig:wsscontourcase3}b and~\ref{fig:wsscontourcase3}e) are in a good agreement with the HF ground truth (Figs.~\ref{fig:wsscontourcase3}a and~\ref{fig:wsscontourcase3}d).  

\begin{figure}[htb]
  \centering
    \subfloat[HF (front view)]{\includegraphics[width=0.32\textwidth]{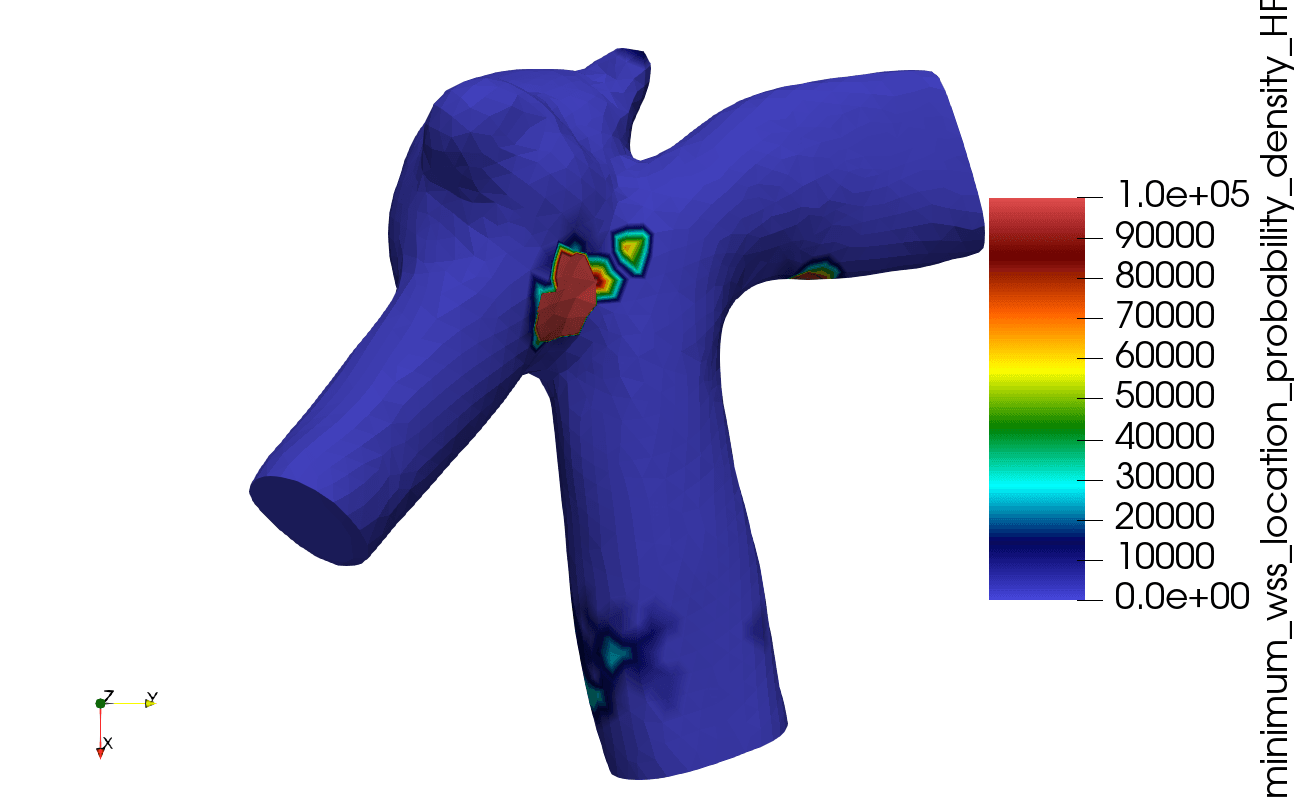}}
    \subfloat[BF (front view)]{\includegraphics[width=0.32\textwidth]{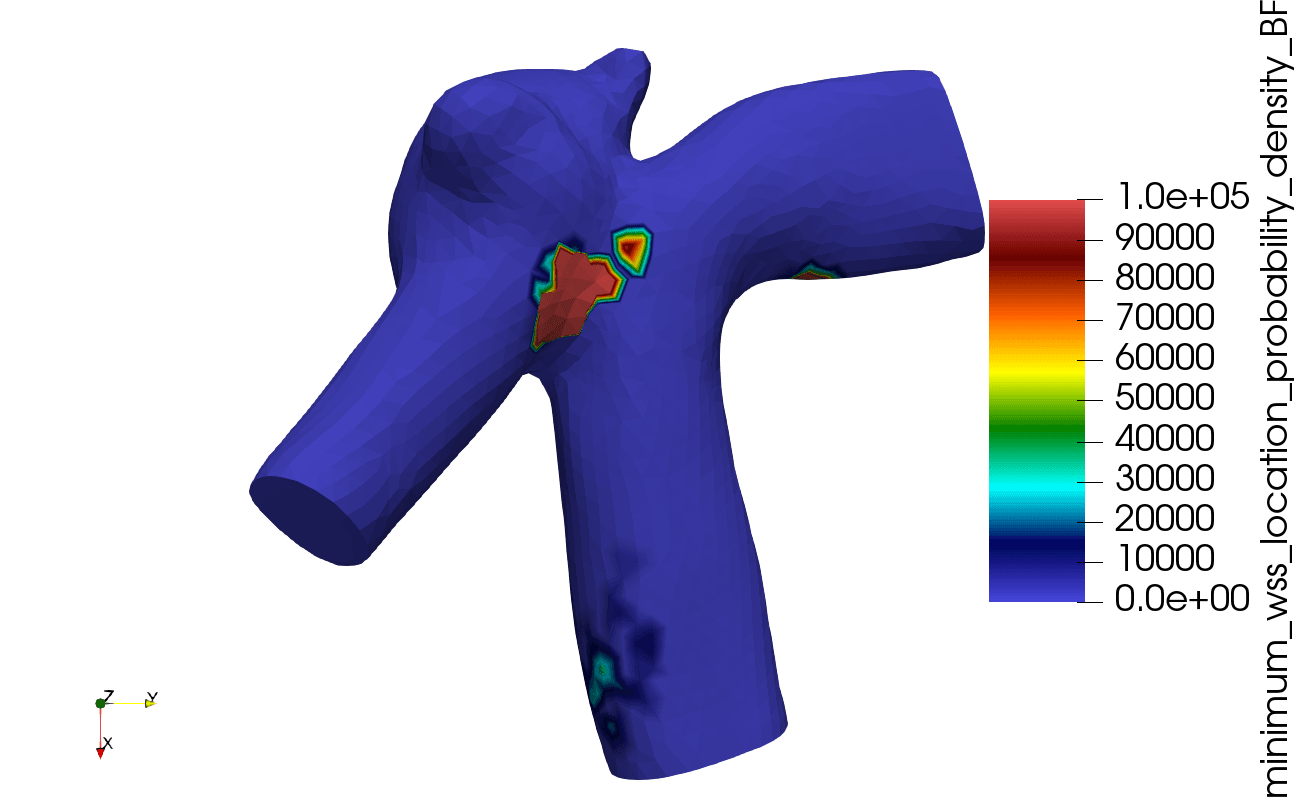}}
    \subfloat[LF (front view)]{\includegraphics[width=0.32\textwidth]{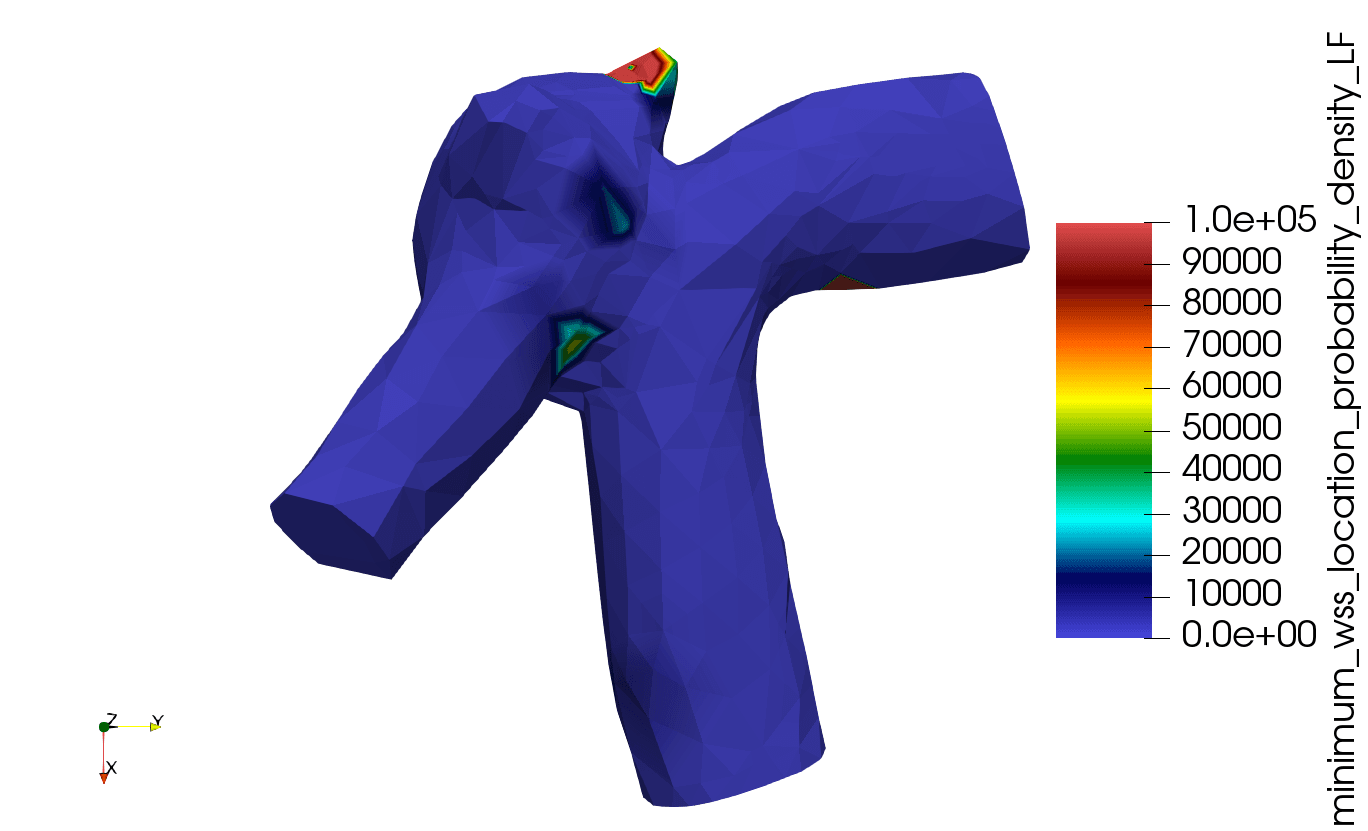}}\\
    \subfloat[HF (back view)]{\includegraphics[width=0.32\textwidth]{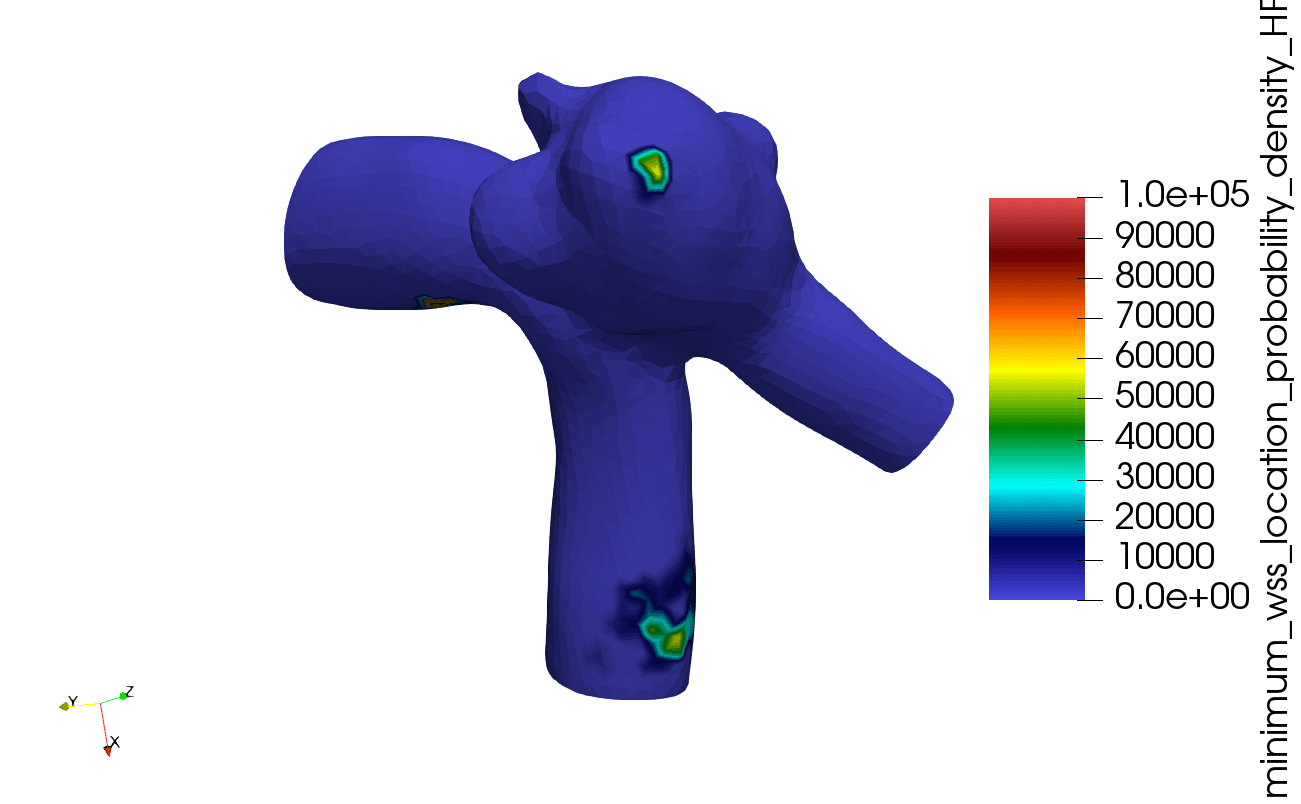}}
    \subfloat[BF (back view)]{\includegraphics[width=0.32\textwidth]{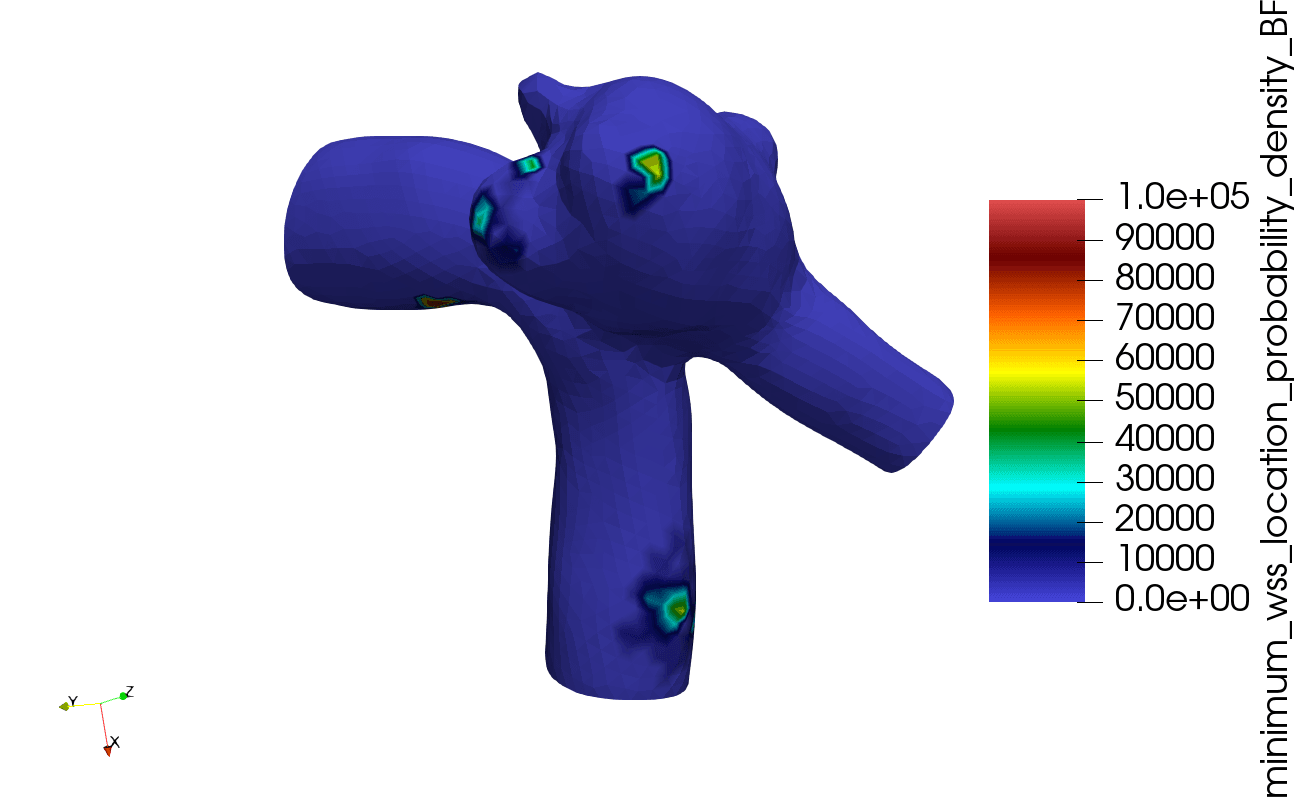}}
    \subfloat[LF (back view)]{\includegraphics[width=0.32\textwidth]{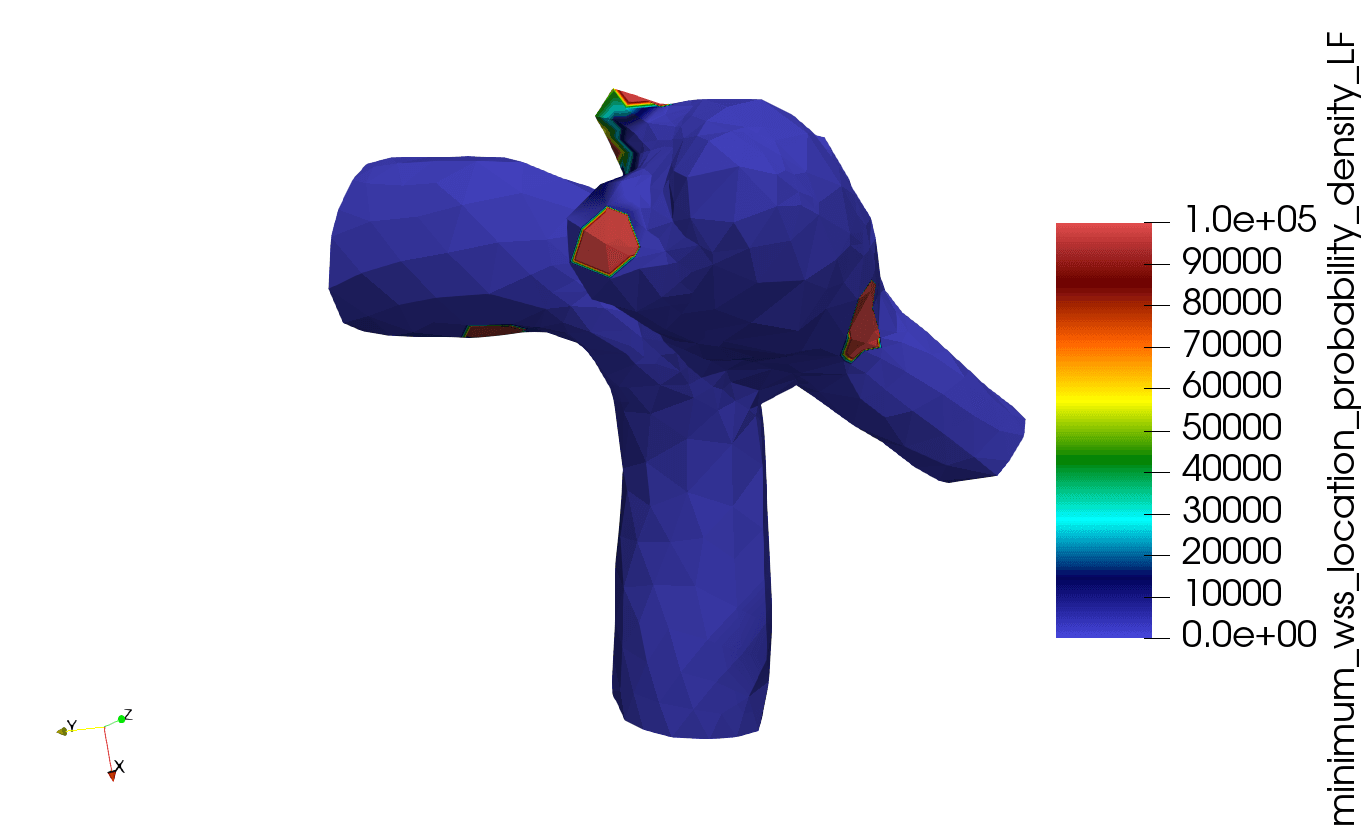}}
    \caption{Probability density distribution of the spatial locations of minimum WSS.}
    \label{fig:rupturepdfcase3}
\end{figure}
Studies have indicated that low WSS causes intimal wall thickening~\cite{dardik2005differential} and might lead to the growth and rupture of aneurysms~\cite{kaminogo2003incidence}. Therefore, it is meaningful to identify the region of the lowest WSS, which is at a high risk of aneurysm rupture. However, this is a challenging task for the surrogate model, since the extreme values are more difficult to capture than the averaged flow quantities are. To better demonstrate the capability of the BF surrogate, we estimated the spatial distributions of the probability density of the minimum WSS locations based on the flow solution ensembles, propagated from the inflow uncertainty by the HF, BF, and LF models, which are shown in Fig.~\ref{fig:rupturepdfcase3}. Based on the HF model results (Figs.~\ref{fig:rupturepdfcase3}a and~\ref{fig:rupturepdfcase3}d), the minimum WSS is located at the lower part of the aneurysm, close to the junction of the parent artery and smaller bifurcation arm. The LF model incorrectly predicts the minimum WSS location, which is at the ridge on the other side of the aneurysm. However, the performance of the BF surrogate model is excellent as the predicted density distribution contour is nearly identical to the HF benchmark results. Note that only 40 HF solutions are used in this case to construct the BF surrogate in a high-dimensional (9-D) parameter space. These comparisons demonstrate that the proposed BF has a great potential for real-world,  UQ problems in cardiovascular applications with high accuracy and remarkable reduction of computational expenses.

\begin{figure}[htb]
\centering
\includegraphics[width=0.5\textwidth]{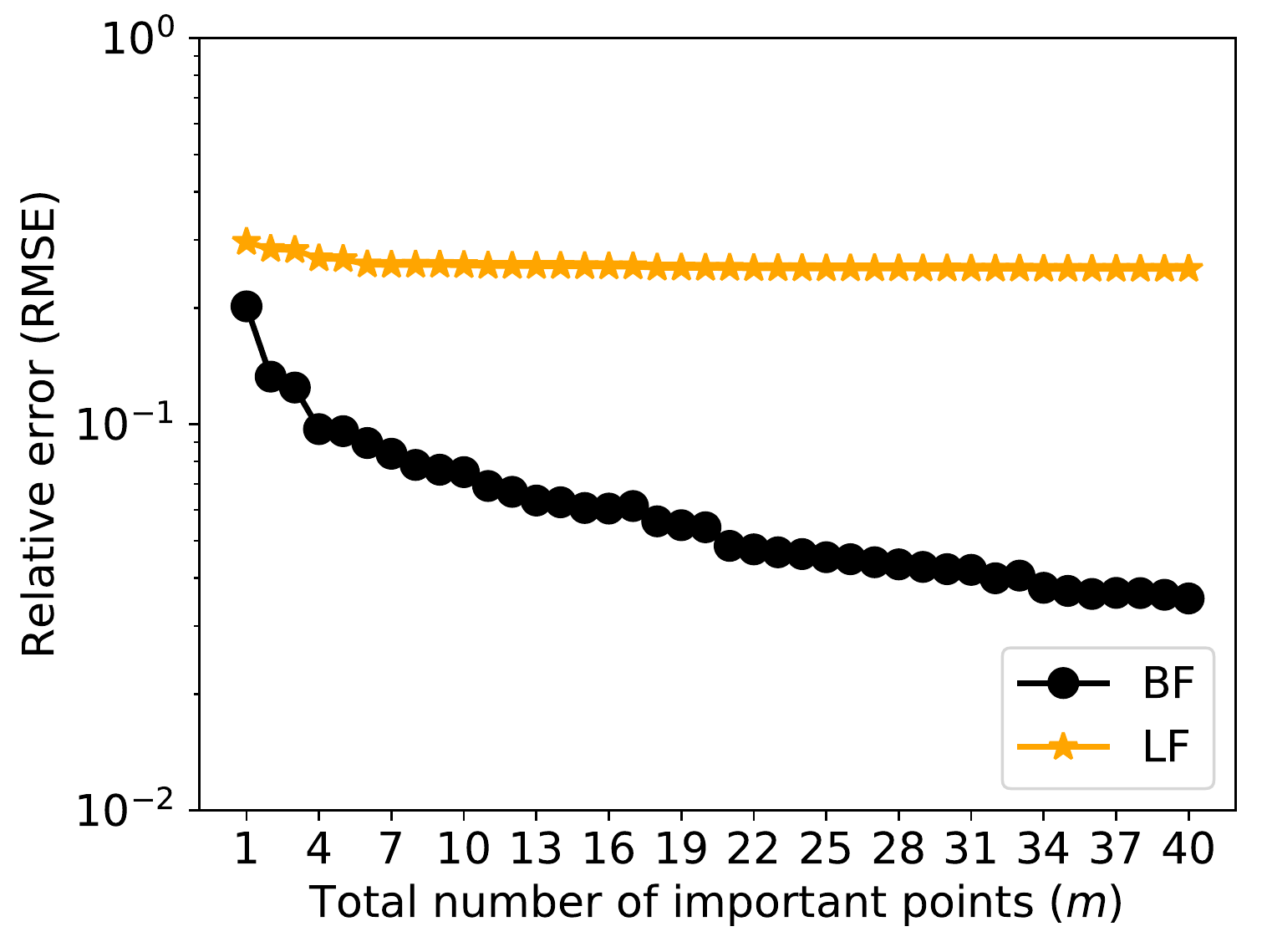}
\caption{The relative root mean squared error (RMSE) of the BF model (black) over 600 test parameter points with respect to the number of important points (HF simulations) in test case 3. The corresponding RMSE of the one with LF basis (orange) are plotted for comparison.}
\label{fig:convergence3}
\end{figure}
Lastly, we present the relative errors of the BF surrogate model with respect to the number ($m$) of HF samples for training in Fig.~\ref{fig:convergence3}, where the LF prediction error is also plotted for comparison. Similar to the previous cases, the LF model has a large relative error, while the BF surrogate with only a few HF solutions can considerably reduce the error by an order of magnitude, though the parameter space is 9-D in this case. Moreover, we observe that the decay of the BF model error is fast even when $m$ is large than 30.     

\section{Discussion on \emph{A Priori} Error Bound Estimation}
\label{sec:discussion}
In practical applications of the BF surrogate, it is useful to know how many important points (i.e., the number of HF simulations) should be involved and what error magnitude is expected on the test set \emph{a priori}. We proposed an empirical approach in Section~\ref{sec:errorEstimate} to assess the model quality and estimate the prediction error of the BF surrogate, where two useful assessment metrics, model similarity $R_s(\mathbf{z})$ (\ref{eqn:relativeDist}) and error component ratio $R_{e}(\mathbf{z})$ (\ref{eqn:relativeDist}), are used. In this section, \emph{a priori} error bound estimations in our three test cases of vascular flows will be discussed based on the proposed assessment method.   
\begin{figure}[htb]
\centering
    {\includegraphics[width=0.5\textwidth]{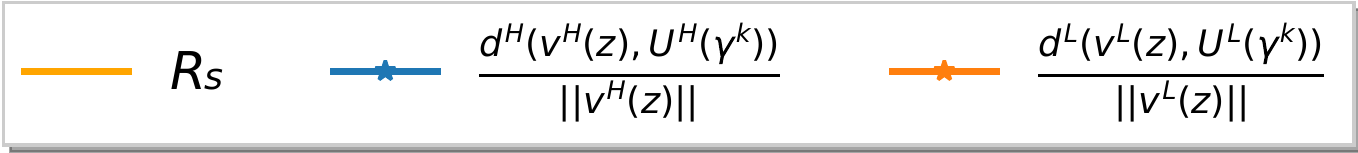}}
    \subfloat[Distance (case 1)]{\includegraphics[width=0.33\textwidth]{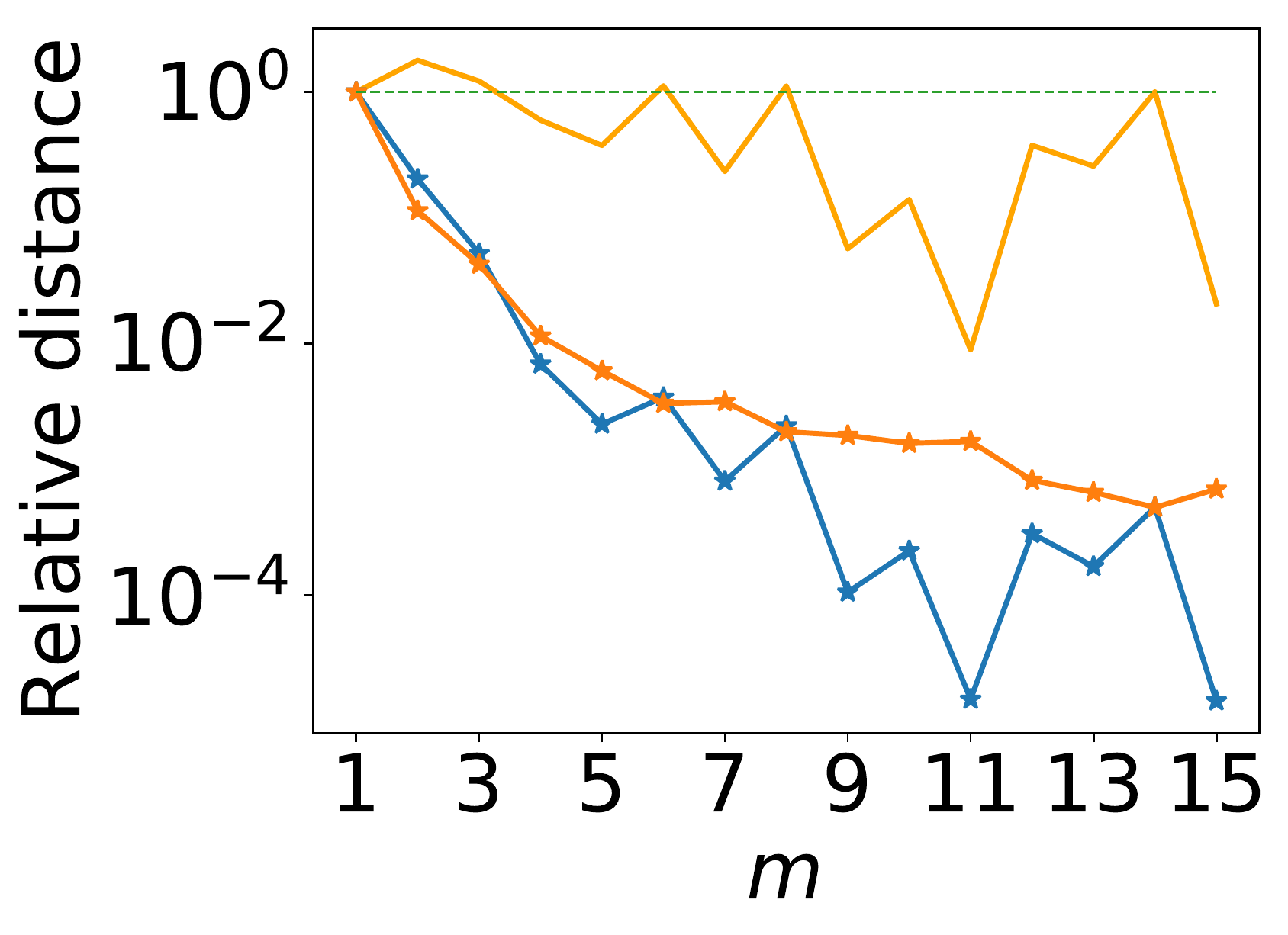}}
    \subfloat[Distance (case 2)]{\includegraphics[width=0.33\textwidth]{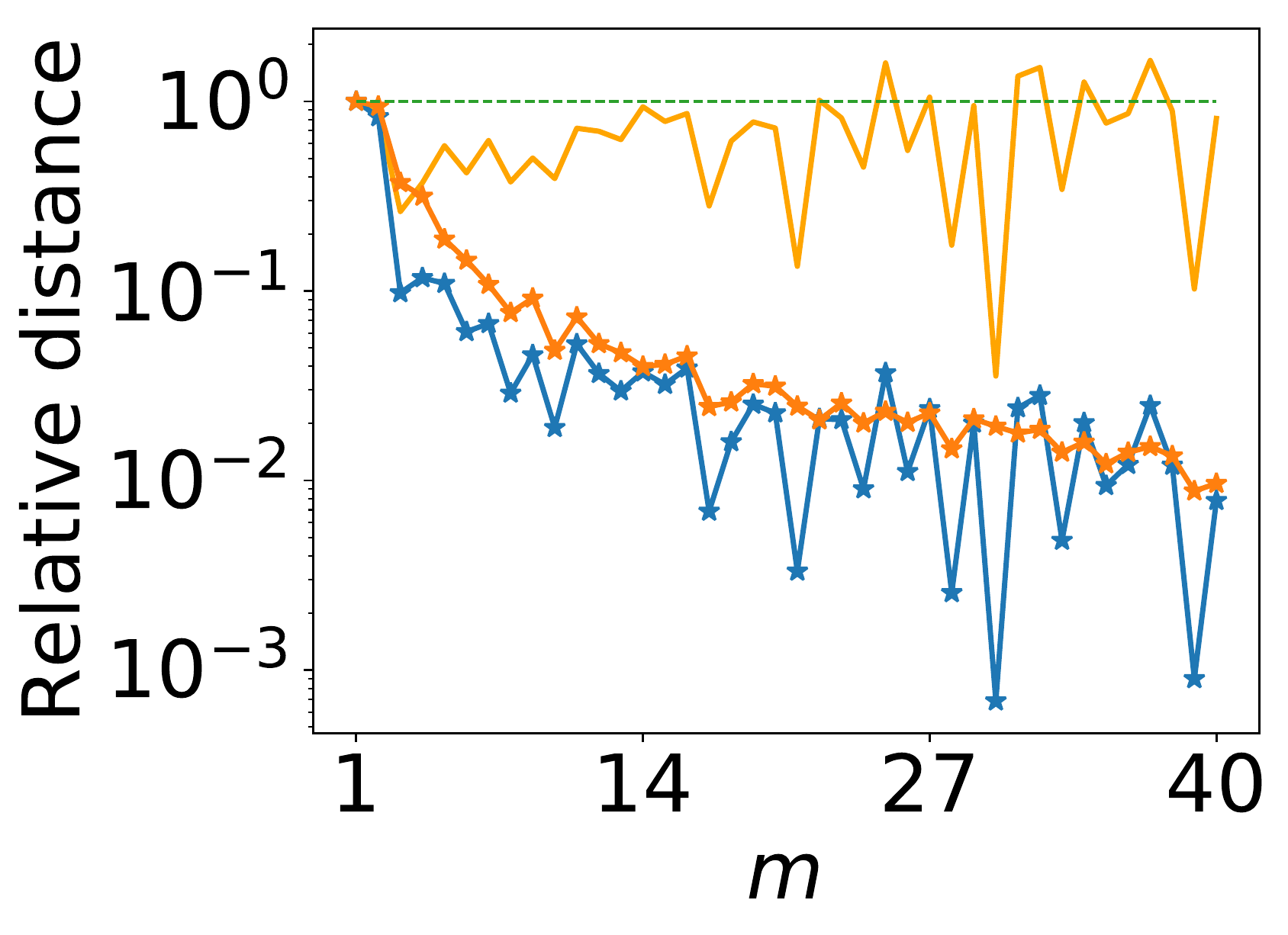}}
    \subfloat[Distance (case 3)]{\includegraphics[width=0.33\textwidth]{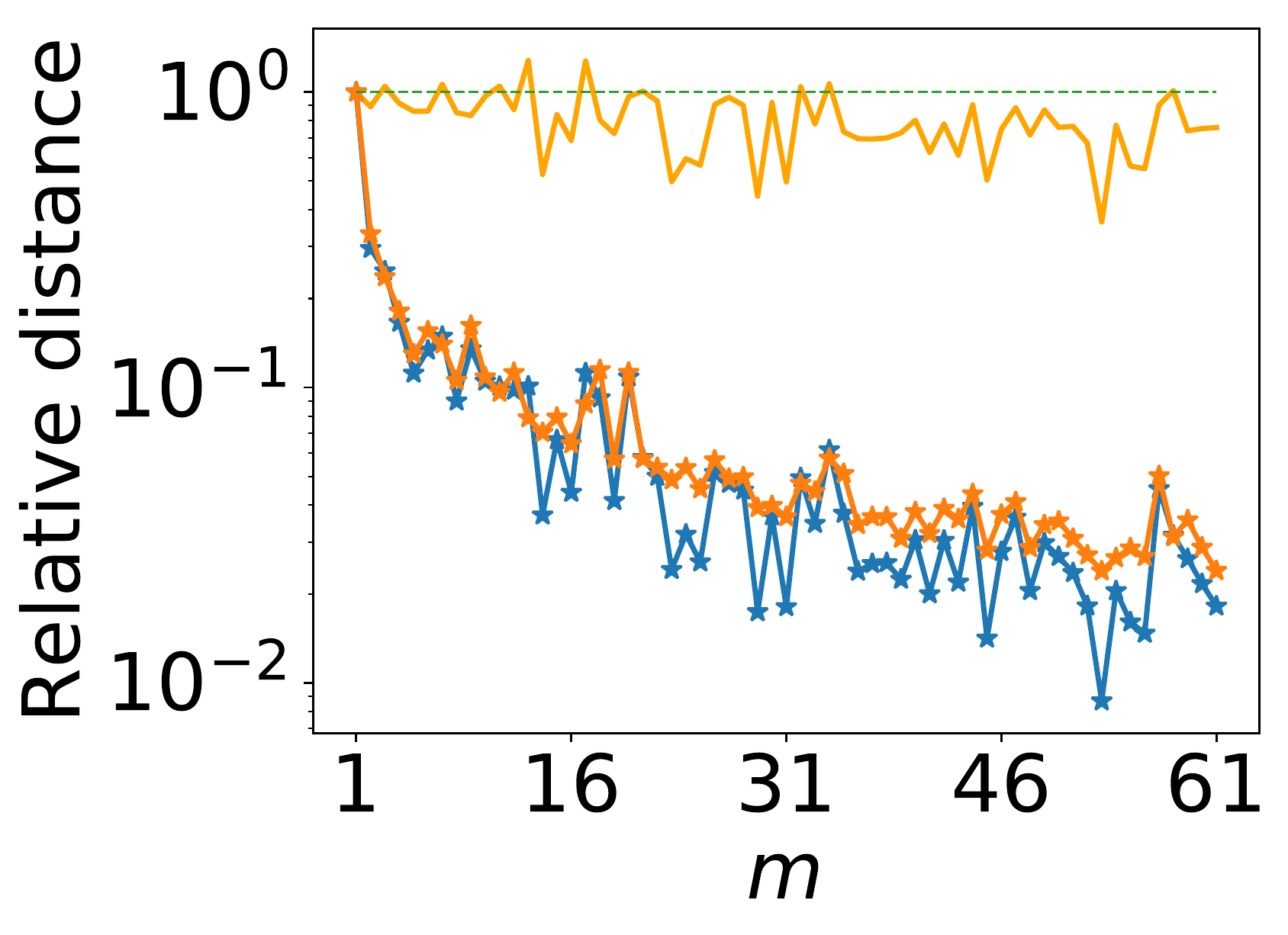}}\\
    \vspace{1.0em}
    {\includegraphics[width=0.5\textwidth]{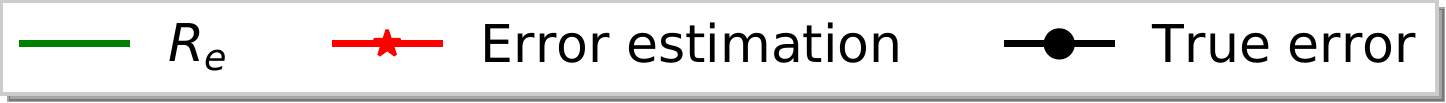}}
    \subfloat[Error bound estimation (case 1)]{\includegraphics[width=0.33\textwidth]{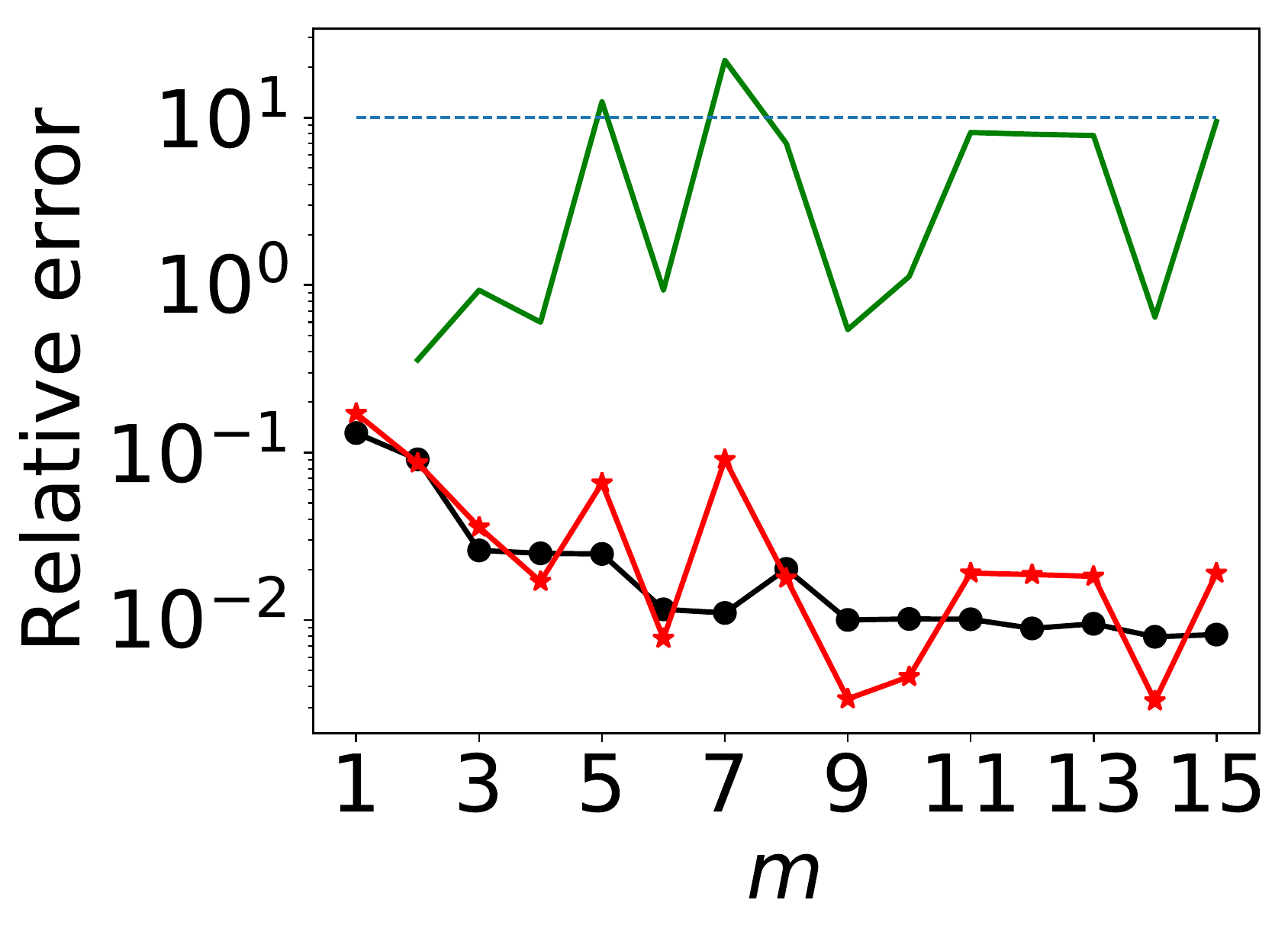}}
    \subfloat[Error bound estimation (case 2)]{\includegraphics[width=0.33\textwidth]{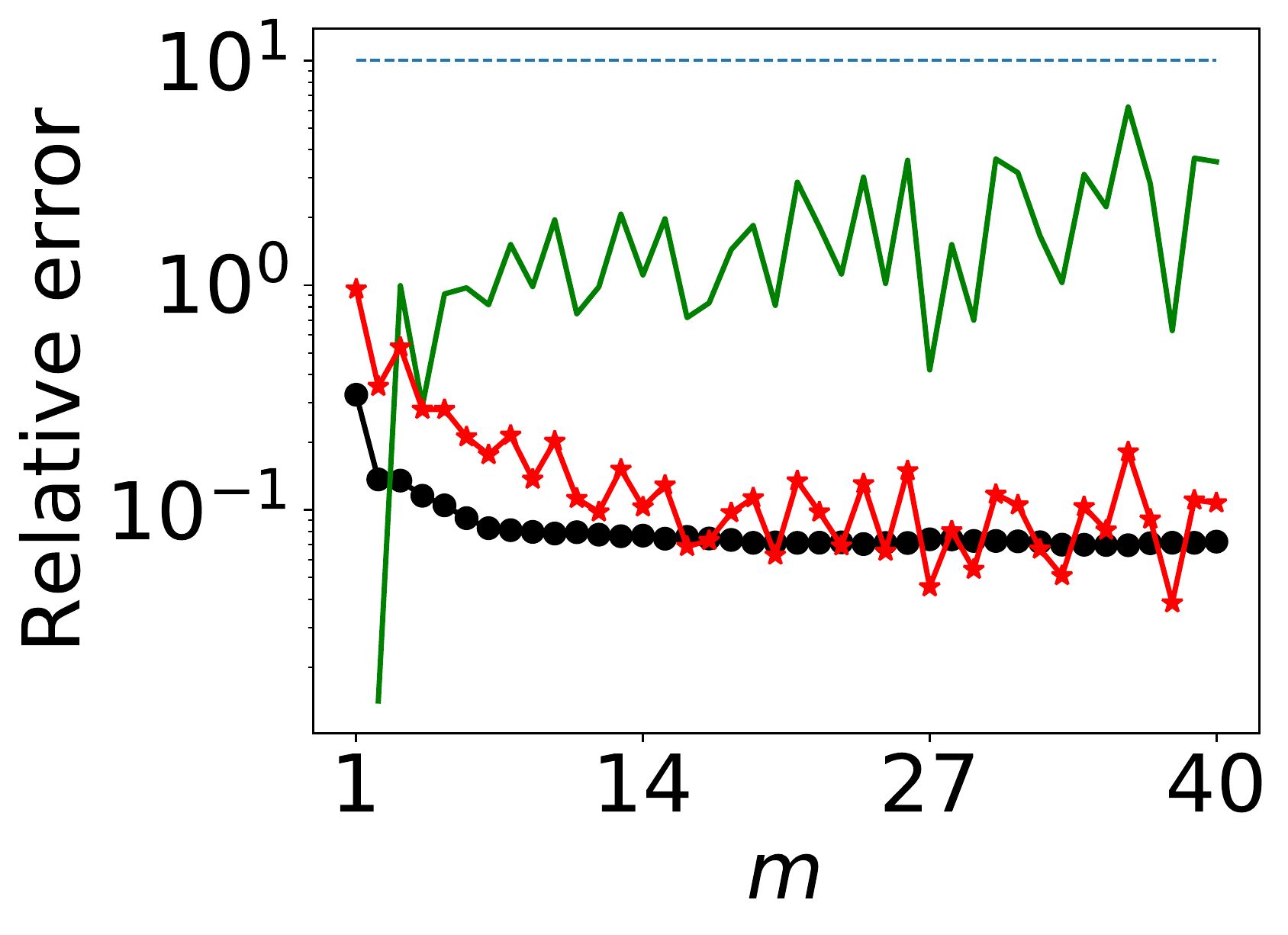}}    
    \subfloat[Error bound estimation (case 3)]{\includegraphics[width=0.33\textwidth]{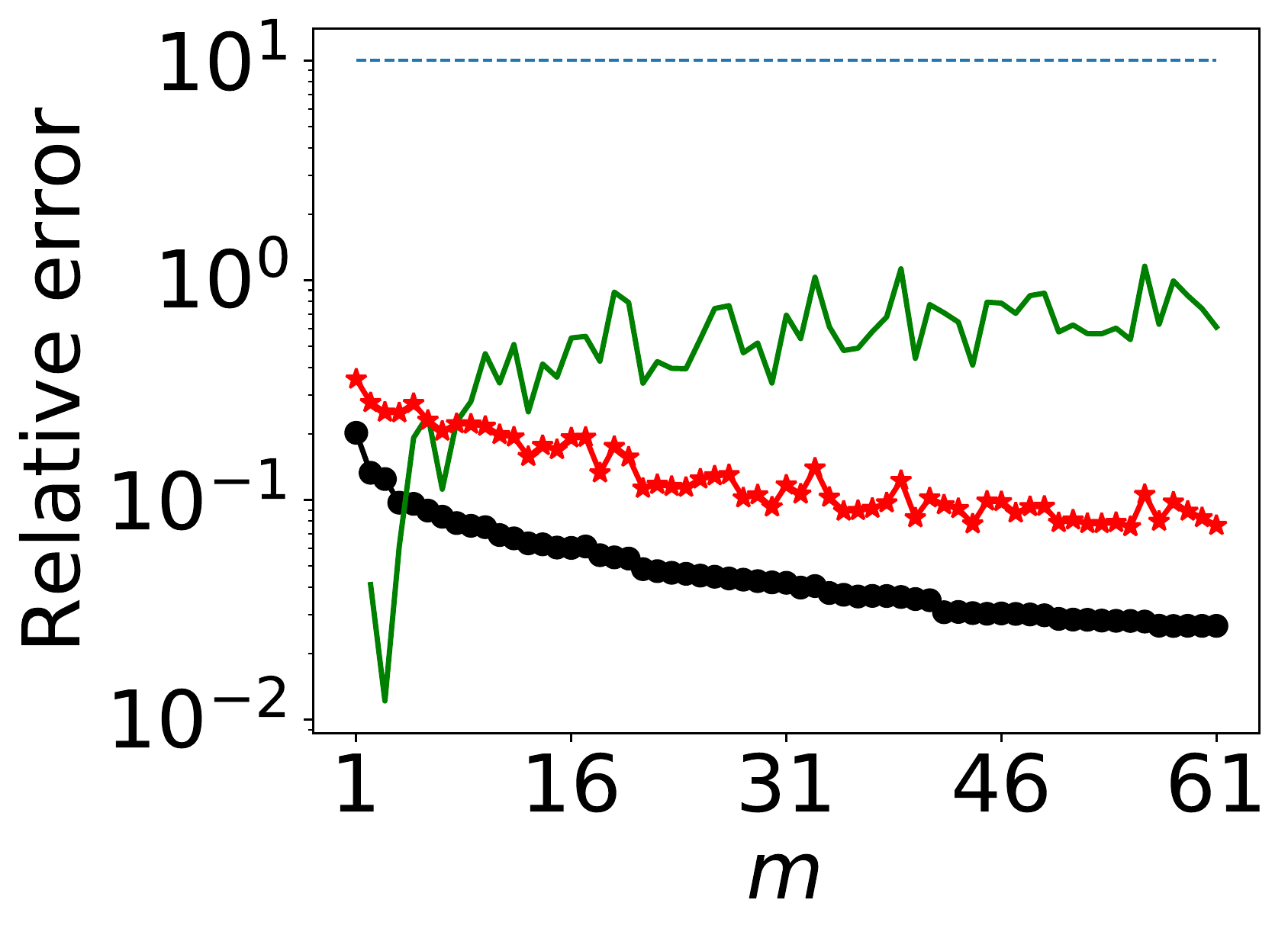}}    
    \caption{ The relative distance (a-c) and error bound estimation (d-e) of the BF models with respect to the number of important points (HF simulations) used for BF reconstruction in cases 1, 2, and 3.}    
    \label{fig:error}
\end{figure}
The model similarity metrics vs. the number of important points used for BF reconstruction over the entire test set are shown in Figs.~\ref{fig:error}(a-c), and the average relative distances from HF/LF solutions on the corresponding test set to the subspace spanned by the previous selected HF solutions are plotted as well. Moreover, the corresponding error bound estimations (\ref{eq:eb2}) of the BF surrogate on the test sets are shown in Figs.~\ref{fig:error}(d-f), where the ground truth of the test errors (black dotted line) are plotted for comparison. For all three cases, the relative distances of both HF and LF solutions decrease as the number of important points grows, which demonstrates the point-selection algorithm is able to pick out important points on which the solutions roughly have the maximum distances to the constructed hi-fidelity approximation space. We can see that the model similarity metric $R_s$ fluctuates under one, especially $R_s$ is close to one for the first a few important points, showing that the LF model is informative for exploring the parameter space, which is one important factor for the success of the bi-fidelity approach observed in the previous sections.

Moreover, the error component ratios $R_{e}$ in three cases mainly remain less than $10$, which indicates that the in-plane error is not dominant over the distance error component, and thus collecting new HF samples based on the max-distance based point selection algorithm is still effective. From Fig.~\ref{fig:error}(d-f), we can see that the empirical error bound estimations can basically capture the trends of the true errors in terms of the number of HF training samples in all cases. Particularly in case 3, where the model similarity $R_s$ is close to one and $R_{e}$ is much less than $10$ (when even $m > 50$), the empirical error estimation bounds the true error well and adding more HF training samples can further reduce the BF prediction error. Compared to case 3, the similarity metric $R_s$ becomes more fluctuating and $R_{e}$ increases significantly in cases 1 and 2 when $m$ grows, and thus the error curves become flat quickly, indicating collecting additional HF simulations does not help to improve the approximation quality of the bi-fidelity surrogate. It is noteworthy that the true error curve (black dotted) are computed based on the HF solutions of the entire test sets (e.g., 600 HF simulations in case 3), while our error bound estimation (red dashed-dotted) is only dependent on the existing pre-selected training HF samples and no  additional high-fidelity samples  are needed. In general, the proposed evaluation metrics and error bound estimation approach provide a practical way to assess the performance of the BF surrogate and enable us to better determine the budget for HF simulations \emph{a priori}.         

\section{Conclusion}
\label{sec:conclusion}
In this work, we investigated the applicability and performance of  uncertainty propagation in 3-D hemodynamics simulations based on a bi-fidelity surrogate proposed and developed in \cite{narayan2014stochastic, zhu2017multi, zhu2014computational}. Unlike the existing work mainly based on coarse meshes, we explored the different options of low-fidelity models, such as 2-D model and unconverged solutions. A  novel empirical error bound estimation is introduced to access the approximation quality of the bi-fidelity surrogate, which is simple to compute and provides more insights and guidance for the practical applications, beyond the target applications.  Three cardiovascular flow cases, including a patient-specific case are investigated to demonstrate the merits of the bi-fidelity approach and the effectiveness of the simple empirical error bound. In the future, we plan to investigate the performance of this approach on the flow cases with more complex physical patterns and improve the error bound estimation.

\section*{Compliance with Ethical Standards}
Conflict of Interest: The authors declare that they have no conflict of interest.
\section*{Aknowledgement}
J.-X Wang would acknowledge partial support from the Defense Advanced Research Projects Agency (DARPA) under the Physics of Artificial Intelligence (PAI) program (contract HR00111890034). X. Zhu was supported by the Simons Foundation (504054).

\section*{Appendix A: Computational Cost for Uncertainty Propagation}
\begin{table}[H]
\begin{center}
\footnotesize{
\begin{tabular}{ |c|cccc| } 
\hline
\multicolumn{5}{|c|} {\textbf{Idealized stenosis (Case 1)}}\\
\hline
m = 6, M =1000, N = 600 & Train (HF) & Train (LF) & Test & Total\\ 
\hline
Bi-fidelity MC&$\approx5.5$&$\approx0.28$&$\approx0.17$&$\approx5.95$\\
\hline
Pure high-fidelity MC& - & - &$\approx 550$&$\approx 550$\\
\hline
\multicolumn{5}{|c|} {\textbf{Idealized bifurcation aneurysm (case 2)}}\\
\hline
m = 25, M =100, N = 250 & Train (HF) & Train (LF) & Test & Total\\ 
\hline
Bi-fidelity MC&$\approx27.1$&$\approx0.7$&$\approx1.74$ &$\approx29.54$\\
\hline
Pure high-fidelity MC & - & - &$\approx 271$&$\approx 271$\\
\hline
\multicolumn{5}{|c|} {\textbf{Patient-specific cerebral aneurysm (case 3)}}\\
\hline
m = 40, M = 2000, N = 600 & Train (HF) & Train (LF) & Test & Total\\ 
\hline
Bi-fidelity MC&$\approx12.22$&$\approx0.33$&$\approx1.74$&$\approx14.29$\\
\hline
Pure high-fidelity MC & - & - &$\approx 183.33$&$\approx 183.33$\\
\hline
\end{tabular}
\caption{Training and testing performance of the BF surrogate for all cases, where m and M are the total numbers of HF and LF samples used for BF training, and N is the number of parameter points (MC samples) for testing. The time unit is the CPU-hour}
\label{tab:apptab1}
}
\end{center}
\end{table}

\end{document}